\documentclass[a4paper,11pt]{article}
\pdfoutput=1 

\usepackage{jheppub} 

\usepackage[T1]{fontenc} 

\usepackage[dvipsnames]{xcolor}

\usepackage{enumitem}

\usepackage{mathrsfs}

\usepackage{subcaption}

\newcommand{\bei}{\begin{itemize}}
\newcommand{\eei}{\end{itemize}}
\newcommand{\bee}{\begin{enumerate}}
\newcommand{\eee}{\end{enumerate}}
\newcommand{\beeL}{\begin{enumerate}[label=(\Alph*)]}
\newcommand{\beel}{\begin{enumerate}[label=(\alph*)]}
\newcommand{\beeR}{\begin{enumerate}[label=(\Roman*)]}
\newcommand{\beer}{\begin{enumerate}[label=(\roman*)]}
\newcommand{\beeLd}{\begin{enumerate}[label=\Alph*.]}
\newcommand{\beeld}{\begin{enumerate}[label=\alph*.]}
\newcommand{\beeRd}{\begin{enumerate}[label=\Roman*.]}
\newcommand{\beerd}{\begin{enumerate}[label=\roman*.]}
\newcommand{\nn}{\nonumber}
\newcommand{\la}{\label}

\def\ka{{\kappa}}
\def\bR{{\mathbb R}}
\def\bZ{{\mathbb Z}}

\def\ka {{\kappa}}
\def\p{\phi}

\def\x {{\text{x}}}
\def\uu {{\text{u}}}

\usepackage{tikz}
\usetikzlibrary{math}
\usetikzlibrary{arrows,shapes,positioning}
\usetikzlibrary{decorations.markings}
\tikzstyle arrowstyle=[scale=1]
\tikzstyle directed=[postaction={decorate,decoration={markings,
    mark=at position .65 with {\arrow[arrowstyle]{stealth}}}}]
\tikzstyle reverse directed=[postaction={decorate,decoration={markings,
    mark=at position .65 with {\arrowreversed[arrowstyle]{stealth};}}}]

\newlength{\mywidth}
\newcommand{\bal}{\begin{equation}\begin{aligned}}
\newcommand{\eal}{\end{aligned}\end{equation}}

\newcommand{\ov}{\over}

\newcommand{\B}{{\scriptscriptstyle\text{B}}}
\newcommand{\F}{{\scriptscriptstyle\text{F}}}
\renewcommand{\L}{{\scriptscriptstyle\text{L}}}
\newcommand{\R}{{\scriptscriptstyle\text{R}}}

\newcommand{\genl}[1]{\mathbf{#1}}
\newcommand{\genr}[1]{\widetilde{\mathbf{#1}}}
\newcommand{\phys}{|\text{phys}\rangle}
\newcommand{\hws}[2]{|\phi_{\scriptscriptstyle\text{#2}}^{\scriptscriptstyle\text{#1}}\rangle}
\newcommand{\lws}[2]{|\varphi_{\scriptscriptstyle\text{#2}}^{\scriptscriptstyle\text{#1}}\rangle}
\newcommand{\repr}[2]{\rho_{\scriptscriptstyle\text{#2}}^{\scriptscriptstyle\text{#1}}}
\newcommand{\de}{\text{d}}
\renewcommand{\x}{\mathsf{x}}
\renewcommand{\uu}{\mathsf{u}}

\newcommand{\phase}{\sigma}

\title{\boldmath On mixed-flux worldsheet scattering in AdS$_3$/CFT$_2$}

\author[1]{Sergey Frolov,}
\author[2,3]{Davide Polvara,}
\author[2,3,4]{Alessandro Sfondrini}

\affiliation[1]{Hamilton Mathematics Institute and School of Mathematics Trinity College, Dublin 2, Ireland.}
\affiliation[2]{Dipartimento di Fisica e Astronomia, Universit\`a degli Studi di Padova, via Marzolo 8,
35131 Padova, Italy.}
\affiliation[3]{Istituto Nazionale di Fisica Nucleare, Sezione di Padova, via Marzolo 8, 35131 Padova,
Italy.}
\affiliation[4]{Institute for Advanced Study, Einstein Drive, Princeton, New Jersey, 08540 USA.}

\emailAdd{frolovs@maths.tcd.ie}
\emailAdd{davide.polvara@unipd.it}
\emailAdd{alessandro.sfondrini@unipd.it}

\abstract{
Strings on $AdS_3\times S^3\times T^4$ with mixed Ramond-Ramond and Neveu-Schwarz-Neveu-Schwarz flux are known to be classically integrable. This is a crucial property of this model, which cannot be studied by conventional worldsheet-CFT techniques. Integrability should carry over to the quantum level, and the worldsheet S~matrix in the lightcone gauge is known up to the so-called dressing factors.
In this work we study the kinematics of mixed-flux theories and  consider a relativistic limit of the S~matrix whereby we can complete the bootstrap program, including the dressing factors for fundamental particles and bound states. This provides an important test for the dressing factors of the full worldsheet model, and offers new insights on the features of the model when the amount of NSNS flux is low.
}

\begin{document} 
\maketitle
\flushbottom

\section{Introduction}
\label{sec:introduction}
The holographic correspondence between strings on $AdS_3$ spaces and two-dimensional superconformal field theories stands out from other AdS/CFT setups~\cite{Maldacena:1997re}. In this case, it is possible to continuously interpolate between supergravity backgrounds supported by a Neveu-Schwarz-Neveu-Schwarz (NSNS) $B$-field and field strength $H=dB$, but without any Ramond-Ramond (RR) fluxes, to a supergravity background with RR fluxes but no $B$-field.%
\footnote{See refs.~\cite{Larsen:1999uk,OhlssonSax:2018hgc} for a discussion of moduli in the AdS3/CFT2 correspondence.}
This gives rise to a one-parameter family of ``mixed-flux'' backgrounds; all of them are related to each other by S-duality, which is non-perturbative in the string coupling~$g_s$. In fact, the perturbative-string description of observables such as the string spectrum (for generic, non-protected states) is quite different as one tunes the ratio of NSNS and RR fluxes.

The simplest setup, which we will consider in this paper, is given by the $AdS_3\times S^3\times T^4$ geometry. The case of NSNS fluxes only is the best understood, as the worldsheet CFT is given by a level-$k$ (supersymmetric) Wess-Zumino-Novikov-Witten model on the worldsheet which can be solved~\cite{Maldacena:2000hw}. The energy levels can be written in a closed form and, like for \textit{e.g.}~flat-space strings, are very degenerate. Only in these cases the dual CFT is understood~\cite{Giribet:2018ada,Eberhardt:2018ouy,Eberhardt:2021vsx}. Turning on the RR flux is believed to lift these degeneracies, though it makes it extremely difficult to study the spectrum by worldsheet-CFT techniques, as the worldsheet model becomes nonlocal (or is accompanied by an intricated system of ghost fields which do not decouple). 

Very remarkably, the classical string non-linear sigma model (NLSM) is intergable for any combination of the fluxes~\cite{Cagnazzo:2012se}. In fact, it is believed that this integrability carries over to the quantum level too when the string is quantised in a suitable lightcone gauge, see \textit{e.g.}~\cite{Sfondrini:2014via}. This is what happens for $AdS_5$ and $AdS_4$ strings, see~\cite{Arutyunov:2009ga,Beisert:2010jr}. The study of quantum integrability is typically done in several steps. The first step is to consider the gauge-fixed model on a plane (the decompactified string worldsheet) and fix its scattering matrix there through symmetries. If the model is integrable, it is sufficient to fix the two-to-two S~matrix, as higher processes follow from it.
Then, much of the S~matrix can be fixed by considering the linearly-realised symmetries of the gauge-fixed model. This was done in~\cite{Hoare:2013ida,Lloyd:2014bsa} building also on perturbative and semiclassical considerations~\cite{Hoare:2013pma,Hoare:2013lja}. This leaves some pre-factors, the so-called \textit{dressing factors}, undetermined. They can only be fixed under some assumptions on the analytic structure of the theory, and by imposing unitarity and crossing symmetry. While this process is relatively simple for relativistic models, it is rather subtle for non-relativistic ones, such as the ones arising on the string worldsheet, see~\cite{Arutyunov:2004vx,Janik:2006dc,Beisert:2006ez}. 

In the case of $AdS_3\times S^3\times T^4$, there is a recent proposal for the dressing factors for pure-RR~\cite{Frolov:2021fmj} and pure-NSNS~\cite{Baggio:2018gct} theories, but not for the generic mixed-flux ones. The main reason is that the underlying analytic structure is quite mysterious and unique. In this paper we study the mixed-flux models in a limit where they become relativistic. Then, we are able to uniquely fix the S~matrix including the dressing factors by imposing analyticity and consistency with the bound-state content of the model. A similar limit was studied in the past in~\cite{Fontanella:2019ury}. However, both the detailed definition of the limit and the conclusions reached differ quite substantially. In that case, the authors constructed a theory of only massless relativistic excitations. Here instead we find a model of massive and massless particles. More specifically, we have $(k-1)$ distinct massive particle representations, where $k$ is the (quantised) NSNS flux. These massive particles correspond to the limit of the $AdS_3\times S^3$ massive excitations and to their bound states. Two more representations are massless, and they are related to the $T^4$~modes. The resulting model is of interest in and of itself, and closely related to that of Fendley and Intriligator~\cite{Fendley:1991ve,Fendley:1992dm}; importantly, it provides a check for future proposals of the mixed-flux dressing factors of the full worldsheet model.

This paper is structured as follows. In section~\ref{sec:fulltheory} we review the main properties of mixed-flux theories, including some features which were not previously discussed in the literature, such as their bound states and the analytic structure of the rapidity plane. That structure is further detailed in appendix~\ref{app:kappaZhukovsky}, while appendix~\ref{app:Smatrices} contains a list of the S-matrix elements of~\cite{Lloyd:2014bsa} in the conventions of this paper. In section~\ref{sec:relativistic} we discuss the relativistic limit, and carry it out at the level of the algebra and of the S-matrix elements. By using crossing and analyticity we then fix the dressing factors in section~\ref{sec:crossing}. In appendix~\ref{app:relativisticS} we perform the S-matrix bootstrap for the relativistic model assuming only its symmetries and particle contents (without taking the limit of the S-matrix elements); interestingly, for specific processes, this allows for more general solutions. We conclude in section~\ref{sec:conclusions}.

\section{Worldsheet scattering for mixed-flux theories}
\label{sec:fulltheory}
The all-loop scattering for $AdS_3\times S^3\times T^4$ strings supported by a mixture of RR and NSNS fluxes was studied in~\cite{Lloyd:2014bsa} and the results were comprehensively summarised in~\cite{Eden:2021xhe}. Here we follow the notation of~\cite{Eden:2021xhe} and refer the reader there for further details.

\subsection{Symmetries and fundamental particles}
\label{sec:fulltheory:symmetries}
The dispersion relation of the model is~\cite{Hoare:2013lja}
\begin{equation}
\label{eq:dispersion}
    E(m,p)= \sqrt{\left(m+\frac{k}{2\pi}p\right)^2+4h^2\sin^2\frac{p}{2}}\,.
\end{equation}
Here $k=0,1,2,\dots$ and $h\geq0$ are parameters of the theory, corresponding to the strength of the NSNS and RR background fluxes, respectively. The parameters $m,p$ identify the various fundamental particles of the model as it can be found in a near-pp-wave expansion~\cite{Berenstein:2002jq,Hoare:2013pma} (that is, at small~$p$ and large string tension). In particular we have two bosons on $AdS_3$, which have $m=\pm1$, two bosons on $S^3$, which also have $m=\pm1$, and four bosons on $T^4$, all of which have $m=0$. We also have as many fermions, with the same values of~$m$. In fact, these particles arrange themselves in supermultiplets. Let us briefly review this structure. 

The full model has a bosonic $so(2,2)\oplus so(4)$ symmetry from $AdS_3\times S^3$, which can be split as $(su(1,1)_{\L}\oplus su(1,1)_{\R})\oplus(su(2)_{\L}\oplus su(2)_{\R})$.%
\footnote{The $T^4$ part also enjoys four $so(2)$ shift-symmetries and a local $so(4)\cong su(2)_{\bullet}\oplus su(2)_{\circ}$ rotation symmetry.} For future convenience, let us introduce the four Cartan elements of $(su(1,1)_{\L}\oplus su(1,1)_{\R})\oplus(su(2)_{\L}\oplus su(2)_{\R})$,
\begin{equation}
    \genl{L}_0,\quad\genr{L}_0,\qquad
    \genl{J}^3,\quad\genr{J}^3.
\end{equation}
This factorisation between ``left'' and ``right'' extends to the whole superalgebra, which takes the form $psu(1,1|2)_{\L}\oplus psu(1,1|2)_{\R}$.
In the notation of~\cite{Eden:2021xhe}, the BPS bounds of this algebra are%
\footnote{
Here $\genl{L}_0$ is the Cartan element of $su(1,1)_{\L}$; the minus sign is such that its eigenvalues are bounded from below, rather than from above, on unitary representations. Similarly for $\genr{L}_0$.
}
\begin{equation}
    \genl{H}\equiv-\genl{L}_0-\genl{J}^3\geq0\,,\qquad
    \genr{H}\equiv-\genr{L}_0-\genr{J}^3\geq0\,.
\end{equation}
In total we have eight supercharges (with dimension $+1/2$) which we will denote by $\genl{Q}$ and $\genr{Q}$, and eight superconformal generators (with dimension $-1/2$), denoted by $\genl{S}$ and $\genr{S}$.
We are mostly interested in the symmetries that commute with the lightcone Hamiltonian
\begin{equation}
    \genl{E}= \genl{H}+\genr{H}\,.
\end{equation}
These include the Cartan elements, which we arrange in the following combinations%
\footnote{Equivalently, instead of considering $\Delta\genl{J}$ we could have considered $\genl{B}\equiv-(\genl{L}_0-\genr{L}_0)+(\genl{J}^3-\genr{J}^3)$ which is orthogonal to $\genl{M}$.}
\begin{equation}
    \genl{M}\equiv \genl{H}-\genr{H}\,,\qquad
    \Delta\genl{J}\equiv-\genl{J}^3+\genr{J}^3\,,
\end{equation}
and the lightcone momentum
\begin{equation}
    \genl{P_+}= -\genl{L}_0-\genr{L}_0+\genl{J}^3+\genr{J}^3\,,
\end{equation}
which is related to the worldsheet length~\cite{Arutyunov:2009ga} and decouples in the limit where the worldsheet is a plane (which is where we will work to discuss the S~matrix).
Half of the supercharges (for a total of eight) commutes with~$\genl{E}$ and forms the algebra~\cite{Borsato:2013qpa}
\begin{equation}
\label{eq:lcalgebra}
\begin{aligned}
\{\genl{Q}^A,\,\genl{S}_B\}= \delta^A_B\,\genl{H}\,,\qquad
\{\genl{Q}^A,\,\genr{Q}_B\}= \delta^A_B\,\genl{C}\,,\\
\{\genr{Q}^A,\,\genr{S}_B\}= \delta^A_B\,\genr{H}\,,\qquad
\{\genl{S}_A,\,\genr{S}^B\}= \delta^A_B\,\genl{C}^\dagger\,.
\end{aligned}
\end{equation}
The generator $\Delta\genl{J}$ (or equivalently $\genl{B}$) acts as an automorphism on the fermionic generators, because they carry spin.
The new central charges $\genl{C},\genl{C}^\dagger$ are akin to Beisert's central extension~\cite{Beisert:2005tm,Arutyunov:2006ak} and in this case couple the left- and right-superalgebras. We stress that these supercharges are a feature of the lightcone-gauge-fixed model; they act nontrivially on unphysical states (\textit{e.g.}, on a single-particle state of momentum $p$) and must annihilate physical states. More precisely, a physical state~$\phys$ must obey
\begin{equation}
\label{eq:physicalstate}
\begin{gathered}
    \genl{E}\phys= E\phys\quad \text{with}\ E\geq0\,,\qquad
    \genl{M}\phys= M\phys\quad \text{with}\ M\in\mathbb{Z}\,,\\
    \genl{C}\phys=\genl{C}^\dagger\phys=0\,.
\end{gathered}
\end{equation}
Generic (multi-particle) physical states are build out of several (single-particle) \textit{unphysical} states. From a near-pp-wave~\cite{Berenstein:2002jq,Hoare:2013pma} and semiclassical analysis~\cite{Hoare:2013lja,Lloyd:2014bsa} we expect that the central charges can be written in terms of the worldsheet momentum $\genl{p}$ as
\begin{equation}
    \genl{C}=\frac{ih}{2}\left(e^{i\genl{p}}-1\right),\qquad
    \genl{C}^\dagger=\frac{ih}{2}\left(1-e^{-i\genl{p}}\right),\qquad
    \genl{M}=\frac{k}{2\pi}\genl{p}+m\,.
\end{equation}
In this formula, $m\in\mathbb{Z}$ distinguishes different representations.
This is nicely consistent with~\eqref{eq:physicalstate} by using the physical-state condition which comes from level matching,
\begin{equation}
    \genl{p}\phys= p \phys\,,\qquad
    \text{with}\ p=0\mod2\pi\,.
\end{equation}
The form of the energy~\eqref{eq:dispersion} finally follows from the shortening condition~\cite{Borsato:2012ud}
\begin{equation}
\label{eq:shortening}
    \genl{H}\,\genr{H}=\genl{C}^\dagger\genl{C}\,.
\end{equation}
Finally, remark that the value of $m$ can be read off by considering a zero-momentum state, in which case $M=m$. However, we immediately notice a possible subtlety due to the fact that the eigenvalues $M$ of $\genl{M}$ (and indeed of all central charges) are invariant under a simultaneous shift of $p$ and $m$,
\begin{equation}
\label{eq:periodicity}
    M(m,p)=M(m+k,p-2\pi)\,.
\end{equation}
This leads to an ambiguity. In fact, while fundamental particles have $m=\pm1,0$ we expect that \textit{bound states} may exist, with larger values of~$|m|$. Eq.~\eqref{eq:periodicity} seems to suggest that most of the possible bound states (indeed, all but $k$ particles) can be obtained by ``boosting''%
\footnote{We use the term ``boost'' loosely, as the model in not relativistic on the worldsheet.}
the momentum of a finite set of particles. This was observed in the $h=0$ limit of this model~\cite{Sfondrini:2020ovj}, where it can be described by a level-$k$ WZNW model. However, at generic $k$ and $h$ it is not immediately obvious whether $p$ should be allowed to take any real value, or should be bounded in the interval $[0,2\pi]$ as it seems to be the case semiclassically~\cite{Hoare:2013lja}.%
\footnote{In~\cite{Hoare:2013lja} the range of the momentum is actually taken to be~$[-\pi,\pi]$ but from their eq.~(3.32), using the principal branch of the logarithm, the domain actually is~$[0,2\pi]$.}
To understand better this issue, we review the structure of the fundamental representations and of possible bound-state representations.

\subsection{Fundamental-particle representations}
\label{sec:fulltheory:fundamental}

\begin{table}[t]
\centering
 \begin{tabular}{|l |l| c|} 
 \hline
 $(m=+1)$& \textbf{State} &  $\Delta \genl{J}$ \\ [0.5ex] 
 \hline
 $S^3$ bos. & $|Y(p)\rangle$ & $+1$  \\[0.5ex]  
 ferm. & $|\Psi^{A}(p)\rangle$ & $+\tfrac{1}{2}$  \\[0.5ex] 
 $AdS_3$ bos. & $|Z(p)\rangle$ & $0$\\[0.5ex] 
 \hline
 \end{tabular}
  \begin{tabular}{|l |l |c|} 
 \hline
 $(m=-1)$& \textbf{State} &  $\Delta \genl{J}$ \\[0.5ex] 
 \hline
 $AdS_3$ bos. & $|\bar{Z}(p)\rangle$ & $0$  \\[0.5ex]  
 ferm. & $|\bar{\Psi}^{A}(p)\rangle$ & $-\tfrac{1}{2}$  \\[0.5ex] 
 $S^3$ bos. & $|\bar{Y}(p)\rangle$ & $-1$\\[0.5ex] 
 \hline
 \end{tabular}
  \begin{tabular}{|l |l |c|} 
 \hline
 $(m=0)$& \textbf{State} &  $\Delta \genl{J}$ \\[0.5ex]
 \hline
 ferm. & $|\chi^{\dot{A}}(p)\rangle$ & $+\tfrac{1}{2}$  \\[0.5ex]  
 $T^4$ bos. & $|T^{\dot{A}A}(p)\rangle$ & $0$  \\[0.5ex] 
 ferm & $|\tilde{\chi}^{\dot{A}}(p)\rangle$ & $-\tfrac{1}{2}$\\[0.5ex] 
 \hline
 \end{tabular}
 \caption{A summary of the representations under which the eight bosons and eight fermions trasfer. In each table, the top state is the highest-weight state of the representation and the bottom is the lowest-weight state. We see that $\Delta\genl{J}$ decreases along the representation, while $M=m+\tfrac{k}{2\pi}p$ is constant for a given worldsheet momentum~$p$. As the particles are identified perturbatively, we implicitly take $p$~small.}
 \label{tab:fundamentalparticles}
\end{table}

In light-cone gauge we expect the theory to feature eight bosons and eight fermions. They fit into four four-dimensional representations of the algebra~\eqref{eq:lcalgebra} and are described in table~\ref{tab:fundamentalparticles}. The lowering operators of the algebra, in this notation, are $\genl{Q}^A$ and $\genr{S}^A$. Their action is proportional, $\genl{Q}^A\sim\genr{S}^A$, because the particles transform in a short representation, \textit{cf.}~\eqref{eq:shortening}.
In fact, all short representations of~\eqref{eq:lcalgebra} are necessarily four-dimensional. This means that supersymmetric bound-state representations, if they exist, should have a form similar to the representations of table~\ref{tab:fundamentalparticles} up to tweaking the values of~$m$ and the eigenvalue of~$\Delta \genl{J}$.

In practice, to discuss the precise form of the representations it is convenient to introduce a smaller algebra, generated by four supercharges,
\begin{equation}
\label{eq:lcalgebrasmall}
\{\genl{q},\,\genl{s}\}= \genl{H}\,,\quad
\{\genr{q},\,\genr{s}\}= \genr{H}\,,\qquad
\{\genl{q},\,\genr{q}\}= \genl{C}\,,\quad
\{\genl{s},\,\genr{s}\}= \genl{C}^\dagger.
\end{equation}
Clearly
\begin{equation}
    \genl{Q}^1=\genl{q}\otimes\genl{1},\quad
    \genl{Q}^2=\Sigma\otimes\genl{q},\qquad
    \genl{S}_1=\genl{s}\otimes\genl{1},\quad
    \genl{S}_2=\Sigma\otimes\genl{s},
\end{equation}
where $\Sigma$ is the fermion sign, and similarly for $\genr{Q}_A$ and~$\genr{S}^A$.
The four-dimensional short representations of~\eqref{eq:lcalgebra} arise as tensor products of two-dimensional representations of~\eqref{eq:lcalgebrasmall}. These smaller representations will depend on the values of $m$, $p$, and on whether the highest-weight state is a boson or a fermion. We define
\begin{equation}
    \hws{*}{*}=\text{highest-weight state},\qquad
    \lws{*}{*}=\text{lowest-weight state},
\end{equation}
where $*$ will be used to distinguish whether the state is  a boson (``B'') or a fermion (``F''), and to indicate its kinematics, which we will denote by left or right (``L'' or~``R''); the label left will be reserved to $m>0$, while right will be reserved to $m<0$.
The representation $\repr{B}{L}(m,p)=(\hws{B}{L},\lws{F}{L})$ (suppressing the $m,p$ dependence of the states) is given by
\begin{equation}
\label{eq:reprBL}
\begin{aligned}
\genl{q}\,\hws{B}{L}= a_{\L}(m,p)\,\lws{F}{L}\,,\qquad
\genr{s}\,\hws{B}{L}= \bar{b}_{\L}(m,p)\,\lws{F}{L}\,,\\
\genl{s}\,\lws{F}{L}= \bar{a}_{\L}(m,p)\,\hws{B}{L}\,,\qquad
\genr{q}\,\lws{F}{L}= b_{\L}(m,p)\,\hws{B}{L}\,,
\end{aligned}
\end{equation}
and $\repr{F}{L}(m,p)=(\hws{F}{L},\lws{B}{L})$ is given by 
\begin{equation}
\label{eq:reprFL}
\begin{aligned}
\genl{q}\,\hws{F}{L}= a_{\L}(m,p)\,\lws{B}{L}\,,\qquad
\genr{s}\,\hws{F}{L}= \bar{b}_{\L}(m,p)\,\lws{B}{L}\,,\\
\genl{s}\,\lws{B}{L}= \bar{a}_{\L}(m,p)\,\hws{F}{L}\,,\qquad
\genr{q}\,\lws{B}{L}= b_{\L}(m,p)\,\hws{F}{L}\,,
\end{aligned}
\end{equation}
with precisely the same representation coefficients. The representations $\repr{B}{R}(m,p)$ and $\repr{F}{R}(m,p)$ have a similar form, up to replacing the representation coefficients with $a_{\R}$, $b_{\R}$ and so on. 
To write the representation coefficients explicitly it is convenient to define the Zhukovsky variables
\begin{equation}
\label{eq:zhukovsky}
\begin{aligned}
    x^\pm_{\L} (m,p)&\equiv&
    \frac{+M(m,p)+E(m,p)}{2h\sin(p/2)}e^{\pm i \frac{p}{2}}\,,\qquad(m\ge 0)\,,\\
    x^\pm_{\R} (m,p)&\equiv&
    \frac{-M(m,p)+E(m,p)}{2h\sin(p/2)}e^{\pm i \frac{p}{2}}\,,\qquad(m<0)\,.
\end{aligned}
\end{equation}
This notation has the advantage of reducing to the usual RR notation~\cite{Borsato:2012ud} if~$k=0$.
The Zhukovsky variables satisfy (omitting the dependence on $m,p$)
\begin{equation}
    \frac{x^+_{*}}{x^-_{*}}=e^{ip},\qquad
    x^+_{*}-\frac{1}{x^+_{*}}-x^-_{*}+\frac{1}{x^-_{*}}=\frac{2i}{h}E\,,
\end{equation}
and
\begin{equation}
    x^+_{\L}+\frac{1}{x^+_{\L}}-x^-_{\L}-\frac{1}{x^-_{\L}}=+\frac{2i}{h}M\,,
    \qquad
    x^+_{\R}+\frac{1}{x^+_{\R}}-x^-_{\R}-\frac{1}{x^-_{\R}}=-\frac{2i}{h}M\,,
\end{equation}
To define the representation coefficients we introduce
\begin{equation}
    \eta_{*}(m,p)=\sqrt{\frac{ih}{2}\left(x^-_*(p)-x^+_*(p)\right)}\,,
\end{equation}
and
\begin{equation}
\begin{aligned}
&a_{\L}=\eta_{\L}\,,\qquad b_{\L}=-\frac{\eta_{\L}}{x^-_{\L}}\,,\qquad
\bar{a}_{\L}=\eta_{\L}\,,\qquad \bar{b}_{\L}=-\frac{\eta_{\L}}{x^+_{\L}}\,,\\
&b_{\R}=\eta_{\R}\,,\qquad a_{\R}=-\frac{\eta_{\R}}{x^-_{\R}}\,,\qquad
\bar{b}_{\R}=\eta_{\R}\,,\qquad \bar{a}_{\R}=-\frac{\eta_{\R}}{x^+_{\R}}\,.
\end{aligned}
\end{equation}
It is easy to verify that this defines the representations introduced before. It remains to define the representations with $m=0$. This can be equivalently done by taking $m\to0$ in either the left or the right representations. The result of this limit is not identical, but it yields two isomorphic representations.

Such short representations may be defined for any~$m\in\mathbb{Z}$ and~$p\in\mathbb{R}$. Not all of these representations, however, appear in the string model. In fact, the representations to which the fundamental particles of the full theory belong are%
\footnote{It should be noted that there are many different (isomorphic) ways to obtain the massless representations. Here we are following the notation of~\cite{Eden:2021xhe}.}
\begin{equation}
\label{eq:fundamentalrepr}
\begin{aligned}
    m=+1:&\qquad
    \repr{B}{L}(+1,p)\otimes\repr{B}{L}(+1,p)\,,\\
    m=-1:&\qquad
    \repr{F}{R}(-1,p)\otimes\repr{F}{R}(-1,p)\,,\\
    m=0:&\qquad
    \Big(\repr{B}{L}(0,p)\otimes
    \repr{F}{L}(0,p)\Big)\oplus\Big(\repr{F}{L}(0,p)\otimes\repr{B}{L}(0,p)\Big).
\end{aligned}
\end{equation}
We will also see that it is natural to restrict $p\in[0,2\pi]$. Still, these are not all representations of the model. Additional ones, which can be constructed as bound states, can be obtained by taking $m=+2,+3,\dots$ for bound states of left particles, and $m=-2,-3,\dots$ for bound states of right particles. We will construct these representations in section~\ref{sec:fulltheory:boundstates}.

\subsection{Multi-particle representations}
\label{sec:fulltheory:multiparticle}
The states of the theory will in general feature several particles over the Fock vacuum. The most important case will be that of two-particle representations, which we will use to construct the S~matrix. Multi-particle representations may be constructed out of the single-particle ones by means of a coproduct. For two-particle states we have
\begin{equation}
\label{eq:coproductfull}
\begin{aligned}
    \genl{Q}^A(p_1,p_2)&= \genl{Q}^A(p_1)\otimes \mathbf{1}+e^{+\frac{i}{2}p_1}\Sigma\otimes \genl{Q}^A(p_2)\,,\\
    \genr{Q}_A(p_1,p_2)&= \genr{Q}_A(p_1)\otimes \mathbf{1}+e^{+\frac{i}{2}p_1}\Sigma\otimes \genr{Q}_A(p_2)\,,\\
    \genl{S}_A(p_1,p_2)&= \genl{S}_A(p_1)\otimes \mathbf{1}+e^{-\frac{i}{2}p_1}\Sigma\otimes \genl{S}_A(p_2)\,,\\
    \genr{S}^A(p_1,p_2)&= \genr{S}^A(p_1)\otimes \mathbf{1}+e^{-\frac{i}{2}p_1}\Sigma\otimes \genr{S}^A(p_2)\,.
\end{aligned}
\end{equation}
This induces the co-product for the central charges, and it clearly trivial for $\genl{E},\genl{M}$ and non-trivial for $\genl{C},\genl{C}^\dagger$. In fact, this coproduct is necessary to ensure that the eigenvalues of $\genl{C},\genl{C}^\dagger$ depend only on the \textit{total} momentum, even in the case of multi-particle states. Here again $\Sigma=(-1)^F$ is the Fermion sign. Clearly these two-particle representations may be defined out of any pair of fundamental particle representations~\eqref{eq:fundamentalrepr}. It is also possible to iterate the construction to obtain three- and more-particle representations, which we will not be needing here.

Using the two-particle representation of the supercharges, it is possible to constrain the two-particle S~matrix~$S(p_1,p_2)$, up to an overall dressing factor for each pair of irreducible representations. The explicit form of the S-matrix elements was derived in~\cite{Lloyd:2014bsa} in slightly different conventions. For completeness, we report it in our conventions in appendix~\ref{app:Smatrices}.

\subsection{The rapidity plane}
\label{sec:fulltheory:uplane}
In order to discuss bound states it is convenient to introduce the rapidity variable~$u$~\cite{Hoare:2013lja}
\begin{equation}
\label{eq:uofx}
    u(x,\kappa)=x+\frac{1}{x}-\frac{\kappa}{\pi}\log x\,,
\end{equation}
from which we can implicitly define $x(u,\kappa)$. 
\begin{figure}
\begin{center}
\includegraphics*[width=1\textwidth]{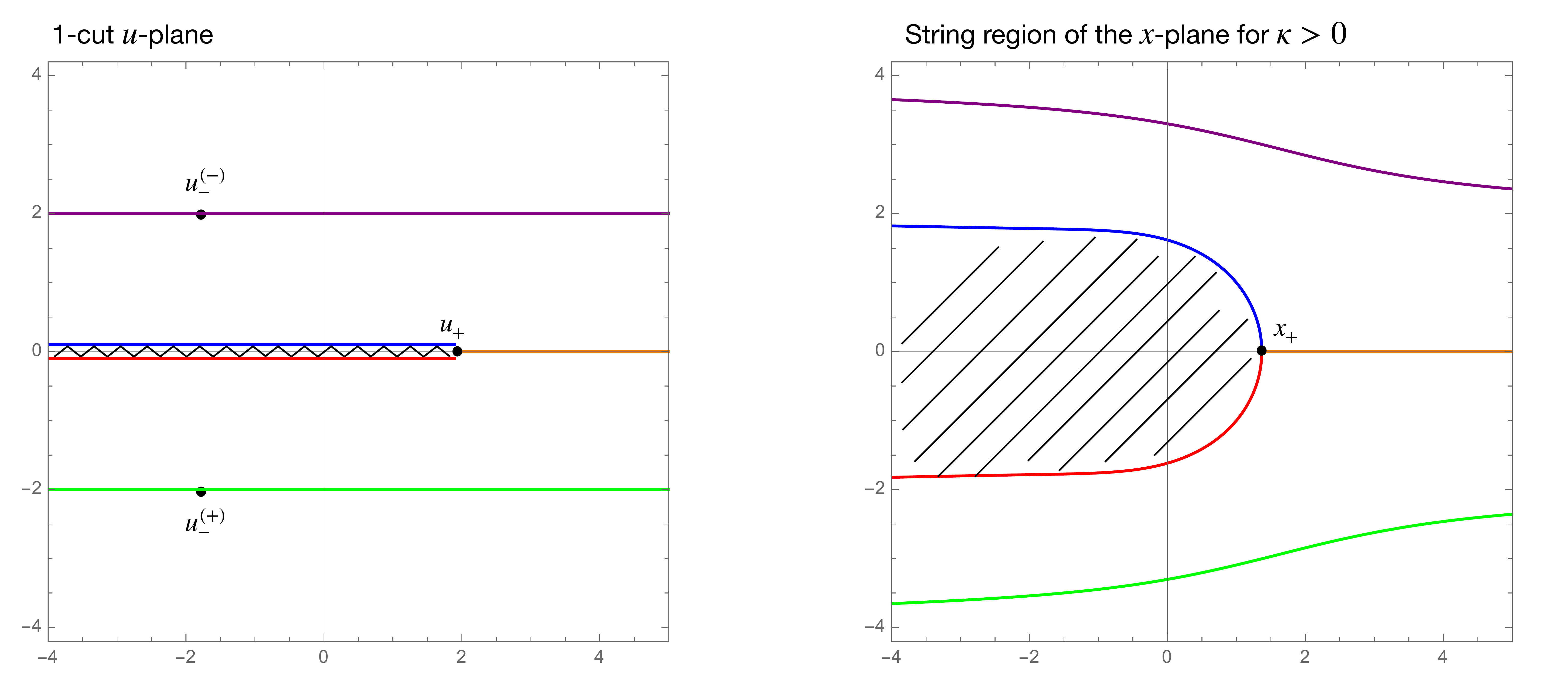}
\end{center}
\caption{The physical region of the $x$-plane (on the right) and associated one-cut $u$-plane (on the left) for $\kappa>0$. The shaded regions are removed from the $x$-plane. 
The different colours show how the $u$-plane on the left is mapped to the corresponding $x$-plane on the right. The zigzag line in the $u$-plane corresponds to a cut. By crossing this cut we end up in the $u$-plane shown in figure~\ref{antistring_u_x_planes_positive_k}, which is mapped to the antistring region of the $x$-plane.}
\label{string_u_x_planes_positive_k}
\end{figure}
In what follows we will use the principal branch $\ln x$ of $\log x$ (with the cut on the negative half-line). 
\begin{figure}
\begin{center}
\includegraphics*[width=1\textwidth]{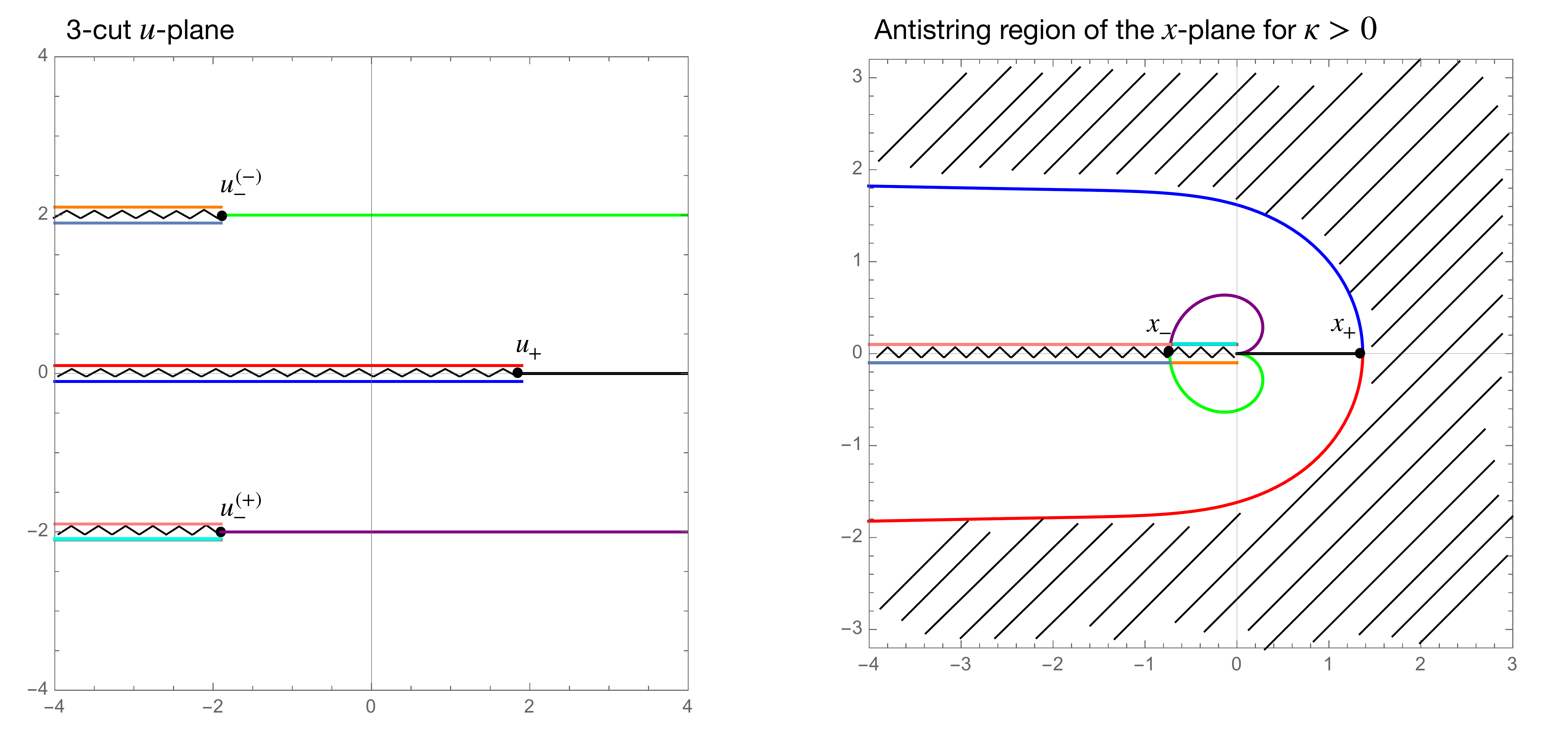} 
\end{center}
\caption{The antistring or crossed region  in the $x$-plane region (on the right) and associated three-cut $u$-plane (on the left) for $\kappa>0$. As in the previous figure, the shaded regions are removed from the $x$-plane and the different colours show the map between the $u$ and $x$ plane. By crossing the cut $(-\infty, u_+)$ in the $u$-plane we return to the $u$-plane depicted in figure~\ref{string_u_x_planes_positive_k}. Similar conventions are used to show the map between $u$ and $x$ planes in figures~\ref{string_u_x_planes_negative_k} and~\ref{antistring_u_x_planes_negative_k}.}
\label{antistring_u_x_planes_positive_k}
\end{figure}
Going $n$~times around the branch point~$x=0$ leads to a monodromy
\begin{equation}
    x^{(n)}(u,\kappa)= x(u+2i\kappa\, n\,,\,\kappa)\,.
\end{equation}
By comparison with the shortening condition~\eqref{eq:shortening} it is clear that $\kappa$ is real, and we may parametrise 
\begin{equation}
    x^\pm_{\L}=x(u\pm\tfrac{i}{h},+\tfrac{k}{h})\,,\qquad
    x^\pm_{\R}=x(u\pm\tfrac{i}{h},-\tfrac{k}{h})\,.
\end{equation}
In analogy with the usual case ($k=\kappa=0$), we define the string reality condition
\begin{equation}
    x(u,\kappa)^*=x(u^*,\kappa)\,,\qquad
    \kappa^*=\kappa\,\qquad\text{(string).}
\end{equation}
It is also possible to define a mirror conjugation rule, which however swaps ``left'' and ``right'' particles
\begin{equation}
    x(u,\kappa)^*=\frac{1}{x(u^*,-\kappa)}\,,\qquad
    \kappa^*=\kappa\,\qquad\text{(mirror).}
\end{equation}
This is a sign that the mirror theory is not unitary (something that can easily be seen from the dispersion relation too). Here we will mostly focus on the string theory.
It is also worth noting that with these definitions, in string theory
\begin{equation}
    p=-i(\ln x^+_* - \ln x^-_*)\in[0,2\pi)\,.
\end{equation}

\begin{figure}
\includegraphics*[width=1\textwidth]{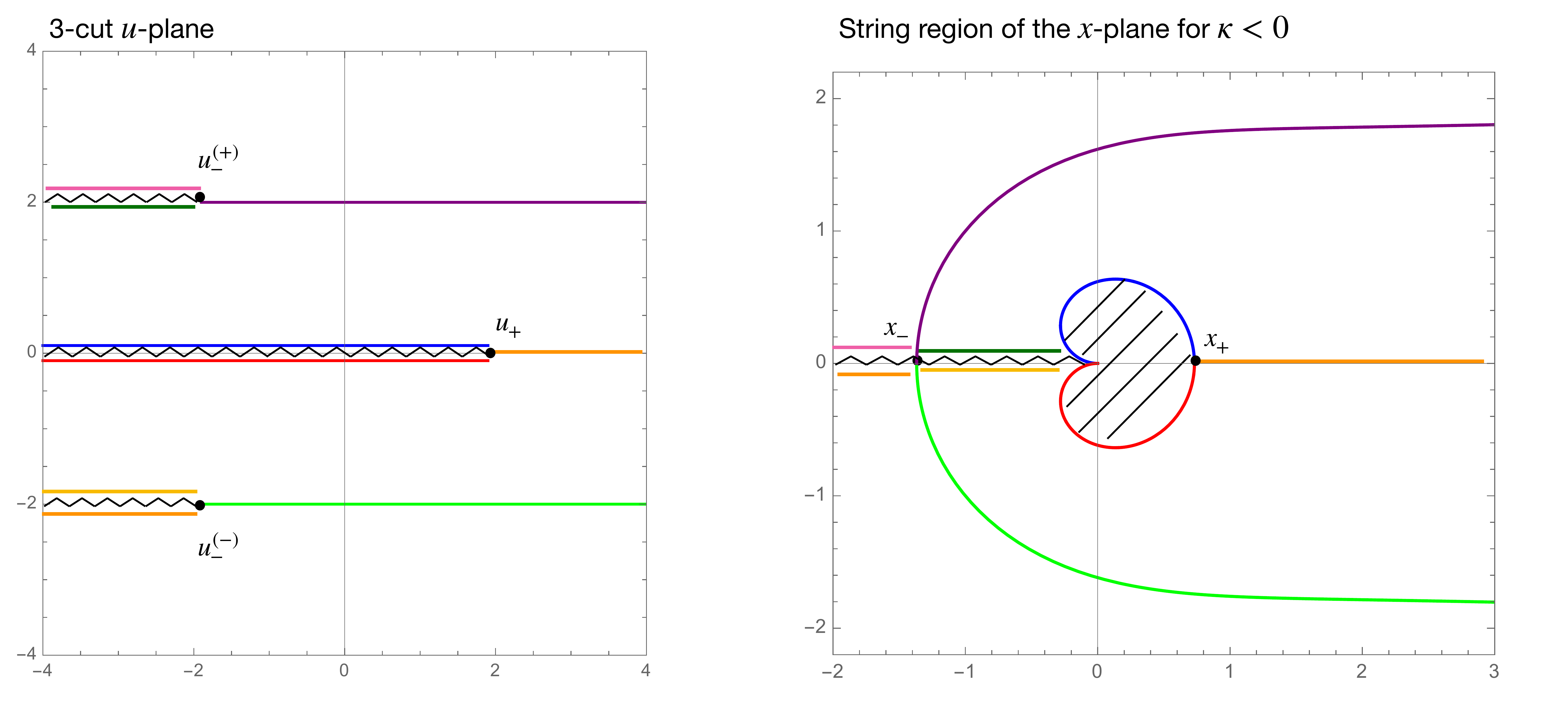}
\caption{The physical region of the $x$-plane  (on the right) and associated three-cut $u$-plane (on the left) for $\kappa<0$. Notice that the number of cuts in the $u$-plane are different when $\kappa<0$ with respect to~$\kappa>0$.}
\label{string_u_x_planes_negative_k}
\end{figure}

\begin{figure}
\includegraphics*[width=1\textwidth]{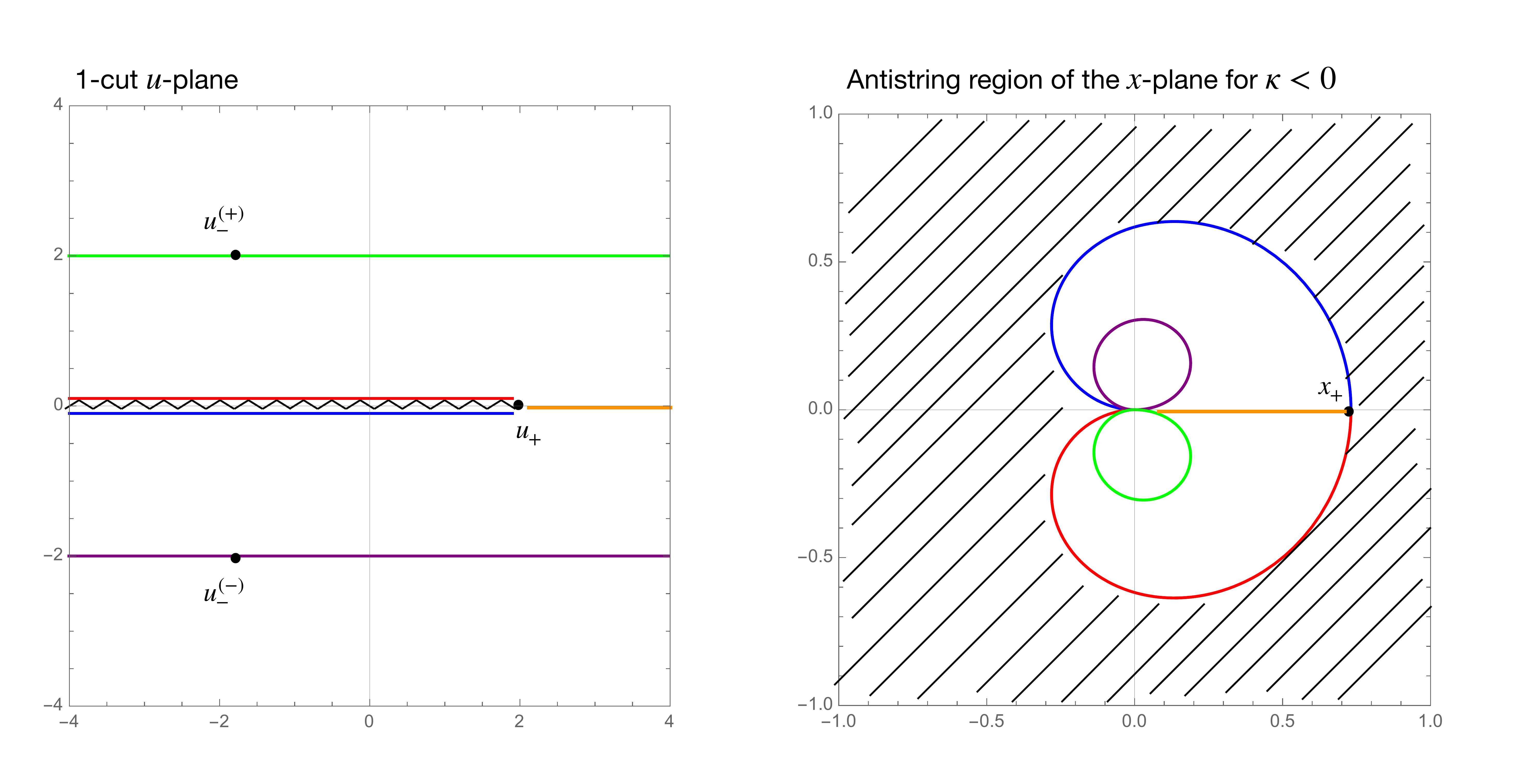} 
\caption{The antistring region of the $x$-plane  (on the right) and associated one-cut $u$-plane (on the left) for $\kappa<0$.}
\label{antistring_u_x_planes_negative_k}
\end{figure}

The branch points of the map $x(u,\kappa)$ can be found from the differential
\begin{equation}
    \frac{\de x}{\de u}=
    \frac{x^2}{(x-\x_+)(x-\x_-)}\,,\qquad
    \x_{\pm}=\frac{\kappa}{2\pi}\pm\sqrt{1+\frac{\kappa^2}{4\pi^2}}\,.
\end{equation}
Similarly to the $\kappa=0$ case, there appear to be two branch points~$\x_\pm$. However, they have \textit{three} images on the $u$-plane.
One, since $\x_+>0$, is
\begin{equation}
    \uu_+=\x_++\frac{1}{\x_+}-\frac{\kappa}{\pi}\ln\x_+ >0\,.
\end{equation}
The remaining pair is
\begin{equation}
\uu_-^{\pm}=-\uu_+ \mp i\kappa\,,
\end{equation}
where the imaginary shift is due to the branch cut of the logarithm. 
Moreover, there is a branch point at $u=\infty$, of logarithmic type.
By taking the branch cuts to be parallel to the real-axis on the $u$-plane, we can easily find their images on the $x$-plane. Recall that in the case $\kappa=k=0$, the images of the branch cuts coincided with the unit circle. Now the picture is more involved, and it depends on whether $\kappa>0$ or $\kappa<0$, as shown in figures~\ref{string_u_x_planes_positive_k}, \ref{antistring_u_x_planes_positive_k}, \ref{string_u_x_planes_negative_k} and~\ref{antistring_u_x_planes_negative_k}. A more detailed description of the map between the $x$ and $u$-plane is reported in appendix~\ref{app:kappaZhukovsky}.

\subsection{Bound states}
\label{sec:fulltheory:boundstates}
Let us now discuss the kinematics of the bound states of this model. 
Let us start by recalling the picture when $\kappa=k=0$. In that case, the eigenvalues of $\genl{M}$ do not depend on the worldhseet momentum, $M=m\in\mathbb{Z}$. Left fundamental particles with $m_1=m_2=+1$ and complex momenta $p_1,p_2$ (and rapidity $u_1, u_2$) can form bound states with $m=+2$, and total momentum $p$ (rapidity $u$). The bound-state condition can be read off the poles of the S~matrix and it is~\cite{Borsato:2013hoa,Seibold:2022mgg}
\begin{equation}
\label{eq:bsconditionkiszero}
x(u_1+\tfrac{i}{h},0)=x(u_2-\tfrac{i}{h},0),\qquad
u=u_1+\tfrac{i}{h}=u_2-\tfrac{i}{h}\,,
\end{equation}
so that the expressions
\begin{equation}
\label{eq:totalEp}
\begin{aligned}
    E(p,m)=&\,\frac{h}{2i}\sum_{j=1}^2\left(
    x(u_j+\tfrac{i}{h},0)-\frac{1}{x(u_j+\tfrac{i}{h},0)}-x(u_j-\tfrac{i}{h},0)+\frac{1}{x(u_j-\tfrac{i}{h},0)}
    \right),\\
    e^{ip}=&\,\frac{x(u_1+\frac{i}{h},0)}{x(u_1-\frac{i}{h},0)}\frac{x(u_2+\frac{i}{h},0)}{x(u_2-\frac{i}{h},0)}\,,
\end{aligned}
\end{equation}
depend simply on $u$ and satisfy the shortening condition with $k=0$ and $m=+2$. This procedure can be iterated to construct bound states with any $m=+2,+3,+4,\dots$, and in a similar way one can construct bound states of right particles with $m=-2,-3,-4,\dots$.
Importantly, left- and right-particles do not form bound states, and neither do particles with $m=0$.

Let us now consider how this picture changes if $\kappa>0$ or $\kappa<0$, so that we can express it in terms of
\begin{equation}
    x(u\pm m \tfrac{i}{h},\kappa)\,.
\end{equation}
The bound-state condition coming from the S~matrix is still the same as~\eqref{eq:bsconditionkiszero}, only it is now expressed in terms of deformed Zhukovsky variables~\eqref{eq:uofx}, namely~\cite{Lloyd:2014bsa}
\begin{equation}
\label{eq:bsconditionkgreater}
x(u_1+\tfrac{i}{h},\pm\tfrac{k}{h})=x(u_2-\tfrac{i}{h},\pm\tfrac{k}{h}),\qquad
u=u_1+\tfrac{i}{h}=u_2-\tfrac{i}{h}\,.
\end{equation}
Notice that either both particles are left, or both particles are right, so that we pick always the same sign for~$\kappa=k/h$, and therefore the condition on $u_1, u_2$ is the same as before, regardless of~$k$. Similarly, the condition for the total energy and total momentum takes a form similar to~\eqref{eq:totalEp}.
This procedure too can be iterated to construct bound states with $m=+2,+3,\dots$ (starting from left particles) or with $m=-2,-3,\dots$ (starting from right particles). The only case which requires some care is when $m=- k$, because  for real values of $u$ such that $u<-\mathsf{u}_+$ the points $u\pm ik/h$ are on the cuts, see figures~\ref{antistring_u_x_planes_positive_k} and~\ref{string_u_x_planes_negative_k}. In this special case the semi-line $u\ge -\mathsf{u}_+$ is in one-to-one correspondence with the momentum interval $[0,2\pi]$, and the corresponding bound state is physical. Starting with a complex value of $u$ and approaching the cuts, one finds that the points $u\pm i k/h$ end up on similar sides of the cuts (upper/upper, or lower/lower). As a result, the momentum and energy for values of $u$ on the cuts are purely imaginary, and those values of $u$ despite being real do not correspond to a physical bound state. 
With this caveat in mind, it is possible to define the bound-state representations for particles of any~$m\in\mathbb{Z}$, keeping in mind that now the eigenvalue of $\genl{M}$ is $M=m+\tfrac{k}{2\pi}p$ with $p\in[0,2\pi]$. Like in the pure-RR case~\cite{Seibold:2022mgg}, the bound-state representations take the same form as the fundamental particles, up to redefining the Zhukovsky variable --- in particular, unlike the case of $AdS_5\times S^5$, the dimension of the representation does not grow with~$|m|$.

In conclusion we find that, similarly to the case $k=0$, the mixed-flux model has infinitely many particles, labelled by $m\in\mathbb{Z}$, with momentum~$p\in[0,2\pi]$. It is possible to construct the bound state representation starting from the two-particle representation. We can essentially repeat the considerations of~\cite{Seibold:2022mgg}.

Let us consider the bound states which we have discussed above, which satisfy~\eqref{eq:bsconditionkgreater} and appear in the string (as opposed to mirror) model. It is easier to discuss everything in terms of the two-dimensional representations of the ``smaller'' algebra~\eqref{eq:lcalgebrasmall}. For left fundamental particles ($m=+1$), we work with the two-dimensional representation $\rho_{\L}^{\B}(+1,p_1)$ and $\rho_{\L}^{\B}(+1,p_2)$, where the momenta are complex. Let the highest-weight state of each of these representations be $|\phi^{\B}_{\L}(+1,p_1)\rangle$ and $|\phi^{\B}_{\L}(+1,p_2)\rangle$, respectively. The bound state representation will have highest-weight state
\begin{equation}
    |\phi^{\B}_{\L}(+2,p)\rangle=|\phi^{\B}_{\L}(+1,p_1)\rangle\otimes|\phi^{\B}_{\L}(+1,p_2)\rangle\,,
\end{equation}
where $p$ is the total bound-state momentum. This is the highest-weight state of a short representation with $m=+2$ and momentum~$p$. Hence, the representation contains one (and only one) other state, proportional to $\genl{q}\,|\phi^{\B}_{\L}(+2,p)\rangle$. In fact, the bound state representation which we constructed is isomorphic to~$\rho^{\B}_{\L}(+2,p)$. Clearly, this procedure may be iterated to construct any representation $\rho^{\B}_{\L}(m,p)$ with $m\geq+1$~integer.

Let us now look at the bound states of two $m=-1$ particles. Here we deal with  two representations $\rho_{\R}^{\F}(-1,p_1)$ and $\rho_{\R}^{\F}(-1,p_2)$. The bound-state representation will have as lowest-weight state
\begin{equation}
    |\varphi^{\B}_{\R}(-2,p)\rangle=|\varphi^{\B}_{\R}(-1,p_1)\rangle\otimes|\varphi^{\B}_{\R}(-1,p_2)\rangle\,.
\end{equation}
Again, this is a short representation, with $m=-2$ and total momentum~$p$. The other state of the representation is proportional to $\genl{s}\,|\varphi^{\B}_{\R}(-2,p)\rangle$ and in fact the whole representation is isomorphic to~$\rho_{\R}^{\F}(m,p_1)$, with $m\leq-1$~integer. It is worth noting that this discussion does not apply to \textit{mirror} bound states, which do not satisfy~\eqref{eq:bsconditionkgreater}. In fact, we would expect mirror bound states to transform in \textit{antisymmetric} representations, whose lowest-weight state is given, for instance, by $|\varphi^{\B}_{\L}(+2,p)\rangle=|\varphi^{\F}_{\L}(+1,p)\rangle\otimes |\varphi^{\F}_{\L}(+1,p)\rangle$. A full analysis of the mirror theory, including its bound states, would be interesting and we hope to carry it out in future work.

\subsection{Crossing equations}
\label{sec:fulltheory:crossing}
The linearly-realised symmetries do not fix entirely the S~matrix. It remains to constrain the ``dressing factors''. This can be done by using unitarity and crossing symmetry. In the following, we write down crossing equations for the different particle sectors of the theory: massive-massive, massive-massless and massless-massless.

\paragraph{Massive-massive crossing equations.}
First of all, let us pick a convention for dressing factors. Let us set, for the scattering of $m=\pm 1$ particles,\footnote{We used the following standard definitions for the highest and lowest weight states: $\big|Y_{u}\big\rangle \equiv |\phi^{\B}_{\L}(1;u)\rangle \otimes |\phi^{\B}_{\L}(1;u)\rangle$, $\big|\bar{Z}_{u}\big\rangle \equiv |\phi^{\F}_{\R}(-1;u)\rangle \otimes |\phi^{\F}_{\R}(-1;u)\rangle$, $\big|\bar{Y}_{u}\big\rangle \equiv |\varphi^{\B}_{\R}(-1;u)\rangle \otimes |\varphi^{\B}_{\R}(-1;u)\rangle$, $\big|Z_{u}\big\rangle \equiv |\varphi^{\F}_{\L}(1;u)\rangle \otimes |\varphi^{\F}_{\L}(1;u)\rangle$.}
\begin{equation}
\label{eq:massivenorm}
    \begin{aligned}
    \mathbf{S}\,\big|Y_{u_1}Y_{u_2}\big\rangle&=&
    {\color{Violet}\frac{x^-_{\text{\tiny L}1}-x^+_{\text{\tiny L}2}}{x^+_{\text{\tiny L}1} - x^-_{\text{\tiny L}2}}}
  \big(\phase^{\bullet\bullet}_{\L\L}\big)^{-2}\,
    \big|Y_{u_2}Y_{u_1}\big\rangle,\\
    \mathbf{S}\,\big|Y_{u_1}\bar{Z}_{u_2}\big\rangle&=&
     {\color{Violet}{\frac{x_{\R2}^+}{x_{\R2}^-}} \frac{1-x^+_{\text{\tiny L}1} x^-_{\text{\tiny R}2} }{1-x^-_{\text{\tiny L}1}x^+_{\text{\tiny R}2}}\frac{1-x^-_{\text{\tiny L}1} x^-_{\text{\tiny R}2} }{1-x^+_{\text{\tiny L}1}x^+_{\text{\tiny R}2}} } \big(\tilde{\phase}^{\bullet\bullet}_{\L\R}\big)^{-2}\,
    \big|\bar{Z}_{u_2}Y_{u_1}\big\rangle,\\
    \mathbf{S}\,\big|\bar{Z}_{u_1}Y_{u_2}\big\rangle&=&
 {\color{Violet}{x_{\R1}^-\ov x_{\R1}^+} \frac{1-x^+_{\text{\tiny R}1} x^-_{\text{\tiny L}2} }{1-x^-_{\text{\tiny R}1}x^+_{\text{\tiny L}2}}\frac{1-x^+_{\text{\tiny R}1} x^+_{\text{\tiny L}2} }{1-x^-_{\text{\tiny R}1}x^-_{\text{\tiny L}2}} }\big(\widetilde{\phase}^{\bullet\bullet}_{\R\L}\big)^{-2}\,
    \big|Y_{u_2}\bar{Z}_{u_1}\big\rangle,\\
    \mathbf{S}\,\big|\bar{Z}_{u_1}\bar{Z}_{u_2}\big\rangle&=&
   {\color{Violet}{x_{\R1}^-\ov x_{\R1}^+} {x_{\R2}^+\ov x_{\R2}^-} \frac{x^+_{\text{\tiny R}1} - x^-_{\text{\tiny R}2}}{x^-_{\text{\tiny R}1}-x^+_{\text{\tiny R}2}}}\big(\phase^{\bullet\bullet}_{\R\R}\big)^{-2}\,
    \big|\bar{Z}_{u_2}\bar{Z}_{u_1}\big\rangle,
    \end{aligned}
\end{equation}
where the dressing factors are the $\phase_{**}$. We have highlighted some rational pre-factors. This is because we want to make it explicit that the scattering of two $Y$ excitations produces a bound-state pole (while $\sigma^{\bullet\bullet}_{\L\L}$ has no physical poles or zeros). Once that factor is fixed, the remaining pre-factors follow by requiring the remaining $\sigma_{**}$s not to have zeros and poles and using crossing, unitarity, and a ``left-right'' symmetry between barred and unbarred particles (\textit{e.g.} $Y\Leftrightarrow\bar{Y}$). Note that the scattering between left and right particles has also potential poles; we will see that one of these has the interpretation of a $t$-channel pole.

The crossing transformation $u\to \bar{u}$ should flip the sign of momentum and energy, as well as flip $m\to-m$~\cite{Lloyd:2014bsa}. In other words,
\begin{equation}
    E_{\L}(\bar{u})=-E_{\R}(u)\,,\qquad
    p_{\L}(\bar{u})=-p_{\R}(u)\,.
\end{equation}
This can be realised by the following map on the Zhukovsky variables
\begin{equation}
    x^\pm_{\L}(\bar{u})=\frac{1}{x^\pm_{\R}(u)}\,,\qquad
    x^\pm_{\R}(\bar{u})=\frac{1}{x^\pm_{\L}(u)}\,.
\end{equation}
Hence, the crossing transformation on the~$u$-plane is
\begin{equation}
    u\to\bar{u}=u\,,\qquad\text{through a cut of the }u\text{-plane.}
\end{equation}
In terms of the $x$-plane, and in analogy with the $k=0$ case, this suggests that physical left (resp.~right) particles live in the part of the $x$-plane outside the image of the cuts of the $u$-plane, see figures~\ref{string_u_x_planes_positive_k} and~\ref{string_u_x_planes_negative_k}.

Using this prescription, we obtain the following crossing equations
\begin{equation}
\label{first_pair_of_massive_crossing_equations_before_the_limit}
\begin{aligned}
\sigma^{\bullet\bullet}_{\text{\tiny LL}} (u_1,u_2)^{2}\tilde\sigma^{\bullet\bullet}_{\text{\tiny RL}} (\bar u_1,u_2)^{2}&=
 {x_{\L2}^+\ov x_{\L2}^-}
\Bigl(\frac{x_{\text{\tiny L}1}^+ - x_{\text{\tiny L}2}^-}{x_{\text{\tiny L}1}^+ - x_{\text{\tiny L}2}^+} \Bigl)^2  {\color{Violet}{x_{\L1}^-\ov x_{\L1}^+}  \frac{x_{\text{\tiny L}1}^+ - x_{\text{\tiny L}2}^+}{x_{\text{\tiny L}1}^- - x_{\text{\tiny L}2}^-}}
\,,\\
 \sigma^{\bullet\bullet}_{\text{\tiny LL}} (\bar u_1,u_2)^{2}\tilde\sigma^{\bullet\bullet}_{\text{\tiny RL}} (u_1,u_2)^{2}&= {x_{\L2}^+\ov x_{\L2}^-} \Bigl( \frac{1-x^-_{\text{\tiny R}1}x^-_{\text{\tiny L}2}}{1-x^-_{\text{\tiny R}1}x^+_{\text{\tiny L}2}} \Bigl)^2  {\color{Violet} \frac{1 - x_{\text{\tiny R}1}^+ x_{\text{\tiny L}2}^+}{1 - x_{\text{\tiny R}1}^- x_{\text{\tiny L}2}^-}} \,.
\end{aligned}
\end{equation}
while the remaining two equations take the form
\begin{equation}
\label{second_pair_of_massive_crossing_equations_before_the_limit}
\begin{aligned}
\sigma^{\bullet\bullet}_{\text{\tiny RR}} (\bar u_1,u_2)^{2}\tilde\sigma^{\bullet\bullet}_{\text{\tiny LR}} (u_1,u_2)^{2}&=
{x_{\R2}^-\ov x_{\R2}^+}
\frac{\big(1-x^+_{\text{\tiny L}1}x^+_{\text{\tiny R}2}\big)^2}{\big(1-x^+_{\text{\tiny L}1}x^-_{\text{\tiny R}2}\big)^2}  {\color{Violet}\Bigl({x_{\R2}^+\ov x_{\R2}^-}\Bigl)^2 \Bigl( \frac{1-x^+_{\text{\tiny L}1}x^-_{\text{\tiny R}2}}{1-x^-_{\text{\tiny L}1}x^+_{\text{\tiny R}2}} \Bigl)^2 \frac{1-x^-_{\text{\tiny L}1}x^-_{\text{\tiny R}2}}{1-x^+_{\text{\tiny L}1}x^+_{\text{\tiny R}2}} }
 \,,\\
 \sigma^{\bullet\bullet}_{\text{\tiny RR}} (u_1,u_2)^{2}\tilde\sigma^{\bullet\bullet}_{\text{\tiny LR}} (\bar u_1,u_2)^{2}&={x_{\R2}^-\ov x_{\R2}^+}
 \Bigl(\frac{x_{\text{\tiny R}1}^- - x_{\text{\tiny R}2}^+}{x_{\text{\tiny R}1}^- - x_{\text{\tiny R}2}^-} \Bigl)^2 {\color{Violet} {x_{\R1}^-\ov x_{\R1}^+}  \Bigl( {x_{\R2}^+\ov x_{\R2}^-} \Bigl)^2 \Bigl(\frac{x_{\text{\tiny R}1}^+ - x_{\text{\tiny R}2}^-}{x_{\text{\tiny R}1}^- - x_{\text{\tiny R}2}^+}  \Bigl)^2 \frac{x_{\text{\tiny R}1}^- - x_{\text{\tiny R}2}^-}{x_{\text{\tiny R}1}^+ - x_{\text{\tiny R}2}^+} }  \,.
\end{aligned}
\end{equation}
Here too we have highlighted some expressions. The  contributions in black in~\eqref{first_pair_of_massive_crossing_equations_before_the_limit} and~\eqref{second_pair_of_massive_crossing_equations_before_the_limit} provide the crossing equations that we would have obtained absorbing the highlighted factors of~\eqref{eq:massivenorm} (\textit{i.e.}, the poles) in the $\sigma_{**}$s. We keep track of these highlighted contributions since we want to match them with those arising in  the relativistic limit explained in the next section.

\paragraph{Mixed mass crossing equations.}
Analogously to what happens in the pure Ramond-Ramond case the dispersion relation~\eqref{eq:dispersion} is non-analytic whenever $m=0 \, \mod k$. In these cases, it is necessary to split particles into a chiral and an antichiral sector, characterised by having $M>0$ and $M<0$ respectively. We label these sectors using superscript signs: `$+$' and `$-$'. For $m=0$ these sectors correspond to the regions of positive and negative momenta. Below we normalise the S-matrix elements associated with the scattering of massive-massless and massless-massive highest-weight states\footnote{We use the following convention for the massless highest and lowest weight states: $\big| \chi^{\dot{1}}_{u}\big\rangle \equiv |\phi^{\B}_{\L}(0,u)\rangle \otimes |\phi^{\F}_{\L}(0,u)\rangle$, $\big| \chi^{\dot{2}}_{u}\big\rangle \equiv |\phi^{\F}_{\L}(0,u)\rangle \otimes |\phi^{\B}_{\L}(0,u)\rangle$, $\big| \tilde{\chi}^{\dot{1}}_{u}\big\rangle \equiv |\varphi^{\F}_{\L}(0,u)\rangle \otimes |\varphi^{\B}_{\L}(0,u)\rangle$, $\big| \tilde{\chi}^{\dot{2}}_{u}\big\rangle \equiv |\varphi^{\B}_{\L}(0,u)\rangle \otimes |\varphi^{\F}_{\L}(0,u)\rangle$.} assuming massless particles coming from the left to be chiral and massless particles coming from the right to be antichiral
\begin{equation}
\label{eq:massivemassless_normalization}
    \begin{aligned}
    \mathbf{S}\,\big|Y_{u_1}\chi^{\dot{\alpha}}_{u_2}\big\rangle&=&
    {\color{Violet} e^{\frac{i}{2} p_1} e^{-\frac{3 i}{2} p_2}  \frac{x_{\text{\tiny L}1}^- - x_{\text{\tiny L}2}^+}{x_{\text{\tiny L}1}^+ - x_{\text{\tiny L}2}^-}} 
  \big(\phase^{\bullet -}_{\L\L}\big)^{-2}\,
    \big|\chi^{\dot{\alpha}}_{u_2}Y_{u_1}\big\rangle,\\
    \mathbf{S}\,\big|\bar{Z}_{u_1}\chi^{\dot{\alpha}}_{u_2}\big\rangle&=&
     {\color{Violet} e^{-\frac{i}{2} p_1} e^{-\frac{3 i}{2} p_2}  \frac{1-x_{\text{\tiny R}1}^+ x_{\text{\tiny L}2}^+}{1-x_{\text{\tiny R}1}^- x_{\text{\tiny L}2}^-} }
  \big(\phase^{\bullet -}_{\R\L}\big)^{-2}\,
    \big| \chi^{\dot{\alpha}}_{u_2} \bar{Z}_{u_1} \big\rangle,\\
    \mathbf{S}\,\big|\chi^{\dot{\alpha}}_{u_1} Y_{u_2} \big\rangle&=&
 {\color{Violet}  e^{\frac{3i}{2} p_1} e^{-\frac{i}{2} p_2}  \frac{x_{\text{\tiny L}1}^- - x_{\text{\tiny L}2}^+}{x_{\text{\tiny L}1}^+ - x_{\text{\tiny L}2}^-}}   \big(\phase^{+ \bullet}_{\L\L}\big)^{-2}\,
    \big|Y_{u_2} \chi^{\dot{\alpha}}_{u_1} \big\rangle,\\
    \mathbf{S}\,\big|\chi^{\dot{\alpha}}_{u_1} \bar{Z}_{u_2} \big\rangle&=&
   {\color{Violet}  e^{\frac{3i}{2} p_1}  e^{\frac{i}{2} p_2}  \frac{1-x_{\text{\tiny L}1}^- x_{\text{\tiny R}2}^-}{1-x_{\text{\tiny L}1}^+ x_{\text{\tiny R}2}^+} } \big(\phase^{+ \bullet}_{\L\R}\big)^{-2}\,
    \big|\bar{Z}_{u_2}\chi^{\dot{\alpha}}_{u_1} \big\rangle \,.
    \end{aligned}
\end{equation}
Working with these conventions we obtain the following crossing equations connecting the different mixed-mass dressing phases: 
\begin{equation}
\label{mixed_mass_full_theory_crossing_eq_1}
\begin{split}
&\big(\phase^{\bullet -}_{\L\L}(u_1, u_2) \big)^{2} \big(\phase^{\bullet -}_{\R\L} (\bar{u}_1, u_2) \big)^{2} = {\frac{x_{\L2}^+}{ x_{\L2}^-}} \Bigl(\frac{x^+_{\text{\tiny L}1}-x^-_{\text{\tiny L}2}}{x^+_{\text{\tiny L}1}-x^+_{\text{\tiny L}2}} \Bigl)^2 {\color{Violet}  e^{-3i p_2}      \frac{x^-_{\text{\tiny L}1}-x^+_{\text{\tiny L}2}}{x^+_{\text{\tiny L}1}-x^-_{\text{\tiny L}2}} \frac{x^+_{\text{\tiny L}1}-x^+_{\text{\tiny L}2}}{x^-_{\text{\tiny L}1}-x^-_{\text{\tiny L}2}} } \, ,\\
&\big(\phase^{\bullet -}_{\L\L}(\bar{u}_1, u_2) \big)^{2} \big(\phase^{\bullet -}_{\R\L} (u_1, u_2) \big)^{2} = {\frac{x_{\L2}^+}{x_{\L2}^-}} \Bigl(\frac{1-x^-_{\text{\tiny R}1} x^-_{\text{\tiny L}2}}{1-x^-_{\text{\tiny R}1} x^+_{\text{\tiny L}2}} \Bigl)^2 {\color{Violet}  e^{-3i p_2} \frac{1-x^-_{\text{\tiny R}1} x^+_{\text{\tiny L}2}}{1-x^+_{\text{\tiny R}1} x^-_{\text{\tiny L}2}} \frac{1-x^+_{\text{\tiny R}1} x^+_{\text{\tiny L}2}}{1-x^-_{\text{\tiny R}1} x^-_{\text{\tiny L}2}}  } \,,
\end{split}
\end{equation}
\begin{equation}
\label{mixed_mass_full_theory_crossing_eq_2}
\begin{split}
&\big(\phase^{+ \bullet}_{\L\L}(u_1, u_2) \big)^{2} \big(\phase^{+ \bullet}_{\L \L} (\bar{u}_1, u_2) \big)^{2} = {\frac{x_{\L2}^+}{ x_{\L2}^-}} \Bigl(\frac{x^+_{\text{\tiny L}1}-x^-_{\text{\tiny L}2}}{x^+_{\text{\tiny L}1}-x^+_{\text{\tiny L}2}} \Bigl)^2 {\color{Violet}   e^{-i p_2} \frac{x^-_{\text{\tiny L}1} - x^+_{\text{\tiny L}2}}{x^+_{\text{\tiny L}1} - x^-_{\text{\tiny L}2}} \frac{x^+_{\text{\tiny L}1} - x^+_{\text{\tiny L}2}}{x^-_{\text{\tiny L}1} - x^-_{\text{\tiny L}2}}  }  ,\\
&\big(\phase^{+ \bullet}_{\L\R}(u_1, u_2) \big)^{2} \big(\phase^{+ \bullet}_{\L \R} (\bar{u}_1, u_2) \big)^{2} = {\frac{x_{\R2}^-}{ x_{\R2}^+}} \Bigl(\frac{1-x^+_{\text{\tiny L}1}x^+_{\text{\tiny R}2}}{1-x^+_{\text{\tiny L}1}x^-_{\text{\tiny R}2}} \Bigl)^2 {\color{Violet}  e^{i p_2}   \frac{1-x^-_{\text{\tiny L}1} x^-_{\text{\tiny R}2}}{1-x^+_{\text{\tiny L}1} x^+_{\text{\tiny R}2}}  \frac{1-x^+_{\text{\tiny L}1} x^-_{\text{\tiny R}2}}{1-x^-_{\text{\tiny L}1} x^+_{\text{\tiny R}2}}  } .
\end{split}
\end{equation}
The normalization in~\eqref{eq:massivemassless_normalization} has been chosen in such a way as to reduce to the one used in~\cite{Frolov:2021fmj} in the limit $k=0$.
Depending on whether the left Zhukovsky variables in the equations above are associated with particles with $m=+1$ or $m=0$ we define  $x^{\pm}_{\text{\tiny L}} \equiv x^{\pm}_{\text{\tiny L}}(+1,p)$ or  $x^{\pm}_{\text{\tiny L}} \equiv x^{\pm}_{\text{\tiny L}}(0,p)$. The right Zhukovsky are instead always defined as $x^{\pm}_{\text{\tiny R}} \equiv x^{\pm}_{\text{\tiny R}}(-1,p)$.
As already mentioned, to define the scattering in the physical region the massless particles should be taken with the correct chirality; in particular, we should define the velocity of the first particle in~\eqref{eq:massivemassless_normalization} to be positive and the velocity of the second particle to be negative. However, by braiding unitarity, the dressing phases in~\eqref{eq:massivemassless_normalization} can be analytically continued to all values of momenta in the complex plane.

\paragraph{Massless-massless crossing equations.}

The normalisation for the scattering between massless highest-weight states with positive chirality is chosen as follows
\begin{equation}
\label{chiral_chiral_normalization_massless_particles}
\mathbf{S}\,\big|\chi^{\dot{\alpha}}_{u_1} \chi^{\dot{\beta}}_{u_2} \big\rangle=\big(\phase^{+ +}(u_1, u_2) \big)^{-2}\big|\chi^{\dot{\beta}}_{u_2} \chi^{\dot{\alpha}}_{u_1} \big\rangle
\end{equation}
while the scattering between mixed chirality highest-weight states is defined by
\begin{equation}
\label{chiral_antichiral_normalization_massless_particles}
\mathbf{S}\,\big|\chi^{\dot{\alpha}}_{u_1} \chi^{\dot{\beta}}_{u_2} \big\rangle=\big(\phase^{+ -}(u_1, u_2) \big)^{-2}\big|\chi^{\dot{\beta}}_{u_2} \chi^{\dot{\alpha}}_{u_1} \big\rangle \,.
\end{equation}
Following the conventions of~\cite{Frolov:2021fmj} we label these dressing factors by
\begin{equation}
\phase^{+ +}(u_1, u_2)\equiv \phase^{\circ \circ}(u_1, u_2) \hspace{3mm} \text{and} \hspace{3mm} \phase^{+ -}(u_1, u_2) \equiv \tilde{\phase}^{\circ \circ}(u_1, u_2) \,.
\end{equation}
The dressing factors $\phase^{- -}$ and $\phase^{- +}$, associated with the scattering of massless particles with negative-negative and negative-positive chiralities, can be obtained from the expression above by using braiding unitarity.
The crossing equations for the massless-massless dressing factors are given by
\begin{subequations}
\label{massless_massless_crossing_equations_full_theory}
\begin{align}
\label{eq:massless_massless_crossing_equations_full_theory_1}
    &\bigl( \phase^{\circ \circ}(u_1, u_2) \bigl)^2 \bigl( \phase^{\circ \circ}(\bar{u}_1, u_2) \bigl)^2 =  {\frac{x_{\L2}^+}{x_{\L2}^-}} \Bigl(\frac{x^+_{\text{\tiny L}1}-x^-_{\text{\tiny L}2}}{x^+_{\text{\tiny L}1}-x^+_{\text{\tiny L}2}} \Bigl)^2 \,,\\
    \label{eq:massless_massless_crossing_equations_full_theory_2}
&\bigl( \tilde{\phase}^{\circ \circ}(u_1, u_2) \bigl)^2 \bigl( \tilde{\phase}^{\circ \circ}(\bar{u}_1, u_2) \bigl)^2 =  {\frac{x_{\L2}^+}{ x_{\L2}^-}} \Bigl(\frac{x^+_{\text{\tiny L}1}-x^-_{\text{\tiny L}2}}{x^+_{\text{\tiny L}1}-x^+_{\text{\tiny L}2}} \Bigl)^2 \,.
\end{align}
\end{subequations}
In the expressions above we define $x^{\pm}_{\text{\tiny L}} \equiv x^{\pm}_{\text{\tiny L}}(0,p)$, where $x^{\pm}_{\text{\tiny L}}(0,p)$ is given in the first row of~\eqref{eq:zhukovsky}.

\section{Relativistic limit}
\label{sec:relativistic}

In order to obtain a firmer grasp on the worldsheet model, we will consider it in a limit where it becomes a relativistic integrable QFT. This limit has some similarities with the one studied in~\cite{Fontanella:2019ury}, but it differs from it in many important ways which we will point out as we study~it.

\subsection{Limiting procedure}
\label{sec:relativistic:limit}
\begin{figure}
     \centering
     \begin{subfigure}[b]{0.32\textwidth}
         \centering
         \includegraphics[width=\textwidth]{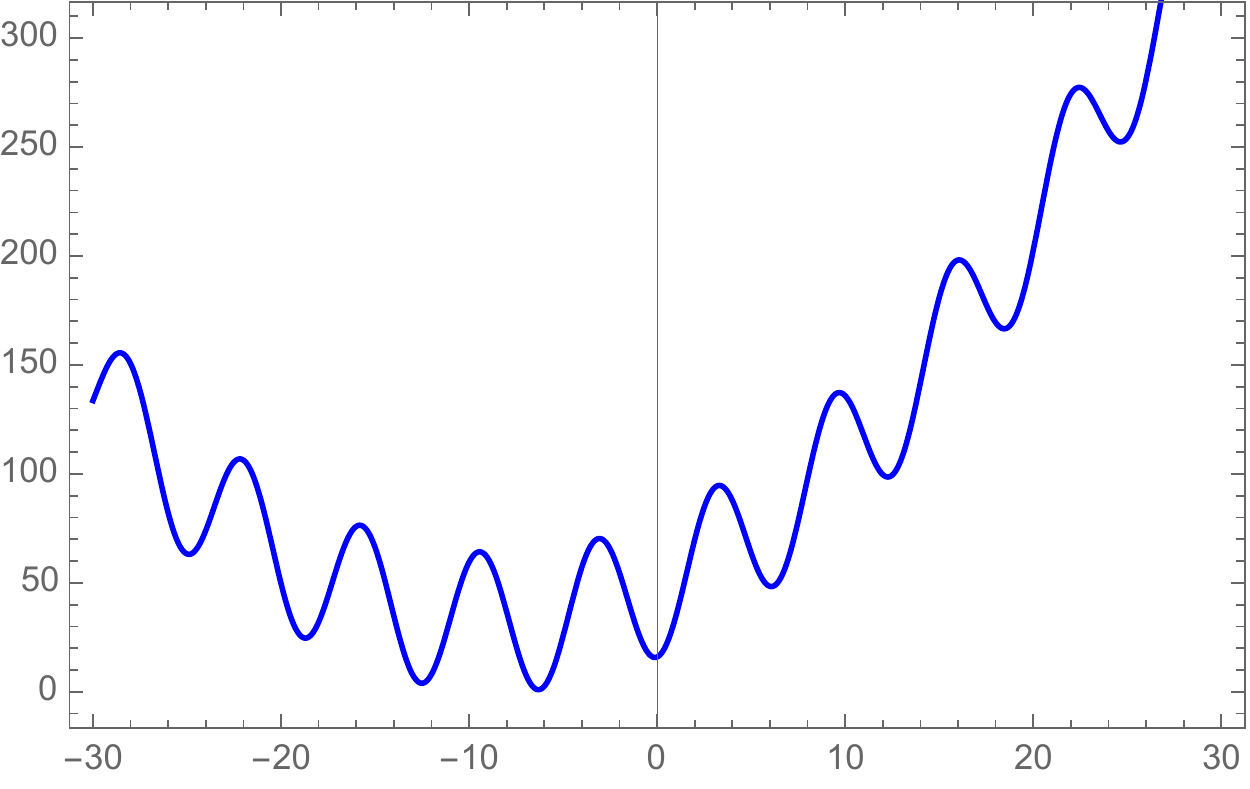}
         \caption{$m=4$, $k=3$, $h=4$}
         \label{fig:three_dispersion_relations_1}
     \end{subfigure}
     \hfill
     \begin{subfigure}[b]{0.32\textwidth}
         \centering
         \includegraphics[width=\textwidth]{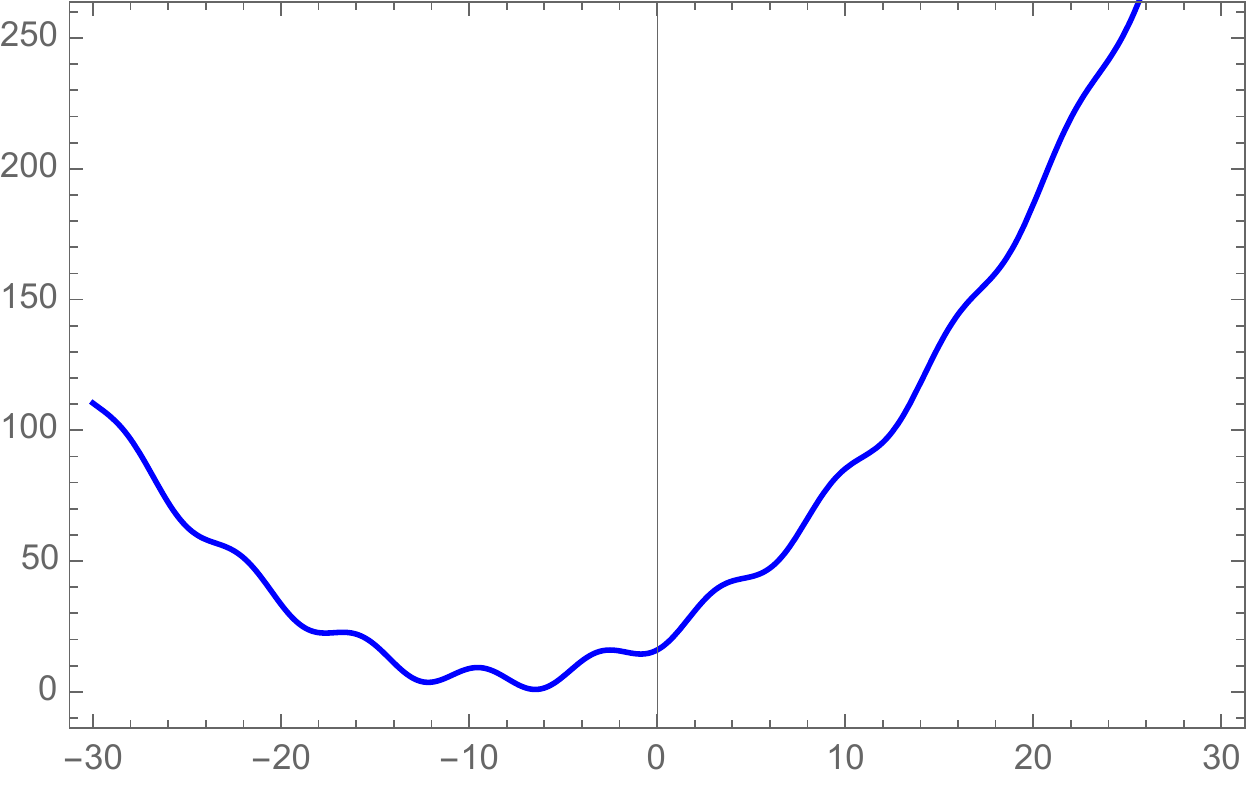}
         \caption{$m=4$, $k=3$, $h=\frac{3}{2}$}
         \label{fig:three_dispersion_relations_2}
     \end{subfigure}
     \hfill
     \begin{subfigure}[b]{0.32\textwidth}
         \centering
         \includegraphics[width=\textwidth]{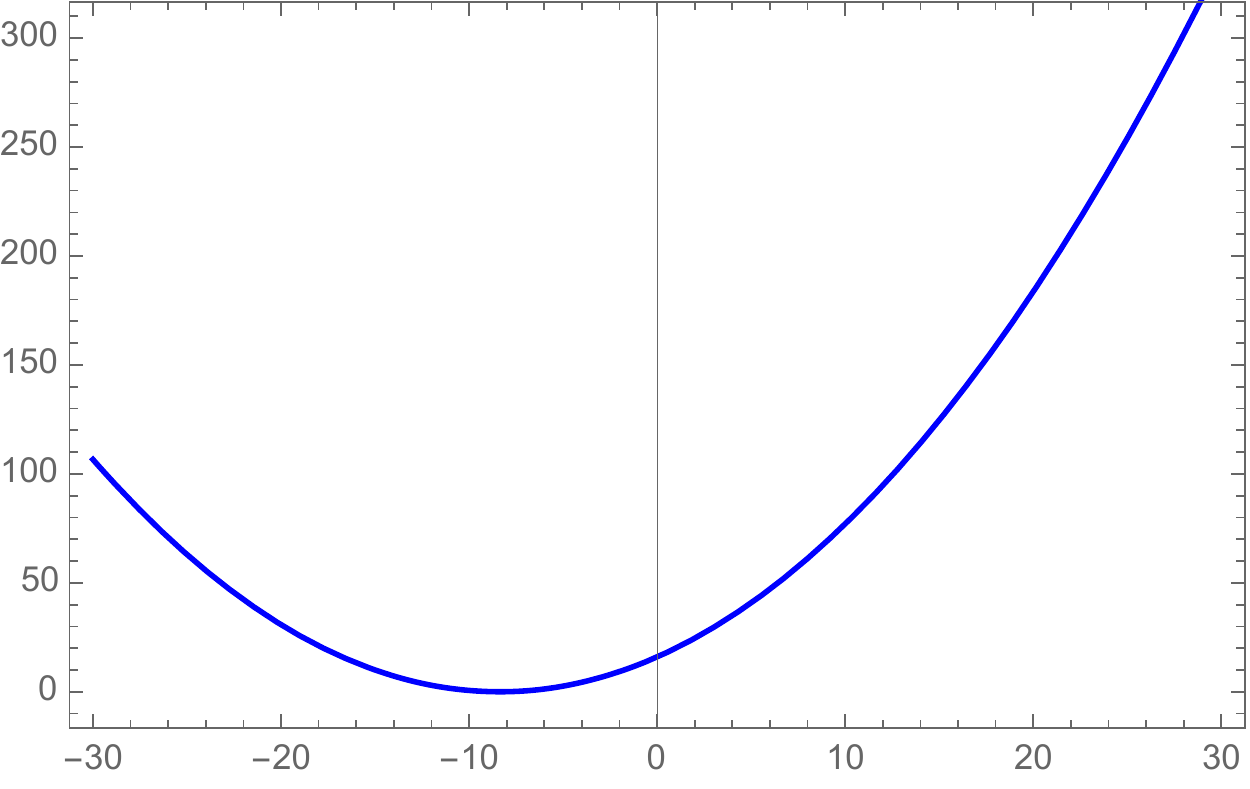}
         \caption{$m=4$, $k=3$, $h=\frac{1}{10}$}
         \label{fig:three_dispersion_relations_3}
     \end{subfigure}
        \caption{Behaviour of $E^2$ at different values of $h$ for $m$ and $k$ fixed. For $\frac{k}{h} \sim 1$ there are many local minima, as shown in figure~\ref{fig:three_dispersion_relations_1}, while increasing $\frac{k}{h}$ these minima disappear (see figures~\ref{fig:three_dispersion_relations_2} and~\ref{fig:three_dispersion_relations_3}) and one global minimum remains.}
        \label{fig:three_dispersion_relations}
\end{figure}

The dispersion relation of our model satisfies
\begin{equation}
\label{eq:dispersion_relation_before_rel_limit}
    E(m,p)^2=\left(m+\frac{k}{2\pi}p\right)^2+4h^2\sin^2\frac{p}{2}\,.
\end{equation}
Depending on the value of~$k/h$, this has a different number of local minima (the smaller $k/h$ is, the more local minima we will find, see figure~\ref{fig:three_dispersion_relations}). 
However, there is always one global minimum at%
\footnote{%
We have seen that particles of the string model have $p\in[0,2\pi]$ and $m\in\mathbb{Z}$. Hence strictly speaking only the minima with $m=0,-1,\dots -k+1,-k$ should be relevant. Nonetheless in what follows it will be convenient to first consider arbitrary values of~$p\in\mathbb{R}$ and any $m\in\mathbb{Z}$. We will see later that indeed, up to equivalences, we may restrict to $(k+1)$ values of~$m$.}
\begin{equation}
\label{eq:pminimum}
    p_{\text{min}} = -\frac{2\pi m}{k}+O(h^2)\,.
\end{equation}
This minimum is regular (quadratic) if $m\neq 0$~mod$k$, so let us restrict to this case at first. The form of eq.~\eqref{eq:dispersion_relation_before_rel_limit} suggests expanding around the minimum using a parameter~$\epsilon(h)$ which goes to zero as~$h\to0$. Therefore, we let
\begin{equation}
\label{eq:pminimum_theta}
    p(\theta) = -\frac{2\pi m}{k}+\epsilon(h)\,\sinh\theta+O(h^2)\,,
\end{equation}
where we will identify~$\theta$ with a relativistic rapidity.
It is easy to check that this yields a relativistic dispersion if and only if~$\epsilon$ is linear in~$h$ at small~$h$. In particular, it is convenient to pick
\begin{equation}
    \epsilon(h)=\frac{4\pi}{k}\left|\sin\frac{m\pi}{k}\right|\,h+O(h^2)\,,\qquad m\neq0~\,\text{mod}\,k\,.
\end{equation}
Then we have
\begin{equation}
    E(m,\theta)= 2h\,\left|\sin\frac{m\pi}{k}\right|\,\cosh\theta+O(h^2)\,,
\end{equation}
while the remaining central charges of the algebra are
\begin{equation}
    M(m,\theta)=2h\,\left|\sin\frac{m\pi}{k}\right|\,\sinh\theta+O(h^2),\qquad
    C(m)=\frac{ih}{2}\left(e^{- i\frac{2\pi m}{k}}-1\right)+O(h^2),
\end{equation}
while $C^\dagger=C^*$. In other words, at leading order we have the massive relativistic relation
\begin{equation}
\label{eq:relativisticdispersion}
    E(m,\theta)^2-M(m,\theta)^2=h^2\,\mu(m)^2\,,\qquad
    \mu(m)=2\,\left|\sin\frac{\pi m}{k}\right|,
\end{equation}
where $\mu(m)$ plays the role of the mass.

In the case where $m=0$~mod$\,k$ we expect~\eqref{eq:relativisticdispersion} to become massless. We therefore need to distinguish between the case where ($M>0$ and $M<0$, respectively). This can be done again by taking linear fluctuations in~$h$, and it gives two branches. At leading order in~$h$,
\begin{equation}
\label{eq:massless-relativistic}
    E^\pm(m,\theta)=h\, e^{\pm\theta},\qquad
    M^\pm(m,\theta)= \pm h\, e^{\pm\theta},\qquad C=0\,,\qquad
    m=0~\text{mod}k\,.
\end{equation}
The superscript sign, plus or minus, in the expressions above labels the branch of the kinematics of the massless particle. In the next sections, we will use the same convention also for the other quantities associated with massless particles. 

Taking the same limit on the Zhukovsky variables we find that the result is regular if $m\neq0$~mod$\,k$,
\bal
\label{relativistic_limit_XLPM_massive}
x_{\L}^\pm& =
  -  \text{sgn}\left(\sin \tfrac{ \pi  m}{k}\right) e^{+\theta \mp\frac{i \pi  m}{k}}  \,,\qquad m\neq 0 \mod k\,,
  \\
x_{\R}^\pm& =
  -  \text{sgn}\left(\sin \tfrac{ \pi  m}{k}\right) e^{-\theta \mp\frac{i \pi  m}{k}}  \,,\qquad m\neq 0 \mod k\,.
  \eal
In the massless case, we obtain instead different (and possibly divergent) scaling limits depending on the chirality:
\begin{equation}
\label{relativistic_limit_XLPM_massless}
x_{\L}^\pm=
\begin{cases}
\frac{\pi ^2 h^2}{k^2} \left(-\frac{k}{\pi  h}\pm i e^{-\theta}\right) \quad \text{for}\ \ M<0\,,\qquad m= 0 \mod k\,,
  \\
\frac{k}{\pi  h}\pm i e^{+\theta} \quad \hspace{15mm} \text{for}\ \ M>0\,,  \qquad m= 0 \mod k \,,
\end{cases}
\end{equation}
and
\begin{equation}
\label{relativistic_limit_XRPM_massless}
x_{\R}^\pm=
\begin{cases}
-\frac{k}{\pi  h}\pm i e^{-\theta} \quad \hspace{9mm} \text{for}\ \ M<0\,,  \qquad m= 0 \mod k \,,\\
\frac{\pi ^2 h^2}{k^2} \left(\frac{k}{\pi  h}\pm i e^{+\theta}\right) \qquad \text{for}\ \ M>0\,,\quad m= 0 \mod k\,.
\end{cases}
\end{equation}
The different behaviour of $x^\pm$ for massive and left and right massless variables indicates that the dynamics of left and right massless particles effectively decouples from the massive ones and from each other. We will see that the corresponding S-matrix elements indeed simplify drastically.

Let us finally comment on the difference between our limiting procedure and that of~\cite{Fontanella:2019ury}. In our case, we have expanded the kinematics around the minimum of the dispersion relation, in such a way as to obtain a relativistic model as~$h\to0$. In~\cite{Fontanella:2019ury} instead the authors take first the limit~$h\to0$, yielding a \textit{gapless} dispersion relation for all particles, and then shift and rescale the resulting kinematics. As far as we can tell, a different identification of the relativistic rapidity with the momenta yields a similar S-matrix eventually, at least for certain processes (even if such processes were interpreted to involve gapless particles in~\cite{Fontanella:2019ury}). However, the fact that the limiting procedure of~\cite{Fontanella:2019ury} produces a gapless dispersion is somewhat artificial and it obscures the presence of bound states. Indeed, despite the poles in the S~matrix of~\cite{Fontanella:2019ury}, the existence of bound states was not discussed by the authors.

\subsection{Representations after the limit}
\label{sec:relativistic:representations}
The limit discussed above can readily be taken on the supercharges which define the various representations of the model. The limit of the left and right representation gives the same result, up to an isomorphism (and up to  specifying whether the highest-weight state is a boson or a fermion). This is not surprising because the main reason to distinguish the left- and right- representations in the worldsheet model was that the physical region of (complex) momenta or rapidity was different in the two cases. This is not an issue here, since we are ``zooming in'' close to a special value of the momentum and obtaining a relativistic model. Up to appropriately defining the one-particle basis, we only have the representations $\repr{B}{rel}(m,\theta)$ and $\repr{F}{rel}(m,\theta)$, which can be specified in terms of the coefficients
\begin{equation}
    a(m,\theta)=\sqrt{\frac{h}{2}}\sqrt{\mu(m)}\,e^{+\frac{\theta}{2}},\qquad
    b(m,\theta)=\sqrt{\frac{h}{2}}\frac{1-e^{-\frac{2\pi m i}{k}}}{i\,\sqrt{\mu(m)}}\,e^{-\frac{\theta}{2}}\,,\qquad m\neq0~\,\text{mod}\,k\,,
\end{equation}
and for real~$\theta$ we have $\bar{a}=a^*=a$ and $\bar{b}=b^*$.
The case of $m=0$~mod$\,k$ must be treated separately and the result depends on the branch, \textit{i.e.}~on the sign of $M$. We have
\begin{equation}
\label{coefficients_massless_representations_rel_limit}
\begin{aligned}
    a(m,\theta)&=\sqrt{h}\,e^{+\frac{\theta}{2}},\qquad &b(m,\theta)&=0\,,\qquad &&m=0~\text{mod}k\,,\quad M>0\,,\\
    a(m,\theta)&=0,\qquad &b(m,\theta)&=\sqrt{h}\,e^{-\frac{\theta}{2}}\,,\qquad &&m=0~\text{mod}k\,,\quad M<0\,.
\end{aligned}
\end{equation}
This confirms that, in this case, the model is chiral.
It is also interesting to note that these one-particle representations are invariant under shifts $m\to m+k$.

The two-particle representation can be constructed with the co-product inherited from eq.~\eqref{eq:coproductfull}. Before the limit, the coproduct of two particles of momentum $p_1,p_2$ and mass $m_1,m_2$ featured a braiding factor of the form $e^{\pm \frac{i}{2}p_1}$. After the limit, this takes the form, for instance
\begin{equation}
\label{eq:coproductlimit}
    \genl{q}(m_1,m_2;\theta_1,\theta_2)=
    \genl{q}(m_1;\theta_1)\otimes \genl{1}+
    e^{i\frac{\pi m_1}{k}}\,\Sigma\otimes\genl{q}(m_2;\theta_2)\,,
\end{equation}
where we recall that $\Sigma$ is the fermion-sign matrix. We see that in this case a shift of $m_1$ by $k$ produces an additional sign,
\begin{equation}
\label{eq:monodromy}
    m_1\to m_1+k\qquad\Leftrightarrow\qquad \Sigma\otimes\genl{1} \to -\Sigma\otimes\genl{1}\,.
\end{equation}
We can interpret this as a change of the grading in the underlying representation. Note that this is true also for shifts of $m_2$, though the coproduct which we used does not make it as evident as it is not symmetric --- it is however sufficient to consider the ``opposite'' coproduct to see it.

\subsection{S~matrix for fundamental particles and bound states}
\label{sec:relativistic:boundstates}
Once we have set up our limiting procedure and understood the particle context after the limit, we can proceed in two ways:
\begin{enumerate}
    \item Construct the representations after the limit and bootstrap the S~matrix, or
    \item Take the limit of the full, nonrelativistic S~matrix.
\end{enumerate}
It is clear that case 2.~must be comprised in case 1., but it is not obvious whether the relativistic bootsrap may yield a more general S-matrix.
We discuss the relativistic bootstrap in appendix~\ref{app:relativisticS},%
\footnote{%
As it turns out, the result of that procedure is slightly more general than taking the limit of the full S~matrix. We will return to this in the conclusions. 
} 
while here we take the limit of the full S~matrix.
This is straightforward, and it yields a relativistic S-matrix of the same form as the one in appendix~\ref{app:relativisticS}.

\paragraph{Left-left fundamental particles.}
In the case of two left ($m=+1$) fundamental particles we have after the limit
\begin{equation}
\begin{aligned}
    A^{\B\B}_{\L\L}&=1,&
    B^{\B\B}_{\L\L}&=\frac{\sinh \frac{\theta}{2}}{\sinh \Bigl(\frac{\theta}{2} + \frac{i \pi}{k}  \Bigr)}\,,\\
    C^{\B\B}_{\L\L}&=\frac{i \sin \tfrac{\pi}{k}}{ \sinh \Bigl(\frac{\theta}{2} + \frac{i \pi}{k} \Bigr)}\,,&
    D^{\B\B}_{\L\L}&=\frac{\sinh \frac{\theta}{2} }{\sinh \Bigl(\frac{\theta}{2} + \frac{i \pi}{k} \Bigr)}\,,\\
    E^{\B\B}_{\L\L}&=\frac{i \sin \tfrac{\pi}{k}}{ \sinh \Bigl(\frac{\theta}{2} + \frac{i \pi}{k}  \Bigr)}\,,\qquad&
    F^{\B\B}_{\L\L}&=- \frac{\sinh \Bigl(\frac{\theta}{2} - \frac{i \pi}{ k} \Bigr)}{\sinh \Bigl(\frac{\theta}{2} + \frac{i \pi}{ k}  \Bigr)}\,,
\end{aligned}
\end{equation}
A first observation is that this defines a \textit{parity-invariant} S~matrix. This was not the case before the limit. The reason why this happens is that our limit expanded around a minimum of the energy up to the quadratic order in fluctuations, thereby discarding all odd terms in the momentum expansion.
Let us now look at the S-matrix element $F^{\B\B}_{\L\L}$, which corresponds to the scattering of two lowest-weight states. Poles or zeros in this element signal possible bound-state channels. We see that $F^{\B\B}_{\L\L}$ vanishes if $\theta=2i\pi /k$, which is inside the physical strip, at least if~$k>2$. This confirms our expectation from section~\ref{sec:fulltheory:boundstates} that left fundamental particles may make bound states with $m=+2$, and that the overall normalisation of the S~matrix must be fixed to provide the necessary poles. Moreover, the bound~state must transform in the limit of the $m=+2$ representation, and may itself create bound-states with other left particles with larger and larger~$m$ (more on this~below).

\paragraph{Right-right fundamental particles.}
A similar structure emerges in the case of two right ($m=-1$) fundamental particles. Here after the limit we find
\begin{equation}
\begin{aligned}
    A^{\F\F}_{\R\R}&=1\,,&
    B^{\F\F}_{\R\R}&=\frac{\sinh \frac{\theta}{2}}{\sinh \Bigl(\frac{\theta}{2} + \frac{i \pi}{k} (k-1)  \Bigr)}\,,\\
    C^{\F\F}_{\R\R}&=\frac{i \sin \tfrac{\pi}{k}}{ \sinh \Bigl(\frac{\theta}{2} + \frac{i \pi}{k} (k-1) \Bigr)}\,,&
    D^{\F\F}_{\R\R}&=\frac{\sinh \frac{\theta}{2} }{\sinh \Bigl(\frac{\theta}{2} + \frac{i \pi}{k} (k-1) \Bigr)}\,,\\
    E^{\F\F}_{\R\R}&=\frac{i \sin \tfrac{\pi}{k}}{ \sinh \Bigl(\frac{\theta}{2} + \frac{i \pi}{k} (k-1) \Bigr)}\,,&\qquad
    F^{\F\F}_{\R\R}&=- \frac{\sinh \Bigl(\frac{\theta}{2} - \frac{i \pi}{ k} (k-1) \Bigr)}{\sinh \Bigl(\frac{\theta}{2} + \frac{i \pi}{ k} (k-1) \Bigr)}\,,
\end{aligned}
\end{equation}
In this case we see that there is a singularity in the physical strip at $\theta=2\pi i/k$, in the form of a pole of~$F^{\F\F}_{\R\R}$. In fact, this is the same type of bound state as before --- it now appears as a pole rather than a  zero because we have swapped the highest and lowest-weight states. This corresponds to a bound state with $m=-2$.

\paragraph{Left-right fundamental particles.}
It is interesting to look at the scattering of left and right fundamental particles (where we do not expect bound states from the discussion of section~\ref{sec:fulltheory:boundstates}). We find
\begin{equation}
\begin{aligned}
    A^{\B\F}_{\L\R}&=1\,,&
    B^{\B\F}_{\L\R}&=\frac{\sinh \Bigl(\frac{\theta}{2} + \frac{i \pi}{2 k} (k-2) \Bigr)}{\sinh \Bigl(\frac{\theta}{2} + \frac{i \pi}{2} \Bigr)}\,,\\
    C^{\B\F}_{\L\R}&=\frac{i \sin \tfrac{\pi}{k}}{ \sinh \Bigl(\frac{\theta}{2} + \frac{i \pi}{2} \Bigr)} e^{\frac{i \pi}{2 k} (2-k)}\,,&
    D^{\B\F}_{\L\R}&=\frac{\sinh \Bigl(\frac{\theta}{2} - \frac{i \pi}{2 k} (k-2) \Bigr)}{\sinh \Bigl(\frac{\theta}{2} + \frac{i \pi}{2 } \Bigr)}\,,\\
    E^{\B\F}_{\L\R}&=\frac{i \sin \tfrac{\pi}{k}}{ \sinh \Bigl(\frac{\theta}{2} + \frac{i \pi}{2} \Bigr)} e^{-\frac{i \pi}{2 k} (2-k)}\,,\qquad&
    F^{\B\F}_{\L\R}&=1\,.
\end{aligned}
\end{equation}
We see that $F^{\B\F}_{\L\R}=1$ and the S matrix never degenerates to a projector for any $\theta$ inside the physical strip, consistently with our expectations.

\paragraph{Arbitrary massive particles.}
Putting together the above observations, and using the fact that after the limit there is no difference between what used to be the ``left'' and ``right'' kinematics, we can write a unified formula for the scattering of particles of arbitrary mass, as long as
\begin{equation}
    m_1\neq 0~\text{mod}\,k\,,\qquad
    m_2\neq 0~\text{mod}\,k\,.
\end{equation}
To obtain these expressions we take the limit of the matrix part of the S~matrix with $m$-dependent Zhukovsky variables. 
We can suppress the L and R labels, which are inconsequential, and only keep track of the statistics. To this end, let us focus on the case of two bosonic highest-weight states, which gives:
\begin{equation}
\label{eq:explicitBBsmatrix}
\begin{aligned}
    A_{12}^{\B\B}=&\,1\,,&
    B_{12}^{\B\B}=&\, \frac{\mathscr{S}_1 e^{\frac{i m_2 \pi}{k}+\theta}-\mathscr{S}_2 e^{\frac{i m_1 \pi}{k}}}{\mathscr{S}_1 e^{\frac{i (m_1+m_2) \pi}{k}+\theta}-\mathscr{S}_2}   \,,\\
    C_{12}^{\B\B}=&\,\frac{i e^{\frac{i m_1 \pi}{k} + \frac{\theta}{2}}\sqrt{\mu(m_1) \mu(m_2)}}{\mathscr{S}_1 e^{\frac{i (m_1+m_2)\pi}{k} + \theta}  - \mathscr{S}_2 }   \,,&
    D_{12}^{\B\B}=&\, \frac{\mathscr{S}_1 e^{\frac{i m_1 \pi}{k}+\theta}-\mathscr{S}_2 e^{\frac{i m_2 \pi}{k}}}{\mathscr{S}_1 e^{\frac{i (m_1+m_2) \pi}{k}+\theta}-\mathscr{S}_2}   \,,\\
    E_{12}^{\B\B}=&\,\frac{i e^{\frac{i m_2 \pi}{k} + \frac{\theta}{2}}\sqrt{\mu(m_1) \mu(m_2)}}{\mathscr{S}_1 e^{\frac{i (m_1+m_2)\pi}{k} + \theta}  - \mathscr{S}_2 }   \,,\qquad&
    F_{12}^{\B\B}=&\, \frac{-\mathscr{S}_1 e^{\theta}+\mathscr{S}_2 e^{\frac{i (m_1+m_2) \pi}{k}}}{\mathscr{S}_1 e^{\frac{i (m_1+m_2) \pi}{k}+\theta}-\mathscr{S}_2}    \,.
\end{aligned}
\end{equation}
Here we introduced
\begin{equation}
    \mathscr{S}_j = \text{sgn}\left[\sin\big(\frac{\pi m_j}{k}\big)\right]\,.
\end{equation}
This is the same formula that one could find by bootstrap in appendix~\ref{app:relativisticS} by considering representations with general~$m$.

In principle, the correct way to obtain the S~matrix involving $m=+2,+3,\dots$  particles is to fuse the $m_1=m_2=+1$ S~matrix above. Similarly, we could have considered the S~matrix acting on $\repr{F}{rel}\otimes\repr{F}{rel}$ with $m_1=m_2=-1$ and fused it to obtain the S~matrices involving $m=-2,-3,\dots$ particles.
However, as it turns out, the two procedures give the same result. This is a consequence of the fact that bound states transform in supersymmetric representations, and that the symmetry constrains the 2-to-2 scattering completely, up of course to a dressing factor.

We postpone the discussion of the fusion properties of these S~matrices until after we solve the crossing equations, as the normalisation of each block is necessary to ensure good fusion. Let us however briefly discuss the pole structure of each block, which is suggestive of the allowed fusion channels. 
We see that the bound-state condition $F_{12}^{\B\B}=0$ has a solution in the physical strip if
\begin{enumerate}
    \item $\mathscr{S}_1=\mathscr{S}_2$ and $2\nu k<m_1+m_2<(2\nu+1)k$, with $\nu\in\mathbb{Z}$, or
    \item $\mathscr{S}_1=-\mathscr{S}_2$ and $(2\nu-1)k<m_1+m_2<2\nu k$, with $\nu\in\mathbb{Z}$.
\end{enumerate}
Recall from our construction of the representations that all the particle content of the model must be $2k$-periodic, see eq.~\eqref{eq:monodromy}.
If we start from two left-particles with $m=+1$ we are in case 1., and we can go on building bound states in this way as long as the masses are sufficiently small with respect to~$k$. In this way, we can go on until we create a particle of mass $m=k-1$. Starting from left fundamental particles, we cannot use rule 2.~to create a bound state, as we never leave the region~$\mathscr{S}_1=\mathscr{S}_2=+1$.
Recall however that, again due to~\eqref{eq:monodromy}, the representations and hence the S-matrix elements have a simple transformation rule under shifting $m_j\to m_j\pm k$: such a shift is tantamount to flipping the statistics of the $j$-th particle. For instance
\begin{equation}
\label{eq:monodromyS}
    \mathbf{S}^{\B\B}_{12}(m_1,m_2;\theta)=
    \mathbf{S}^{\F\B}_{12}(m_1-k,m_2;\theta)\,,\qquad
    \mathbf{S}^{\B\B}_{12}(m_1,m_2;\theta)=
    \mathbf{S}^{\B\F}_{12}(m_1,m_2-k;\theta)\,,
\end{equation}
It follows that, for instance, the relativistic limit of a left-bound state with $m=k-1$ is equivalent to the relativistic limit of a right-particle with $m=-1$.
In fact, up to making the appropriate shifts in $m_1$ and/or $m_2$, eq.~\eqref{eq:explicitBBsmatrix} may be used to describe the limit of the scattering of arbitrary combinations of left/right particles.

We come to the conclusions that in this model we can consider $(k-1)$ distinct massive representations, with
\begin{equation}
    \mu\in \left\{2\sin\left(\tfrac{\pi}{k}\right),2\sin\left(\tfrac{2\pi}{k}\right),\dots,2\sin\left(\tfrac{(k-1)\pi}{k}\right)\right\}\,.
\end{equation}
Clearly for $k$ odd all masses come in pairs, and the interpretation is that the two particles with identical masses are one the antiparticle of the other. In the case where $k$ happens to be even, the particle of mass $\mu=2\sin\tfrac{\pi}{2}=2$ is its own antiparticle.
In this sense, there is no longer any distinction between ``left'' and ``right'' particles when we are considering bound states. At best, we can  single out the fundamental left and right particles as the one having $m=+1$~mod$k$ and $m=-1$~mod$k$, respectively, but in this relativistic limit one will be equivalent to a bound state of the other. It is also intriguing to note that something special happens at $k=1$ and $k=2$. In the first case, there are no massive modes at all. This is in good accord with the fact that the dual theory only features four massless (and free) bosons and fermions~\cite{Eberhardt:2018ouy}. For $k=2$, it appears that left and right modes must be identified, \textit{i.e.}~there are fewer massive modes than one would na\"ively expect from the pp-wave spectrum of $AdS_3\times S^3$. It would be very interesting to understand this fact from a worldsheet CFT/WZNW construction.

We have discussed at some length the bound-state condition $F^{\B\B}_{12}=0$, which we have used to construct the $(k-1)$ massive particles of the model. These are the equivalent of the \textit{string theory bound states}, which transform in \textit{symmetric representations}~\cite{Seibold:2022mgg}. While a detailed analysis of the mirror theory is beyond the scope of this paper, on general grounds and by analogy with the pure-RR ($k=0$) case we would expect to find \textit{mirror bound states} too, and we would expect them to transform in the \textit{anti-symmetric representation}. In other words, such bound states should arise when $A^{\B\B}_{12}=0$ or, up to a normalisation, $F^{\B\B}_{12}=\infty$.
It is easy to see that there are indeed such poles in the physical strip, and that they appear when $k<(m_1+m_2)<2k$ in the bosonic case. An equivalent result holds for $-2k<(m_1+m_2)<-k$ in the fermionic case.
This suggests that, after the relativistic limit, both ``string'' and ``mirror'' bound states live in the same physical strip --- which is in a sense expected for a relativistic theory. These new poles do not generate new types of representation. In fact, let us briefly review the types of representations emerging out of either bound-state pole. For definiteness, we consider $0<m_1,m_2<k$ and the bosonic representation $\repr{B}{rel}$. We have
\begin{equation}
\begin{aligned}
     0<m_1+m_2<k&\quad
    \repr{B}{rel}(m_1)\otimes\repr{B}{rel}(m_2)\supset\repr{B}{rel}(m_1+m_2),\\
    k<m_1+m_2<2k&\quad
    \repr{B}{rel}(m_1)\otimes\repr{B}{rel}(m_2)\supset\repr{F}{rel}(m_1+m_2)\cong\repr{B}{rel}(m_1+m_2-k),
\end{aligned}
\end{equation}
where the first line corresponds to the ``string'' bound state with $F^{\B\B}_{12}=0$ and the second line to the ``mirror'' one with $F^{\B\B}_{12}=\infty$. 

\paragraph{Scattering of massive and massless particles.}
Let us now consider the case where one of the two particles is massless, meaning that it has $m=0$~mod$\,k$. Because of the $2k$-periodicity of~$m$, it is sufficient to consider two cases. Let us set, with a slight abuse of notation
\begin{equation}
    \mathscr{S}_i=\begin{cases}
        +1\,,\qquad m_i=0~\text{mod}(2k)\,,\\
        -1\,,\qquad m_i=k~\text{mod}(2k)\,.
    \end{cases}
\end{equation}
Finally, we need to recall that massless particles can have chirality ``plus'', meaning that they move to the right at the speed of light, or ``minus, meaning they move to the left. Accordingly, the physical scattering processes involving one massive and one massless particle are
\begin{equation}
\begin{aligned}
    A_{12}^{\B_+\B}=&\,1\,,&
    B_{12}^{\B_+\B}=&\, \mathscr{S}_1   \,,\\
    C_{12}^{\B_+\B}=&\,0   \,,&
    D_{12}^{\B_+\B}=&\, e^{-\frac{i\pi m_2}{k}}   \,,\\
    E_{12}^{\B_+\B}=&\,0\,,\qquad&
    F_{12}^{\B_+\B}=&\, -\mathscr{S}_1 e^{-\frac{i\pi m_2}{k}}  \,.
\end{aligned}
\end{equation}
and
\begin{equation}
\begin{aligned}
    A_{12}^{\B\B_-}=&\,1\,,&
    B_{12}^{\B\B_-}=&\, e^{-\frac{i\pi m_1}{k}}   \,,\\
    C_{12}^{\B\B_-}=&\, 0  \,,&
    D_{12}^{\B\B_-}=&\, \mathscr{S}_2  \,,\\
    E_{12}^{\B\B_-}=&\, 0 \,,\qquad&
    F_{12}^{\B\B_-}=&\, -\mathscr{S}_2 e^{-\frac{i\pi m_1}{k}}   \,.
\end{aligned}
\end{equation}
We see that the scattering is particularly simple, without any rotation in isotopic space.  The S-matrix elements of the inverse processes can be found by imposing braiding unitarity, while the statistics can be changed by using the monodromy condition~\eqref{eq:monodromyS}.

\paragraph{Massless scattering.}
Let us come to the case of two massless particles. Like it was the case before taking the limit~\cite{Frolov:2021zyc}, we need examine the scattering depending on the chirality of the particles involved.
The most natural case is the one where they collide head on, which gives a free S~matrix, up to some signs. Namely we find
\begin{equation}
\label{opposite_chirality_massless_scattering_S_matrix_elements}
\begin{aligned}
    A_{12}^{\B_+\B_-}=&\,1\,,&
    B_{12}^{\B_+\B_-}=&\, \mathscr{S}_1   \,,&
    C_{12}^{\B_+\B_-}=&\,0   \,,\\
    D_{12}^{\B_+\B_-}=&\,  \mathscr{S}_2   \,,\qquad&
    E_{12}^{\B_+\B_-}=&\,0   \,,\qquad&
    F_{12}^{\B_+\B_-}=&\, -\mathscr{S}_1 \mathscr{S}_2   \,.
\end{aligned}
\end{equation}
It is also possible that two massless particles have the same chirality, though this is not a perturbative scattering process. Here we have
\begin{equation}
\label{eq:masslessS_after_limit}
\begin{aligned}
    A_{12}^{\B_+\B_+}=&\,1\,,&
    B_{12}^{\B_+\B_+}=&\, -\mathscr{S}_1 \tanh \frac{\theta}{2}  \,,\\
    C_{12}^{\B_+\B_+}=&\, \frac{\mathscr{S}_1 \mathscr{S}_2}{\cosh \frac{\theta}{2}} \,,&
    D_{12}^{\B_+\B_+}=&\, \mathscr{S}_2 \tanh \frac{\theta}{2}  \,,\\
    E_{12}^{\B_+\B_+}=&\, \frac{1}{\cosh \frac{\theta}{2}} \,,\qquad&
    F_{12}^{\B_+\B_+}=&\, \mathscr{S}_1 \mathscr{S}_2 \,.
\end{aligned}
\end{equation}
It is worth noting that for these processes, and only for these processes, the relativistic bootstrap yields a more general solution, see appendix~\ref{app:relativisticS}.

\bigskip

In conclusion, we have found that the poles of the S~matrix suggest that the model should feature $k-1$ massive particles. Moreover, there are two types of massless particles, distinguished by their highest-weight states. Hence, it is sufficient to consider $k+1$ distinct representations, corresponding to the S-matrix blocks by
\begin{equation}
    \genl{S}^{\B\B}(m_1,m_2;\theta)\,,\qquad
    m_i=0,1,\dots, k\,.
\end{equation}

\section{Crossing equations and dressing factors}
\label{sec:crossing}

Above we have fixed the matrix part of the S~matrix by symmetry arguments. We now turn to fixing the normalisations by imposing (relativistic) crossing symmetry as well as compatibility with the bound-state structure. 

\subsection{Crossing equations}
\label{sec:relativistic:crossing}
The construction of the crossing equations in the original model related ``left'' massive particles to ``right'' massive particles (and massless particles to themselves). After the limit, this amounts to relating a representation with bosonic/fermionic highest-weight state of mass~$m$ to one of mass~$-m$ with fermionic/bosonic highest weight state. Clearly, the S~matrices in the normalisation given above (with $A_{m,m'}=1$) are not crossing-symmetric by themselves. It is necessary to multiply each block by an arbitrary dressing factor. In analogy with the original theory let us redefine each block by a multiplicative constant. For the case $0<m_1,m_2<k$, let us set
\begin{equation}
\label{S_matrix_normalization_after_limit}
    \genl{S}^{\B\B}(m_1,m_2,\theta)\quad\to\quad
    \left[\Phi(m_1,m_2;\theta)\right]^{\tfrac{1}{2}}\, \left[\sigma(m_1,m_2;\theta)\right]^{-1}\,\genl{S}^{\B\B}(m_1,m_2,\theta)\,.
\end{equation}
In this manner, the relativistic limit of the full S-matrix associated with the scattering of two particles in the representations $\repr{B}{rel}(m_1,\theta_1) \otimes \repr{B}{rel}(m_1,\theta_1)$ and $\repr{B}{rel}(m_2,\theta_2) \otimes \repr{B}{rel}(m_2,\theta_2)$ is given by
\begin{multline}
\label{complete_S_matrix_after_limit}
   \mathbf{S}_{su(1,1)_{c.e.}^{\oplus 4}}^{\B\B} (m_1, m_2; \theta)\\
   = \Phi(m_1,m_2;\theta) \left[\sigma(m_1,m_2;\theta)\right]^{-2} \Bigl( \genl{S}^{\B\B}(m_1,m_2,\theta) \otimes \genl{S}^{\B\B}(m_1,m_2,\theta) \Bigl) \,.
\end{multline}
Both $\Phi$ and $\sigma$ in~\eqref{complete_S_matrix_after_limit} are phases, \textit{i.e.}
\begin{equation}
    [\Phi(m_1,m_2;\theta^*)]^*\,\Phi(m_1,m_2;\theta)=1\,,\qquad
    [\sigma(m_1,m_2;\theta^*)]^*\,\sigma(m_1,m_2;\theta)=1\,,
\end{equation}
and satisfy braiding unitarity,
\begin{equation}
    \Phi(m_2,m_1;-\theta)\,\Phi(m_1,m_2;\theta)=1\,,\qquad
    \sigma(m_2,m_1;-\theta)^{-2}\,\sigma(m_1,m_2;\theta)^{-2}=1\,.
\end{equation}

Admittedly, splitting the prefactor in two pieces as we have done is a little arbitrary. Here we want to single out a piece that, at least for the scattering of physical magnons, has no zeros or poles in the physical strip. This is the minimal dressing factor $\sigma$, which we want to identify with the limit of a non-trivial dressing factor from the full (non-relativistic) theory. The poles and zeros will come from the CDD factor~$\Phi$, which satisfies a sort of homogeneous crossing equation and should be related to a simple (perhaps rational) prefactor in the full non-relativistic theory.%
\footnote{%
It should be remarked however that the condition that $\sigma$ has no poles or zeros is not stable under fusion; in this sense, the splitting between $\Phi$ and $\sigma$ is even more artificial.}

As discussed, the monodromy under $m\to m+k$ of the representations and hence of the matrix part of the S-matrix, see~\eqref{eq:monodromyS}, indicates that we only need to consider $(k-1)$ massive representations and two massless ones. If we assume this to be the case, it is natural to impose that the dressing factors too are compatible with the monodromy~\eqref{eq:monodromyS}. In this way we can use the crossing equation to relate $m\leftrightarrow(k-m)$, rather than $m\leftrightarrow-m$. 
We find that it must be, for $0<m_1,m_2<k$
\begin{equation}
\label{non_trivial_crossing_eq_arbitrary_m_after_limit}
\begin{split}
     & \sigma(m_1,m_2;\theta)^{-2}\, \sigma(k-m_1,m_2;\theta+i \pi)^{-2}  = \Bigl(  f_{m_1, m_2}(\theta) \Bigl)^2  \,,\\
     & \sigma(m_2,m_1;\theta)^{-2}\, \sigma(m_2,k-m_1;\theta+i \pi)^{-2} =  \Bigl( f_{m_1, m_2}(\theta) \Bigl)^2 \,,
      \end{split}
\end{equation}
with
\begin{equation}
    f_{m_1, m_2}(\theta)=\frac{\sinh \Bigl( \frac{\theta}{2}-\frac{i \pi}{2 k} (m_1-m_2) \Bigr)}{\sinh \Bigl( \frac{\theta}{2}-\frac{i \pi}{2 k} (m_1+m_2) \Bigr)}\,,
    \qquad 0<m_1,m_2<k\,;
\end{equation}
while the CDD factors satisfy, by definition, the homogeneous crossing equations
\begin{equation}
\label{eq:crossingCDD}
    \Phi(m_1,m_2;\theta)\,\Phi(k-m_1,m_2;\theta+i\pi)=1\,,\qquad
    \Phi(m_2,m_1;\theta)\,\Phi(m_2,k-m_1 ;\theta+i\pi)=1
    \,,
\end{equation}

To solve this equation, we start from a process involving the scattering of two fundamental (left) particles, such as
\begin{equation}
    \mathbf{S}\,\big|Y(\theta_1)\,Y(\theta_2)\big\rangle=
    \Phi(1,1;\theta_{12})\, \left[\sigma(1,1;\theta_{12})\right]^{-2}
    \big|Y(\theta_2)\,Y(\theta_1)\big\rangle 
\end{equation}
and demand that $\sigma(1,1;\theta)$ has no zeros and poles (and that it solves crossing). This is enough to fix $\sigma(1,1;\theta)$. Moreover, it follows that $\Phi(1,1;\theta)$ must contain the $s$-channel bound-state pole.
Having determined in such a way the fundamental dressing factors, the remaining $\sigma$s and $\Phi$s follow by fusion.
As we remarked, it so happens that through the fusion procedure $\sigma(m_1,m_2;\theta)$ will develop singularities in the physical strip; this happens for $m_1+m_2>k$ as we shall~see.

\subsection{Massive dressing factor}
\label{sec:relativistic:massivedressing}
Let us consider first the case where both particles are massive, \textit{i.e.}\ $m_j\neq0$~mod$\,k$. 
It is easy to check that the following expression solves the crossing equations.
\begin{equation}
\label{minimal_sigma_ratio_of_R_functions}
    \sigma(m_1,m_2;\theta)^{-2} =  \frac{R\left(\theta-\frac{i\pi(m_1+m_2)}{k}\right)^2\,R\left(\theta+\frac{i\pi(m_1+m_2)}{k}\right)^2}{R\left(\theta-\frac{i\pi(m_1-m_2)}{k}\right)^2 \,R\left(\theta+\frac{i\pi(m_1-m_2)}{k}\right)^2} \,,
\end{equation}
where $R(\theta)$ can be expressed in terms of $\Gamma$-functions or Barnes~$G$-function%
\footnote{The $G$-function obeys $G(z+1)=\Gamma(z)\,G(z)$ with $G(1)=1$. The function $\psi(z)$ is the Digamma function, defined by $\psi(z)=\tfrac{\de}{\de z}\log\Gamma(z)$.}
\begin{equation}
\label{R_function_definition}
     R(\theta)\equiv \frac{G(1- \frac{\theta}{2\pi i})}{G(1+ \frac{\theta}{2\pi i}) } =   \left(\frac{e}{2\pi}\right)^{+\frac{\theta}{2\pi i}}\prod_{\ell=1}^\infty \frac{\Gamma(\ell+\frac{\theta}{2\pi i})}{\Gamma(\ell-\frac{\theta}{2\pi i})}\,e^{-\frac{\theta}{\pi i}\,\psi(\ell) }\,.
\end{equation}
The function~$R(\theta)$ obeys the properties
\begin{equation}
    R (-\theta)\,R (\theta)=1\,,\qquad
    [R(\theta^*)]^*\,R(\theta)=1\,,
\end{equation}
as well as the monodromy property
\begin{equation}
\label{eq:R_function_monodromy}
    R(\theta-2\pi i) =i\,  \frac{\pi}{ \sinh{\tfrac{\theta}{2}}}\,R(\theta)\,,\qquad  R (\theta+\pi i) =   \frac{\cosh{\tfrac{\theta}{2}}}{\pi} \,R(\theta-\pi i)\,,
\end{equation}
which can be used to prove the crossing equation.
In~\cite{Fontanella:2019ury}, where a similar relativistic limit was considered (albeit with a different dispersion relation and no bound-states), a dressing factor was also proposed for processes related to some special cases of what we considered here, namely~$m_1=1$ and $m_2=1$ or $m_2=k-1$ (this is the case related to the scattering of fundamental particles). Though part of such dressing factors are given only implicitly as a Fourier transform, it is possible to check numerically that they agree with~\eqref{minimal_sigma_ratio_of_R_functions}.%
\footnote{We thank the authors of~\cite{Fontanella:2019ury} for pointing this out to us.} 
Similar dressing factors were considered by Fendley and Intriligator in~\cite{Fendley:1991ve,Fendley:1992dm}, for any $m_1$ and $m_2$. The main difference between our discussion and that of~\cite{Fendley:1991ve,Fendley:1992dm} is in the structure of the representations: we will deal with four-dimensional $\rho\otimes\rho$ representations, while~\cite{Fendley:1991ve,Fendley:1992dm} dealt with~$\rho$. This will also affect the poles and ``CDD factors'' of the model: while~\cite{Fendley:1991ve,Fendley:1992dm} normalised their S~matrix by a CDD factor~$\Phi(\theta)$ so that~$\Phi(\theta)\,\sigma^{-1}(\theta)\,\mathbf{S}(\theta)$ has the correct pole structure, we will rather use $\Phi^{1/2}(\theta)\,\sigma^{-1}(\theta)\,\mathbf{S}(\theta)$ so that $\Phi(\theta)\,\sigma^{-2}(\theta)\,[\mathbf{S}\otimes \mathbf{S}](\theta)$ has the right poles, as we shall discuss below.

\paragraph{Poles.}
While the functions~$\sigma(m_1,m_2;\theta)^{-2}$ solve the crossing equations, they do not have the correct pole structure to account for the bound states of the theory. For instance, we expect a pole in the physical strip for the scattering of two particles with $m_1=m_2=1$. 
We see that $R(\theta)$ has the following singularities
\begin{equation}
    \text{poles:}\ \theta=-2\pi i\, n\,,\qquad
    \text{zeros:}\  \theta=+2\pi i\, n\,,\qquad
    n=1,2,\dots\,.
\end{equation}
Hence the expression in~\eqref{minimal_sigma_ratio_of_R_functions} has no poles in the physical strip $(0, i\pi)$; however, it does contain a zero in the strip when $m_1+m_2>k$. In fact
\begin{equation}
    R \bigl(\theta+\frac{i\pi(m_1+m_2)}{k} \bigl)=0\quad \text{at}\quad \theta=\frac{i \pi}{k} (2k-m_1-m_2)
\end{equation}
and the numerator of~$\sigma(m_1,m_2;\theta)^{-2}$ has a zero of order two. This zero is in the physical strip when $m_1+m_2>k$ and it is necessary to cancel a second-order pole at $\theta=\frac{i \pi}{k} (2k-m_1-m_2)$ arising in $ \genl{S}^{\B\B}(m_1,m_2,\theta) \otimes \genl{S}^{\B\B}(m_1,m_2,\theta) $ from the denominators of the coefficients of the S~matrix~\eqref{eq:explicitBBsmatrix} when $\mathscr{S}_1=\mathscr{S}_2=+1$ .
Taking the zero of~$\sigma(m_1,m_2;\theta)^{-2}$ into account, all the S-matrix elements have no poles within the strip; all poles will be contained in $\Phi(m_1,m_2;\theta)$.

\paragraph{``CDD'' factors.}
The pole structure will necessarily come from the pre-factor~$\Phi(1,1;\theta)$. Because of eq.~\eqref{eq:crossingCDD}, it is necessary to simultaneously modify~$\Phi(k-1, 1;\theta)$ in an appropriate way.
To this end, following~\cite{Braden:1989bu}, let us introduce the building block
\begin{equation}
\label{definition_building_block}
    [m]_\theta\equiv \frac{\sinh{\Bigl( \frac{\theta}{2} + \frac{i \pi m}{2k} \Bigl)}}{\sinh{\Bigl( \frac{\theta}{2} - \frac{i \pi m}{2k} \Bigl)}} \,,
\end{equation}
which almost satisfies a trivial crossing equation, up to a sign:
\begin{equation}
\label{eq:crossing-buildingblock}
    [m]_{\theta}\,[k-m]_{\theta\pm i\pi}=-1\,.
\end{equation}
The CDD factors for the $m_1=1$ and $m_2=k-1$ can then be defined as
\begin{subequations}
\label{eq:CDD-fundamental}
\begin{align}
&\Phi(1,1;\theta)=\Phi(k-1,k-1;\theta)=[2]_{\theta} [0]_{\theta} \,,\\
&\Phi(1, k-1;\theta)=\Phi(k-1, 1;\theta)=[k]_{\theta}[k-2]_{\theta}  \,.
\end{align}
\end{subequations}
From~\eqref{eq:crossing-buildingblock} we see that pairs of building blocks $[m]_{\theta}$ and $[k-m]_{\theta}$ satisfy the homogeneous equation~\eqref{eq:crossingCDD} but for a minus sign; this minus sign plays no role since all the CDD factors in~\eqref{eq:CDD-fundamental} contain an even number of building blocks. By fusing these fundamental building blocks it is possible to obtain a universal formula valid for any $m_1$ and $m_2 \in \{1, \dots, k-1\}$: 
\begin{equation}
\label{general_formula_for_CDD_factor}
     \Phi(m_1,m_2;\theta)=
     \frac{\prod_{n=0}^{N}\left(\big[|m_1-m_2|+2n\big]_\theta\right)^2}{\big[|m_1-m_2|\big]_\theta\,\big[m_1+m_2\big]_\theta}\,,\qquad
     N=
     \begin{cases}
         m_1 &\quad m_1\leq m_2\,,\\
         m_2 &\quad m_2<m_1\,.
     \end{cases}
\end{equation}
This formula corresponds to the minimal S-matrix of Toda theories of $A_{k-1}$ type, with $k$ playing the role of the Coxeter number of the Lie algebra (see for example \cite{Arinshtein:1979pb,Braden:1989bu}). 

\paragraph{Fusion.} Let us now discuss why this S~matrix, with the given normalisation, behaves well under fusion. If we consider two particles with quantum numbers $m_1$ and $m_2$ such that $0<m_1+m_2<k$ then there is a pole in the S-matrix. This pole appears in the S-matrix element involving the scattering of $|\phi^{\B}\otimes\phi^{\B}\rangle$ with $|\phi^{\B}\otimes\phi^{\B}\rangle$, and it comes from the prefactor $\Phi(m_1,m_2;\theta)$. The singularity is located at rapidity
\begin{equation}
\theta= \frac{i \pi}{k} (m_1+m_2) \,.
\end{equation}
By contrast, the scattering of $|\varphi^{\F}\otimes\varphi^{\F}\rangle$ with $|\varphi^{\F}\otimes\varphi^{\F}\rangle$ vanishes, because the S-matrix element $F^{\B\B}_{12}$ has a zero there.
This suggests that a bound state in the symmetric representation must exist. In each copy of the tensor product there should be contained the state
\begin{equation}
\label{eq:fusion_bosons_bound_state}
    |\phi^{\B}(m_b, \theta_b)\rangle=|\phi^{\B}(m_1,\theta_b-\frac{i \pi}{k} m_2)\rangle\otimes|\phi^{\B}(m_2,\theta_b+\frac{i \pi}{k} m_1)\rangle\,,
\end{equation}
with associated fermionic partner $|\varphi^{\F}_{\L}(m_b, \theta_b)\rangle$ obtained by acting on $|\phi^{\B}(m_b, \theta_b)\rangle$ with $\genl{q}$ or $\genr{s}$. If the doublet of states defined in this manner is a bound state representation, then $\forall \ m_3 \in \{1, \dots, k-1\}$ and $\theta=\theta_b-\theta_3$ the projection of
\begin{equation}
\label{S_matrix_fusion_tensor_form}
\Bigl(\genl{S}^{\B\B}(m_1,m_3,\theta-\frac{i \pi}{k} m_2) \otimes 1  \Bigl) \cdot \Bigl(1 \otimes \genl{S}^{\B\B}(m_2,m_3,\theta+\frac{i \pi}{k} m_1)  \Bigl) 
\end{equation}
onto the vector space spanned by
\begin{equation}
\label{fused_basis_of_states}
\begin{split}
    &|\phi^{\B}(m_b, \theta_b)\,\phi^{\B}(m_3, \theta_3)\rangle \,, \quad |\phi^{\B}(m_b, \theta_b)\,\varphi^{\F}(m_3, \theta_3)\rangle,\\
    &|\varphi^{\F}(m_b, \theta_b)\,\phi^{\B}(m_3, \theta_3)\rangle \,, \quad |\varphi^{\F}(m_b, \theta_b)\,\varphi^{\F}(m_3, \theta_3)\rangle \,,
    \end{split}
\end{equation}
needs to be equal to $\genl{S}^{\B\B}(m_b,m_3,\theta)$. This fact can be easily checked by using  the S-matrix elements~\eqref{eq:explicitBBsmatrix}, together with the fusion properties of the dressing factors; these factors satisfy
\begin{equation}
\begin{split}
&\sigma(m_3,m_b;\theta)^{-1}=\sigma(m_3,m_1;\theta- \frac{i \pi}{k} m_2)^{-1}  \sigma(m_3,m_2;\theta+ \frac{i \pi}{k} m_1)^{-1} \,,\\
&\Phi(m_3,m_b;\theta)=\Phi(m_3,m_1;\theta- \frac{i \pi}{k} m_2)  \Phi(m_3,m_2;\theta+ \frac{i \pi}{k} m_1) \,,
\end{split}
\end{equation}
$\forall \ m_1$, $m_2$, $m_3$  and $m_b\equiv m_1+m_2 \in \{1, \dots, k-1\}$.
This substantiates our claim that our construction of $\sigma$'s is indeed compatible with fusion.

If instead $k<m_1+m_2<2k$, the scattering situation is reversed: $F_{12}^{\B\B}$ has a pole for 
\begin{equation}
\theta= \frac{i \pi}{k} (2k-m_1-m_2)
\end{equation}
in the physical strip\footnote{As previously remarked, for $\mathscr{S}_1=\mathscr{S}_2=1$, a pole in the element $F_{12}^{\B\B}$ in~\eqref{eq:explicitBBsmatrix} appears which is cancelled by a zero of $\sigma(m_1,m_2;\theta)^{-1}$ located at the same point. However, a pole is introduced also by $\Phi(m_1, m_2 ;\theta)$ and the element  $F_{12}^{\B\B}$, after having been multiplied by the dressing factor, has a singularity at $\theta= \frac{i \pi}{k} (2k-m_1-m_2)$.}. This singularity suggests the existence of a bosonic bound state involving
\begin{equation}
    |\varphi^{\B}(m_b, \theta_b)\rangle=|\varphi^{\F}(m_1,\theta_b-\frac{i \pi}{k} (k-m_2))\rangle\otimes|\varphi^{\F}(m_2,\theta_b+\frac{i \pi}{k} (k-m_1))\rangle\,,
\end{equation}
with $m_b\equiv m_1+m_2$; similarly to before, the fermionic partner $|\phi^{\F}(m_b, \theta_b)\rangle$ is obtained by acting with $\genl{s}$ or $\genr{q}$ on $|\varphi^{\B}(m_b, \theta_b)\rangle$. However, this is not a new particle: as already discussed in the previous sections we can identify 
\begin{equation}
\label{particle_identification_under_fusion}
(|\varphi^{\B}(m_b, \theta_b)\rangle \,, |\phi^{\F}(m_b, \theta_b)\rangle) = (|\varphi^{\F}(m_b-k, \theta_b)\rangle \,, |\phi^{\B}(m_b-k, \theta_b)\rangle) \,,
\end{equation}
and the bound state can therefore be recognised as a particle of quantum number $k-m_b$ already found in the fusion process. 
Then it is possible to show that the action of
\begin{equation}
\label{S_matrix_fusion_tensor_form_second_equation}
\Bigl(\genl{S}^{\B\B}(m_1,m_3,\theta-\frac{i \pi}{k} (k-m_2) ) \otimes 1  \Bigl) \cdot \Bigl(1 \otimes \genl{S}^{\B\B}(m_2,m_3,\theta+\frac{i \pi}{k} (k-m_1))  \Bigl)
\end{equation}
on the basis
\begin{equation}
\label{second_fused_basis_of_states}
\begin{split}
    &|\phi^{\B}(m_b-k, \theta_b)\,\phi^{\B}(m_3, \theta_3)\rangle \,, \quad |\phi^{\B}(m_b-k, \theta_b)\,\varphi^{\F}(m_3, \theta_3)\rangle,\\
    &|\varphi^{\F}(m_b-k, \theta_b)\,\phi^{\B}(m_3, \theta_3)\rangle \,, \quad |\varphi^{\F}(m_b-k, \theta_b)\,\varphi^{\F}(m_3, \theta_3)\rangle \,,
    \end{split}
\end{equation}
is equal to $\genl{S}^{\B\B}(m_b-k,m_3, \theta)$. This implies that if we start fusing massive particles in the sector $\mathscr{S}_1=\mathscr{S}_2=+1$ we never go outside this sector and we find exactly $k-1$ massive particles.

\begin{figure}
\begin{center}
\begin{tikzpicture}
\tikzmath{\y=1.7;}

\draw[thick,directed] (-0.8*\y,1*\y) -- (0*\y,0*\y);
\draw[thick, directed] (-0.8*\y,-0.8*\y) -- (0*\y,0*\y);
\draw[ultra thick,directed][red] (0*\y,0*\y) -- (1.5*\y,0*\y);
\filldraw[thick, black] (-0.7*\y,1.2*\y)  node[anchor=west] {{$\phi^\B_{m_1}$}};
\filldraw[thick, black] (-0.7*\y,-0.8*\y)  node[anchor=west] {{$\phi^\B_{m_2}$}};
\filldraw[ultra thick, red] (+0.7*\y,0.2*\y)  node[anchor=west] {{$\phi^\B_{m_1+m_2}$}};

\draw[black!90] (-0.23*\y,0.25*\y) arc(120:230:0.25*\y);
\draw[black!90] (0.25*\y,0.05*\y) arc(0:120:0.25*\y);
\draw[black!90] (0.25*\y,-0.05*\y) arc(0:-120:0.25*\y);
\filldraw[black!90] (-1.6*\y,0*\y)  node[anchor=west] {{$\tfrac{i \pi (m_1+m_2)}{k}$}};
\filldraw[black!90] (-0.3*\y,0.5*\y)  node[anchor=west] {{$\tfrac{i \pi (k-m_2)}{k}$}};
\filldraw[black!90] (-0.3*\y,-0.5*\y)  node[anchor=west] {{$\tfrac{i \pi (k-m_1)}{k}$}};

\draw[thick, directed] (-1.27279*\y+5*\y,0.141421*\y) -- (0*\y+5*\y,0*\y);
\draw[ultra thick,directed][red] (0*\y+5*\y,-1.1317*\y) -- (0*\y+5*\y,0*\y);
\draw[thick, directed] (0*\y+5*\y,0*\y) -- (1.06066*\y+5*\y,1.06066*\y);
\draw[black!90] (1.1*\y+4*\y,0.25*\y) arc(60:170:0.25*\y);
\draw[black!90] (0.7*\y+4*\y,-0.05*\y) arc(180:270:0.25*\y);
\draw[black!90] (1.1*\y+4*\y,-0.3*\y) arc(-80:30:0.25*\y);

\filldraw[black] (5.5*\y,-0.2*\y)  node[anchor=west] {{$\tfrac{i \pi (k-m_1)}{k}$}};
\filldraw[black] (4.1*\y,0.6*\y)  node[anchor=west] {{$\tfrac{i \pi (k-m_2)}{k}$}};
\filldraw[black] (3.5*\y,-0.4*\y)  node[anchor=west] {{$\tfrac{i \pi (m_1+m_2)}{k}$}};

\filldraw[black, thick] (-0.7*\y+6*\y,1.2*\y)  node[anchor=west] {{$\phi^\B_{m_1+m_2}$}};
\filldraw[red] (-1*\y+6*\y,-0.8*\y)  node[anchor=west] {{$\phi^\B_{m_2}$}};
\filldraw[black, thick] (+1.2*\y+2*\y,0.4*\y)  node[anchor=west] {{$\phi^\B_{m_1}$}};

\end{tikzpicture}
\caption{Examples of bound states in the symmetric representation (in red) propagating in the $s$-channel (on the left) and $t$-channel (on the right) for $m_1+m_2<k$, and their fusing angles.}
\label{Bound_states_in_different_channels}
\end{center}
\end{figure}
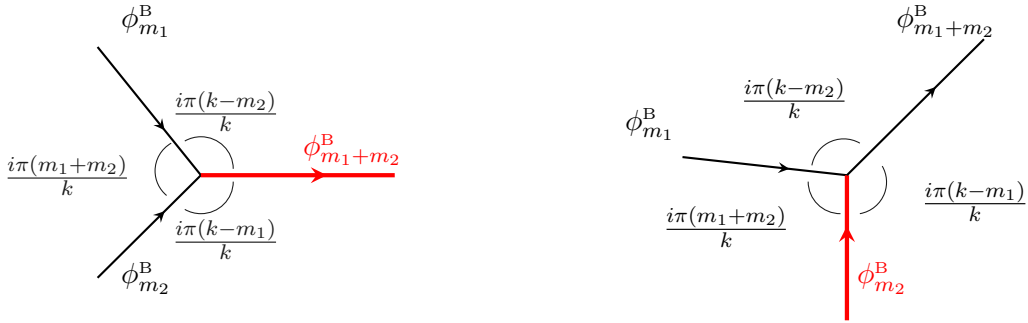

\paragraph{Bound states in the crossed channels.}
From the poles of the S~matrix in the $s$-channel we have read off the masses of the bound states and build all the representations starting from $\repr{B}{rel}(+1,\theta)$ by fusion. Additional poles come from the $t$-channel.
Consider the bound state on the LHS of figure~\ref{Bound_states_in_different_channels} (in red), which obtained by fusing $|\phi^{\B}(m_1)\rangle$ and $|\phi^{\B}(m_2)\rangle$. If we scatter particles in the representations $\repr{B}{rel}(m_1,\theta_1)$ and $\repr{B}{rel}(m_2,\theta_2)$ this bound state will appear as a particle propagating in the $s$-channel. However, the same fusing vertex responsible for the propagation of this bound state in the $s$-channel in the scattering process between $\repr{B}{rel}(m_1,\theta_1)$ and $\repr{B}{rel}(m_2,\theta_2)$ is also responsible for the propagation of the bound state $|\phi^{\B}(m_2)\rangle$
in the $t$-channel of the scattering process between $\repr{B}{rel}(m_1,\theta_1)$ and $\repr{B}{rel}(m_1+m_2,\theta_2)$.
Remark that in the $s$-channel, the residue of the S~matrix at the poles is a projector onto the bound state representation. In our case the residue of the original $4 \times 4$ S-matrix at the pole is a matrix of rank two projecting onto a two-dimensional (short) subrepresentation.
As a consequence of this fact the residue of the full $16 \times 16$  S-matrix in~\eqref{complete_S_matrix_after_limit} has rank~$4$ on the bound state.
On the other hand, the residue of the S~matrix at poles associated with bound states propagating in the $t$-channel has a maximal rank. This can be seen in the figure by turning considering similar diagrams involving $\phi^{\F}_{m_1}$ and $\phi^{\F}_{m_2}$; the corresponding process has no pole in the $s$-channel, but it is still singular in the $t$-channel.

The appearance of $t$ can also be seen from the fusion of the CDD factors. Consider first the scattering two fundamental particles with $m_1=m_2=1$. The CDD factor is simply $\Phi(1,1;\theta)=[2]_\theta$, with a single $s$-channel pole at $\theta=\frac{2i \pi}{k}$. Upon fusion we have
\begin{equation}
\label{eq:CDD_1_2_obtained_by_fusion}
\Phi(1,2;\theta) =\Phi(2,1;\theta)=\Phi(1,1;\theta+\tfrac{i \pi}{k})   \Phi(1,1;\theta-\tfrac{i \pi}{k}) =[3]_{\theta} [1]_{\theta} \,.
\end{equation}
This expression has two poles: one at $\theta=\frac{i \pi}{k} (2+1)=\frac{3 i \pi}{k}$ due to~$[3]_{\theta}$ and associated with the propagation of a particle in the $s$-channel, and one at $\theta=\frac{i \pi}{k} (2-1)=\frac{i \pi}{k}$ and associated with the propagation of a particle in the $t$-channel. It is possible to check that the transmission elements in the S-matrix have residues with opposite signs at the locations of these poles, as expected in unitary S~matrices with particles propagating in different channels.

\paragraph{Comparison with the full, non-relativistic S~matrix.}

We finally compare the dressing factors found so far with the dressing factors of the full theory before the limit.
Using~\eqref{relativistic_limit_XLPM_massive} we can take the relativistic limit on the equations in~\eqref{first_pair_of_massive_crossing_equations_before_the_limit}: in our normalisation, the ``blue terms'' go to $1$ in the limit,
\begin{equation}
\begin{aligned}
\sigma^{\bullet\bullet}_{\text{\tiny LL}} (u_1,u_2)^{2}\tilde\sigma^{\bullet\bullet}_{\text{\tiny RL}} (\bar u_1,u_2)^{2}&\to f_{1, 1}(\theta_{12}) ^{-2}\times \,  {\color{Violet} 1}
\,,\\
 \sigma^{\bullet\bullet}_{\text{\tiny LL}} (\bar u_1,u_2)^{2}\tilde\sigma^{\bullet\bullet}_{\text{\tiny RL}} (u_1,u_2)^{2}&\to f_{k-1, 1}(\theta_{12}) ^{-2} \times \, {\color{Violet} 1} \,.
\end{aligned}
\end{equation}
Comparing these equations with~\eqref{non_trivial_crossing_eq_arbitrary_m_after_limit}, we recognise that the phases $\sigma^{\bullet\bullet}_{\text{\tiny LL}}$ and $\tilde\sigma^{\bullet\bullet}_{\text{\tiny RL}}$ in the limit can be matched to
\begin{subequations}
\begin{align}
    &\sigma^{\bullet\bullet}_{\text{\tiny LL}} (u_1,u_2)^{2} \to \sigma(1,1; \theta_{12})^2 \,,\\
    &\tilde\sigma^{\bullet\bullet}_{\text{\tiny RL}} (u_1,u_2)^{2} \to \sigma(k-1,1; \theta_{12})^2 \,.
\end{align}
\end{subequations}
Performing the same limit on~\eqref{second_pair_of_massive_crossing_equations_before_the_limit} we obtain
\begin{equation}
\begin{aligned}
\sigma^{\bullet\bullet}_{\text{\tiny RR}} (\bar u_1,u_2)^{2}\tilde\sigma^{\bullet\bullet}_{\text{\tiny LR}} (u_1,u_2)^{2}&\to
f_{1, k-1}(\theta_{12}) ^{-2}  \times \, {\color{Violet} ([2]_{\theta_{12}+i \pi})^{-2}}
 \,,\\
 \sigma^{\bullet\bullet}_{\text{\tiny RR}} (u_1,u_2)^{2}\tilde\sigma^{\bullet\bullet}_{\text{\tiny LR}} (\bar u_1,u_2)^{2}&\to f_{k-1, k-1}(\theta_{12})^{-2} \times \, {\color{Violet} ([2]_{\theta_{12}})^{-2} }  \,.
\end{aligned}
\end{equation}
This time the blue factors in~\eqref{second_pair_of_massive_crossing_equations_before_the_limit} do not become trivial in the limit and we identify the solutions of the crossing equations with\footnote{The second solution can also be written as $\sigma^{\bullet\bullet}_{\text{\tiny RR}} (u_1,u_2)^{2} \to  \sigma(-1,-1; \theta_{12})^2$ using the following property of $\sigma$: $\sigma(m_1+k,m_2+k;\theta)^{-1} = - [m_1+m_2]_\theta \, \sigma(m_1,m_2;\theta)^{-1}$.}
\begin{subequations}
\label{relativistic_limit_massive_sLR_sRR}
\begin{align}
    &\tilde\sigma^{\bullet\bullet}_{\text{\tiny LR}} (u_1,u_2)^{2} \to \sigma(1,k-1; \theta_{12})^2 \,,\\
    &\sigma^{\bullet\bullet}_{\text{\tiny RR}} (u_1,u_2)^{2} \to ([2]_{\theta_{12}})^{-2} \sigma(k-1,k-1; \theta_{12})^2 \,.
\end{align}
\end{subequations}
This term may appear a little baffling. To understand why it is necessary, recall that $\sigma(m_1,m_2;\theta)^{-1}$ has a simple zero at $\theta=\frac{i \pi}{k} (2k-m_1-m_2)$ which is inside the physical strip when $k<m_1+m_2<2k$; as a consequence of this fact $\sigma(k-1,k-1; \theta_1- \theta_2)^2$ has a pole of order two at $\theta=\frac{2i \pi}{k}$. However, this pole should not appear in $\sigma^{\bullet\bullet}_{\text{\tiny RR}} (u_1,u_2)^{2}$. Hence, the factor of $([2]_{\theta})^{-2}$ is precisely what is needed. 
We conclude that, with the normalisation used in section~\ref{sec:fulltheory:crossing}, the dressing phases
$\sigma^{\bullet\bullet}_{\text{\tiny LL}}$, $\sigma^{\bullet\bullet}_{\text{\tiny RR}}$, $\tilde{\sigma}^{\bullet\bullet}_{\text{\tiny LR}}$ and $\tilde{\sigma}^{\bullet\bullet}_{\text{\tiny RL}}$ have no poles and zeros in the physical strip after the relativistic limit.

\subsection{Mixed-mass and massless dressing factors}
\label{sec:relativistic:masslessdressing}
The remaining dressing factors are split into two groups: mixed-mass dressing factors (associated with the scattering of a massive and a massless particle) and massless-massless dressing factors. In the relativistic limit the single-particle massless representations are $\repr{B}{rel}(0, \theta)$ and $\repr{B}{rel}(k, \theta)$. Indeed due to the monodromy of the supercharges, after the limit we can identify $\repr{F}{rel}(0, \theta)=\repr{B}{rel}(k, \theta)$ and $\repr{F}{rel}(k, \theta)=\repr{B}{rel}(0, \theta)$, and all representations become $2k$-periodic in $m$. In the following paragraphs, we provide solutions to the crossing equations involving these representations.

\paragraph{Mixed-mass dressing factors.}

Analogously to what we did for the scattering between massive particles let us define the relativistic limit of the complete mixed-mass S~matrices as follows
\begin{subequations}
\label{eq:complete_mixed_mass_S_matrix_after_limit}
\begin{align}
\label{eq:complete_mixed_mass_S_matrix_after_limit_1}
   &\mathbf{S}_{su(1,1)_{c.e.}^{\oplus 4}}^{\B -} (m, a; \theta)
   = \left[\sigma(m,-;\theta)\right]^{-2} \Bigl( \genl{S}^{\B\B_-}(m, a,\theta) \otimes \genl{S}^{\B\B_-}(m, k-a,\theta) \Bigl) \,, \\
\label{eq:complete_mixed_mass_S_matrix_after_limit_2}
   &\mathbf{S}_{su(1,1)_{c.e.}^{\oplus 4}}^{+ \B} (a, m; \theta)
   = \left[\sigma(+, m;\theta)\right]^{-2} \Bigl( \genl{S}^{\B_+\B}(a, m,\theta) \otimes \genl{S}^{\B_+\B}(k-a, m,\theta) \Bigl) \,,
   \end{align}
\end{subequations}
where $a$ can be either $0$ or $k$, $m$ can be any integer $\in \{1, \dots, k-1\}$ and the subscript signs $\pm$ correspond to the chiralities of the massless particle.
The S~matrices in~\eqref{eq:complete_mixed_mass_S_matrix_after_limit_1} and~\eqref{eq:complete_mixed_mass_S_matrix_after_limit_2} describe the scattering between particles in the representations $ \repr{B}{rel}(m, \theta) \otimes \repr{B}{rel}(m, \theta)$ and $\repr{B}{rel}(a, \theta) \otimes \repr{B}{rel}(k-a, \theta) $, and $\repr{B}{rel}(a, \theta) \otimes \repr{B}{rel}(k-a, \theta) $ and $ \repr{B}{rel}(m, \theta) \otimes \repr{B}{rel}(m, \theta)$ respectively. 
Differently from~\eqref{complete_S_matrix_after_limit}, we set $\Phi=1$ in this case; indeed we do not want to introduce additional poles or zeros in the physical strip since bound states are not expected in scattering processes involving massless particles. 

We find the following two independent sets of crossing equations for the mixed-mass dressing factors 
\begin{subequations}
\label{eq:mixed_mass_crossing_relativistic_limit_1}
\begin{align}
&\left[\sigma(m,-;\theta)\right]^{2} \left[\sigma(k-m,-;\theta+i \pi)\right]^{2}=1 \,,\\
&\left[\sigma(m,-;\theta+i \pi)\right]^{2} \left[\sigma(k-m,-;\theta)\right]^{2}=1 \,,
\end{align}
\end{subequations}
and
\begin{subequations}
\label{eq:mixed_mass_crossing_relativistic_limit_2}
\begin{align}
&\left[\sigma(+,m;\theta)\right]^{2} \left[\sigma(+,m;\theta+i \pi)\right]^{2}= e^{-\frac{2 \pi i m}{k}} \,,\\
&\left[\sigma(+, k-m;\theta)\right]^{2} \left[\sigma(+, k-m;\theta+i \pi)\right]^{2}=e^{\frac{2 \pi i m}{k}} \,.
\end{align}
\end{subequations}
Constant solutions to these equations can be found for any value of $m$.
In the following, we consider the case $m=1$ and $m=k-1$ where these equations need to correspond to the relativistic limit of~\eqref{mixed_mass_full_theory_crossing_eq_1} and~\eqref{mixed_mass_full_theory_crossing_eq_2}.
With the normalisation introduced in~\eqref{eq:massivemassless_normalization} in the relativistic limit the crossing equation for the dressing phases $\phase^{\bullet -}_{\L\L}$, $\phase^{+ \bullet}_{\L\L}$, $\phase^{\bullet -}_{\L\R}$ and $\phase^{+ \bullet }_{\R\L}$ become
\begin{equation}
\label{mixed_mass_rel_limit_crossing_eq_1}
\begin{split}
&\big(\phase^{\bullet -}_{\L\L}(\theta) \big)^{2} \big(\phase^{\bullet -}_{\R\L} (\theta+i \pi) \big)^{2} = 1 \times  {\color{Violet} 1 } =1 \, ,\\
&\big(\phase^{\bullet -}_{\L\L}(\bar{u}_1, u_2) \big)^{2} \big(\phase^{\bullet -}_{\R\L} (u_1, u_2) \big)^{2} = 1 \times {\color{Violet}    1 } =1 \,,\\
&\big(\phase^{+ \bullet}_{\L\L}(\theta) \big)^{2} \big(\phase^{+ \bullet}_{\L \L} (\theta+ i \pi) \big)^{2} = e^{-\frac{2i \pi}{k}} \times {\color{Violet}    e^{\frac{2i \pi}{k}}  } =1 ,\\
&\big(\phase^{+ \bullet}_{\L\R}(\theta) \big)^{2} \big(\phase^{+ \bullet}_{\L \R} (\theta+ i \pi) \big)^{2} = e^{\frac{2i \pi}{k}} \times {\color{Violet}    e^{-\frac{2i \pi}{k}}  } =1 \,,
\end{split}
\end{equation}
and admit the constant simple solution $\phase^{\bullet -}_{\L\L} = \phase^{+ \bullet}_{\L\L} = \phase^{\bullet -}_{\L\R} =\phase^{+ \bullet }_{\R\L}=1$.

For a physical process to make sense a massless particle incoming from the left should have positive velocity (i.e. should be chiral) while a massless particle incoming from the right should have negative velocity (i.e. should be antichiral). However, even if this condition is not satisfied solutions
for the mixed-mass dressing factors can be provided. In particular, 
using the same normalisation~\eqref{eq:complete_mixed_mass_S_matrix_after_limit} also for these unphysical scattering processes, by braiding unitarity it needs to hold that
$$
\left[\sigma(m,+;\theta)\right]^{-2} \left[\sigma(+,m;-\theta)\right]^{-2}=1.
$$
In this manner the `unphysical' dressing factors $\left[\sigma(m,+;\theta)\right]^{-2}$ and $\left[\sigma(-,m;\theta)\right]^{-2}$ can be defined in terms of the `physical' dressing factors in~\eqref{eq:complete_mixed_mass_S_matrix_after_limit}.

\paragraph{Massless-massless dressing factors.}
We define the S~matrices associated with the scattering of massless particles in the chiral-chiral and chiral-antichiral sectors to be
\begin{subequations}
\label{eq:complete_massless_mass_S_matrix_after_limit}
\begin{align}
   &\mathbf{S}_{su(1,1)_{c.e.}^{\oplus 4}}^{+ -} (a, b; \theta)
   = \left[\sigma(+,-;\theta)\right]^{-2} \Bigl( \genl{S}^{\B_+ \B_-}(a, b,\theta) \otimes \genl{S}^{\B_+ \B_-}(k-a, k-b,\theta) \Bigl) \,, \\
\label{eq:complete_massless_S_matrix_after_limit_2}
   &\mathbf{S}_{su(1,1)_{c.e.}^{\oplus 4}}^{+ +} (a, b; \theta)
   = \left[\sigma(+,+;\theta)\right]^{-2} \Bigl( \genl{S}^{\B_+ \B_+}(a, b,\theta) \otimes \genl{S}^{\B_+ \B_+}(k-a, k-b,\theta) \Bigl) \,.
   \end{align}
\end{subequations}
The parameters $a$ and $b$ can either be $0$ or $k$, for a total of four possible choices.
These choices correspond to the fact that we expect two distinct $su(2)_\circ$ representations, denoted by indices $\dot{\alpha}=1,2$, \textit{cf.}~\eqref{chiral_chiral_normalization_massless_particles} and~\eqref{chiral_antichiral_normalization_massless_particles}.%
\footnote{%
Remark that here we are assuming that the $su(2)_\circ$ structure of the S~matrix is trivial.  
}
As before, the superscript signs on $\sigma^{\pm\pm}$, \textit{etc.}, label the chiralities of the massless particle.
As we can see from the relations in~\eqref{opposite_chirality_massless_scattering_S_matrix_elements} the S-matrix elements associated with the scattering of massless particles with opposite chirality are trivial. We obtain the following simple crossing equations for the scattering of massless particles with opposite chirality
\begin{equation}
\left[\sigma(+,-;\theta)\right]^{2} \left[\sigma(+,-;\theta+i \pi)\right]^{2}=+1 \,,
\end{equation}
which corresponds to the relativistic limit of~\eqref{eq:massless_massless_crossing_equations_full_theory_2} and has a trivial solution.

In contrast, as shown from the S-matrix elements in~\eqref{eq:masslessS_after_limit}, the scattering between particles of the same chirality is nontrivial; in the case in which the scattered particles are both chiral the crossing equation can be read from the limit of~\eqref{eq:massless_massless_crossing_equations_full_theory_1} and is given by
\begin{equation}
\label{eq:massless_same_chirality_limit_crossing}
\left[\sigma(+,+;\theta)\right]^{-2} \left[\sigma(+,+;\theta+i \pi)\right]^{-2}= \tanh^2{\frac{\theta}{2}} \,.
\end{equation}
which admit as minimal solution
\begin{equation}
\label{eq:solutions_massless_crossing_limit_from_original_theory}
\left[\sigma(+,+;\theta)\right]^{-2}= a(\theta) \biggl( \frac{R(\theta- i \pi) R(\theta+ i \pi)}{R^2(\theta)} \biggl)^2
\end{equation}
The $R$-functions on the RHS of the equality above are provided in~\eqref{R_function_definition}. 
Up to an auxiliary function $a(\theta)$, which needs to satisfy
\begin{equation}
a(\theta) a(\theta+i \pi)=-1 \ \ \ \,, \ \ \ a(\theta) a(-\theta)=1 \,,
\end{equation}
and needs to be a phase for $\theta \in \mathbb{R}$, the expression in~\eqref{eq:solutions_massless_crossing_limit_from_original_theory} is equal to the sine-Gordon dressing factor. A possibility for the function $a(\theta)$ was provided in~\cite{Frolov:2021fmj} in the resolution of the dressing factors of the pure Ramond-Ramond worldsheet theory and is given by
\begin{equation}
\label{eq:definition_auxiliary_function_a}
a(\theta)= -i \tanh{\Bigl( \frac{\theta}{2} - i \frac{\pi}{4} \Bigl)} \,.
\end{equation}
Similar solutions can be obtained for the dressing factors associated with the scattering of massless particles with negative chirality.

We remark that the solution in~\eqref{eq:solutions_massless_crossing_limit_from_original_theory} was obtained by taking the limit of the S-matrix of the full theory first and then solving the associated crossing equations in the relativistic limit. 
It is however important to mention that a different derivation is possible: this derivation consists in constructing the representations after the limit and bootstrapping the S~matrix again from scratch, as mentioned at the beginning of section~\ref{sec:relativistic:boundstates}. In appendix~\ref{app:relativisticS}, following this different derivation, we show that a larger space of solutions is admitted for the scattering of massless particles of the same chirality and~\eqref{eq:solutions_massless_crossing_limit_from_original_theory} corresponds to a particular point in the space of these solutions.
Interestingly, requiring the model to have two irreducible massless representations constructed as tensor products of two-dimensional building blocks~$\repr{B}{rel}$ and $\repr{F}{rel}$, fixes the solution to be precisely~\eqref{eq:solutions_massless_crossing_limit_from_original_theory}.

\section{Conclusions}
\label{sec:conclusions}
In this paper we have studied the worldsheet theory emerging from mixed-flux $AdS_3\times S^3\times T^4$  in lightcone gauge.
In the full non-relativistic theory we have considered the Zhukovsky-plane and rapidity-plane kinematics of the model, and the allowed structure of bound states.
Then, we have studied  a relativistic limit of model.The resulting model is an integrable, supersymmetric and relativistic QFT in two dimensions. As it could have been expected from the discussion of~\cite{Sfondrini:2020ovj}, its particle content is dictated by the WZNW level~$k$. More precisely, it has  $(k-1)$ massive multiplets, with masses
\begin{equation}
    \mu\in\left\{
    2\sin\left(\frac{\pi}{k}\right),
    2\sin\left(\frac{2\pi}{k}\right),\dots,
    2\sin\left(\frac{(k-2)\pi}{k}\right),
    2\sin\left(\frac{(k-1)\pi}{k}\right)
    \right\}\,,
\end{equation}
as well as two massless multiplets. Notice that all masses come in pairs if $k$ is odd, as $\sin\tfrac{\pi}{k}=\sin\tfrac{(k-1)\pi}{k}$. The resulting pairs of representations give particle-antiparticle pairs. The case of $k$~even is a little special, as there is a single representation of mass~$\mu=2\sin\tfrac{\pi}{2}=2$ which is its own charge conjugate. 
The model has a rich structure of bound states, which we have described in some detail, and that allows one to generate all massive multiplets starting from a multiplet of mass~$\mu= 2\sin\tfrac{\pi}{k}$ and using fusion.
The  representations generated in this way have higher and higher mass initially, and then (after considering a bound state of $\sim k/2$ particles) the mass start decreasing. In this sense, the antiparticle of a given excitation is also a bound-state of several such excitations.

An immediate consequence of this discussion is that for $k=1$ there are no massive particles (but only the $T^4$ modes, sitting in two massless representations), as expected from the WZNW model description at the NSNS point. A second observation, which would be important to understand in more detail, is that $k=2$ is special too. In that case there is only one massive representation, which is its own charge-conjugate. In other words, the total number of particles at $k=2$ is lower than what we would expect from just counting the fundamental excitations in a near-pp-wave expansion of the string model (we would expect, na\"ively, \textit{two massive} and two massless fundamental representations, rather than \textit{one massive} and two massless representations). This is not entirely surprising if we consider that, at the NSNS point, the theory can be described by the RSN approach. In partciular, we need to consider a worldsheet-supersymmetric WZW model based on the Ka\v{c}-Moody algebra $sl(2)^{(1)}_{k}\oplus su(2)^{(1)}_{k}$ which requires extra care at~$k=2$, as the bosonic part of $su(2)^{(1)}_{k}$ becomes trivial. This seems to fit with our observation, but it would be interesting to analyse this special case in more detail.

An important result of our work is the construction of the dressing factor of the relativistic models that we considered. Before the relativistic limit, the construction of the dressing factors of the mixed-flux theory is a major obstacle to the construction of the mirror TBA equation and the quantitative study of the theory by integrability. After the limit, relativistic invariance drastically simplifies the analytic structure of the S~matrix and it allows us to solve the crossing equations. The minimal solution for the dressing factors is then expressed in terms of product of Barnes $G$-functions, \textit{cf.}~\eqref{minimal_sigma_ratio_of_R_functions}; the bound-state poles can be taken into account by introducing suitable CDD factors. These results constitute a test for future proposals of the dressing factors of the full theory. This is quite crucial because there are currently no perturbative constraints on the dressing factors at small string tension (unlike what was the case in $AdS_5$ and $\mathcal{N}=4$ SYM) and even at strong tension  it is not clear to what extent the existing perturbative computations can be trusted, due to infrared divergences (\textit{cf.}\ the discussion in~\cite{Frolov:2021fmj}).

The integrable models which we have encountered here in the relativistic limit may be of interest in and of themselves, as supersymmetric integrable QFTs. Indeed, the building blocks of our constructions are closely related to the $\mathcal{N}=2$ models considered in~\cite{Fendley:1991ve,Fendley:1992dm}. It is worth remarking that we encountered some interesting new physics  when considering the collinear scattering of massless particles. In that case, as discussed in appendix~\ref{app:relativisticS}, we find a \textit{one parameter} family of integrable S~matrices, complete with crossing-invariant dressing factors, which to our knowledge were previously unknown.

Finally, it might be interesting to study the TBA of the relativistic model which we constructed. The low-energy relativistic limit which we considered is well defined at the level of the S~matrix. It is not immediately clear what its interpretation may be at the level of the spectrum and of string theory (or of the unknown CFT dual). Nonetheless,  it is a perfectly well-defined relativistic model whose spectrum will capture some of the features of the original theory. 
This could be a stepping stone towards constructing the mirror TBA equations for the full model with mixed-flux, so far only known for the pure-RR~\cite{Frolov:2021bwp,Brollo:2023pkl,Frolov:2023wji} and pure-NSNS~\cite{Dei:2018mfl} cases.%
\footnote{For the pure-RR case, a set of ``quantum spectral curve'' equations has also been proposed~\cite{Ekhammar:2021pys,Cavaglia:2021eqr,Cavaglia:2022xld}. These should encode the same information about the spectrum as the mirror TBA, but currently the relation between the two proposals remains unclear.}
 
We hope to return to some of these questions in the future.

\section*{Acknowledgments}
We thank Matheus Augusto Fabri, Alessio Miscioscia, and Roberto Volpato for useful related discussions.
The authors also thank the participants of the workshop ``Integrability in Low-Supersymmetry Theories'' in Filicudi, Italy, for a stimulating environment where part of this work was carried out.
DP and AS acknowledge support from the European Union – NextGenerationEU, and from the program STARS@UNIPD, under project ``Exact-Holography – A new exact approach to holography: harnessing the power of string theory, conformal field theory, and integrable models.''

\appendix

\section{\texorpdfstring{$\kappa$}{kappa}-deformed Zhukovsky map} 
\label{app:kappaZhukovsky}

In this appendix we discuss the properties of the $\ka$-deformed Zhukovsky map   
defined through the following equation
\bal
\label{uplane}
u(x,\ka)=x+\frac{1}{x}- {\frac{\kappa}{\pi}}\,\log x  \quad\Leftrightarrow \quad x = x(u,\kappa)\,,
\eal
where $\ka$ in general is a complex parameter. Equation \eqref{uplane} defines a map from the Riemann surface of $\log x$  to the $u$-plane, and we want to analyse the inverse map given by $x(u,\kappa)$. Clearly, the function $x(u,\kappa)$ is  multi-valued, and in the limit $\ka\to0$ it becomes the usual inverse Zhukovsky map, and has two branches.  For finite $\ka$ it is sufficient to analyse eq.\eqref{uplane} on any branch  of $\log x$, and we use the principal branch $\ln x$ of $\log x$ with the cut  $(-\infty,0)$ on the $x$-plane.  In what follows we use  the notation $x(u,\ka)$ to denote the multi-valued function satisfying \eqref{uplane} on the principal branch $\ln x$, and then $x^{(n)}(u,\ka)$ satisfying \eqref{uplane} on the $n$-th branch of $\log x$ is given by
\bal
x^{(n)}(u,\ka)= x(u+2i\ka\, n\,,\,\ka)\,.
\eal
The equation  \eqref{uplane}  enjoys a very important  inversion symmetry:
if $x(u,\ka)$ solves \eqref{uplane} then ${1\ov x(u,-\ka)}$ also solves the equation. 
As a result, the $x$-plane with the cut  $(-\infty,0)$ covers the $u$-plane twice, and the function  $x(u,\ka)$  has two branches.
In what follows we will be interested in the case $\ka\in\bR$. Then,  due to this inversion symmetry, it is sufficient to analyse the properties of   $x(u,\ka)$ with $\ka>0$. The function $x(u,-\ka)$ has the properties of 
${1\ov x(u,\ka)}$.

 With our choice of the cut on the $x$-plane the complex conjugate function $x(u,\ka)^*$  satisfies the equation
\bal
\label{eq:uccplane}
x(u,\ka)^*+\frac{1}{x(u,\ka)^*}- {\kappa^*\ov\pi}\,\ln x(u,\ka)^*  = u^*\,,
\eal
and one can impose the following two conjugacy conditions
\bal
x(u,\ka)^* = x(u^*,\ka^*)\,,\quad  x(u,\ka)^* = {1\ov x(u^*, -\ka^*) }\,,
\eal
which can be used to define  two different sets of branches of $x(u,\ka)$.

 In string theory $\ka^*=\ka$, and we want the function $x(u,\ka)$ to satisfy
\bal\la{stringcc}
x(u,\ka)^* = x(u^*,\ka)\,,\quad \ka^*=\ka\,,
\eal
because then the momentum and energy are  real for real $u$. We often refer to the branch satisfying the reality  condition \eqref{stringcc} and  containing the point  $x=+\infty$  as to the string $u$-plane, and to the second branch as to the anti-string $u$-plane.

 In mirror theory if we keep $\ka$ real,  we get the condition
\bal\la{mirrorcc}
x(u,\ka)^* = {1\ov x(u^*, -\ka) }\,,\quad \ka^*=\ka\,,
\eal
and we refer to the branch satisfying the reality  condition \eqref{mirrorcc} and the condition $\Im(x)<0$ as to the mirror $u$-plane, and to the second branch satisfying  the condition $\Im(x)>0$ as to the anti-mirror $u$-plane.
The condition  \eqref{mirrorcc}  relates two different functions, and the mirror theory is not unitary.\footnote{ It might be interesting to assume that in mirror theory $\ka^* = -\ka$, so it is purely imaginary, and get string theory not just by Wick rotation but also by the analytic continuation in $\ka$. Then, $x(u,\ka)$  satisfies
\bal\nn
x(u,\ka)^* = {1\ov x(u^*, \ka) }\,,\quad \ka^*=-\ka
\eal
It is unclear whether it is necessary, and in what follows we assume that $\ka$ is real.}

\medskip

To find the location of the branch points we compute 
\bal
{dx\ov du} = {x^2\ov (x-\x_+) (x-\x_-)}\,,\quad \x_\pm\equiv {\ka\ov2\pi}\pm \sqrt{1+{\ka^2\ov4\pi^2}}
\eal 
These formulae show that a better parametrisation of $\ka$ might be 
\bal
\ka = 2\pi\sinh\eta\quad \Rightarrow \quad\x_\pm = \pm e^{\pm\eta}
\eal
which makes obvious that $\x_\pm(\ka) = 1/\x_\pm(-\ka)$  as expected from the inversion symmetry.

The zeroes and poles of $dx/du$ potentially correspond to branch points on the $u$-plane where a branch of $x(u,\ka)$ is defined. 
Clearly, there may be a branch point located at
\bal
\uu_+ = \x_++\frac{1}{\x_+}- {\kappa\ov\pi}\,\log \x_+ &= + 2\sqrt{{\kappa ^2\ov4\pi^2}+1}-{\kappa\ov\pi}  \ln \left({\ka\ov2\pi}+ \sqrt{1+{\ka^2\ov4\pi^2}}\right) 
\\
&= +2\cosh\eta-2\eta\sinh\eta
\eal
This  branch point $\uu_+$ is of the square-root type, and, as we will see later, going around it  $x(u,\ka)$ transforms  according to the inversion symmetry
\bal
x^\circlearrowleft(u,\ka)={1\ov x(u,-\ka)}\,,
\eal
where $x^\circlearrowleft$ is the result of the analytic continuation along a path $\circlearrowleft$ surrounding the  branch point $\uu_+$. Note also that $\uu_+$ does not depend on the sign of $\ka$: 
$\uu_+(-\ka) = \uu_+(\ka)$.

Since $\x_-$ is negative for $\ka$ real, there may be in fact two branch points located at
\bal
\uu_-^{\pm} = \x_-+\frac{1}{\x_-}- {\kappa\ov\pi}\,\ln \x_- &= -2\sqrt{{\kappa ^2\ov4\pi^2}+1}-{\kappa\ov\pi}  \ln \left({\ka\ov2\pi}- \sqrt{1+{\ka^2\ov4\pi^2}}\right) 
\\
&= -2\cosh\eta+2\eta\sinh\eta \mp i\ka=-\uu_+  \mp i\ka\,,\quad \ka\in\bR
\eal
where $+$ in $\uu_-^{\pm}$ is for $\x_-$ on the upper edge of the cut of $\log x$, and we have used the principal branch of $\log x$. Since the images of 
these two branch points are on the edges of the  cut of $\log x$ moving a point around any of them takes it to a different $x$-plane. These  branch points are also
 of the square-root type, and going around $\uu_-^{\pm}$ along a path $\circlearrowleft_-^{\pm}$ transforms $x(u,\ka)$ as
\bal
x^{{\circlearrowleft }_-^{\pm}}(u,\ka)={ x(u\pm 2i\ka,\ka)}\,.
\eal
 
 In addition there is a branch point at $u=\infty$ corresponding to $x=0$ and $x=\infty$ which is of the logarithmic type  as can be seen by solving \eqref{uplane} for large $u$
 \bal
 x(u,\ka) = u - {\ka\ov\pi}\log u+ \cdots \quad \text{or}\quad x(u,\ka)  = {1\ov u + {\ka\ov\pi}\log u} + \cdots
 \eal
 The result of the analytic continuation along a path   surrounding $u=\infty$ depends on the cut structure of a $u$-plane, the orientation of the path and the initial point of the path. It will be discussed later. 
 
 In what follows we always choose all cuts on a $u$-plane to be horizontal. 
 We will see that there is a branch of $x(u,\ka)$ where there are only two branch points at $\uu_+$ and $-\infty$, and we refer to the branch as the principal branch of  $x(u,\ka)$.
  
To understand where a $u$-plane is mapped onto the $x$-plane, and how cuts can be chosen,  let us find which curves  on the $x$-plane are mapped to horizontal lines of the $u$-plane. We use polar coordinates
 \bal
 x=\rho\, e^{i\p}\,,
 \eal
and rewrite \eqref{uplane} in the form
\bal
\label{eq:uplanepolar}
\left(\rho+\frac{1}{\rho} \right)\cos\p - {\kappa\ov\pi}\,\log \rho  +i\left[ \left(\rho-\frac{1}{\rho} \right)\sin\p -{\ka\ov\pi}\p\right] = u=\mu+i\,\nu\,.
\eal
Thus, we get that the equation of the curves which are mapped to (a segment of) the horizontal line through the point $(0,\nu)$ where $\nu=\Im(u)$  is given by
\bal
\label{eq:horline}
 \left(\rho-\frac{1}{\rho} \right)\sin\p -{\ka\ov\pi}\p = \nu\,, \quad -\pi\le \p\le \pi\,,
\eal
and therefore if $\nu\neq 0$ or $\nu\neq \pm \ka $, the  solution is
\bal\label{eq:rhophiv}
\rho(\phi,\nu)= {{\kappa\ov\pi}\, \phi +\nu \ov 2\sin \phi }+\sqrt{ 1+{({\kappa\ov\pi}\, \phi+\nu)^2  \ov 4\sin^2 \phi }}\,.
\eal
On the $x$-plane the solution is represented by two disconnected curves, one in the lower half-plane and the other in the upper one. Each of the curves is mapped to the whole horizontal line,
as can be seen from the formula
\bal\label{eq:Reuphiv}
 \mu(\phi,\nu) = 2\sqrt{ 1+{({\kappa\ov\pi}\, \phi+\nu)^2  \ov 4\sin^2 \phi } }\, \cos\p  - {\kappa\ov\pi}\,\log \rho(\phi,\nu)\,,\quad \mu=\Re(u)\,.
\eal
We will discuss these curves in more detail later but first let us consider the cases
where  $\nu= 0$ or $\nu= \pm \ka $.  For each of the three cases the corresponding horizontal line goes through a branch point, and analysing which curves on the $x$-plane are mapped to these lines we can understand how to choose cuts.

\beeRd
\item
We begin with the case $\nu=0$, and  get
\bal
\label{eq:ugeubp}
 \nu=0:\qquad \left(\rho-\frac{1}{\rho} \right)\sin\p -{\ka\ov\pi}\p = 0\,, \quad -\pi\le\p\le\pi\,.
\eal
 This equation has two solutions. The first one is
 \bal
 \phi =0\,,\quad \rho\ge 0\,,\quad u = \rho+\frac{1}{\rho}   - {\kappa\ov\pi}\,\log \rho\, \ge\, \uu_+\,.
 \eal
 In fact both intervals $0\le\rho\le\x_+$ and  $\rho\ge\x_+$ are mapped to the semi-line $u\, \ge\, \uu_+$.  If we choose  the semi-line $u\, \ge\, \uu_+$ to be a cut of a $u$-plane and consider the lower half-plane $-\pi<\p<0$ then the lower  edge of the cut ($v=-0$) is mapped to the semi-line $\rho\ge\x_+$ while the upper edge of the cut ($v=+0$) is mapped to the  interval $0\le\rho\le\x_+$  on the $x$-plane, as can be seen from \eqref{eq:rhophiv}. This is a long mirror theory cut which  in the limit $\ka\to0$  becomes a   cut from $+2$ to $+\infty$.

 The second solution of \eqref{eq:ugeubp} is given by
 \bal\la{eq:rhophiv0}
\rho(\phi,0)= {\kappa\, \phi  \ov2\pi\sin \phi }+\sqrt{ 1+{\kappa^2\, \phi^2  \ov 4\pi^2\sin^2 \phi }}\,,\quad \rho(0,0) = \x_+ \,.
\eal
This curve covers the semi-line $u\, \le\, \uu_+$ twice.
Depending on whether $\ka>0$ or $\ka<0$ the plots of the images of the cut are very different but, as expected, they are related by $x(u,-\ka)=1/x(u,\ka)$, see
plots for $\ka=\pm 0.3, \pm 1, \pm 3 $  in the figures below.
\begin{center}
\includegraphics*[width=0.5\textwidth]{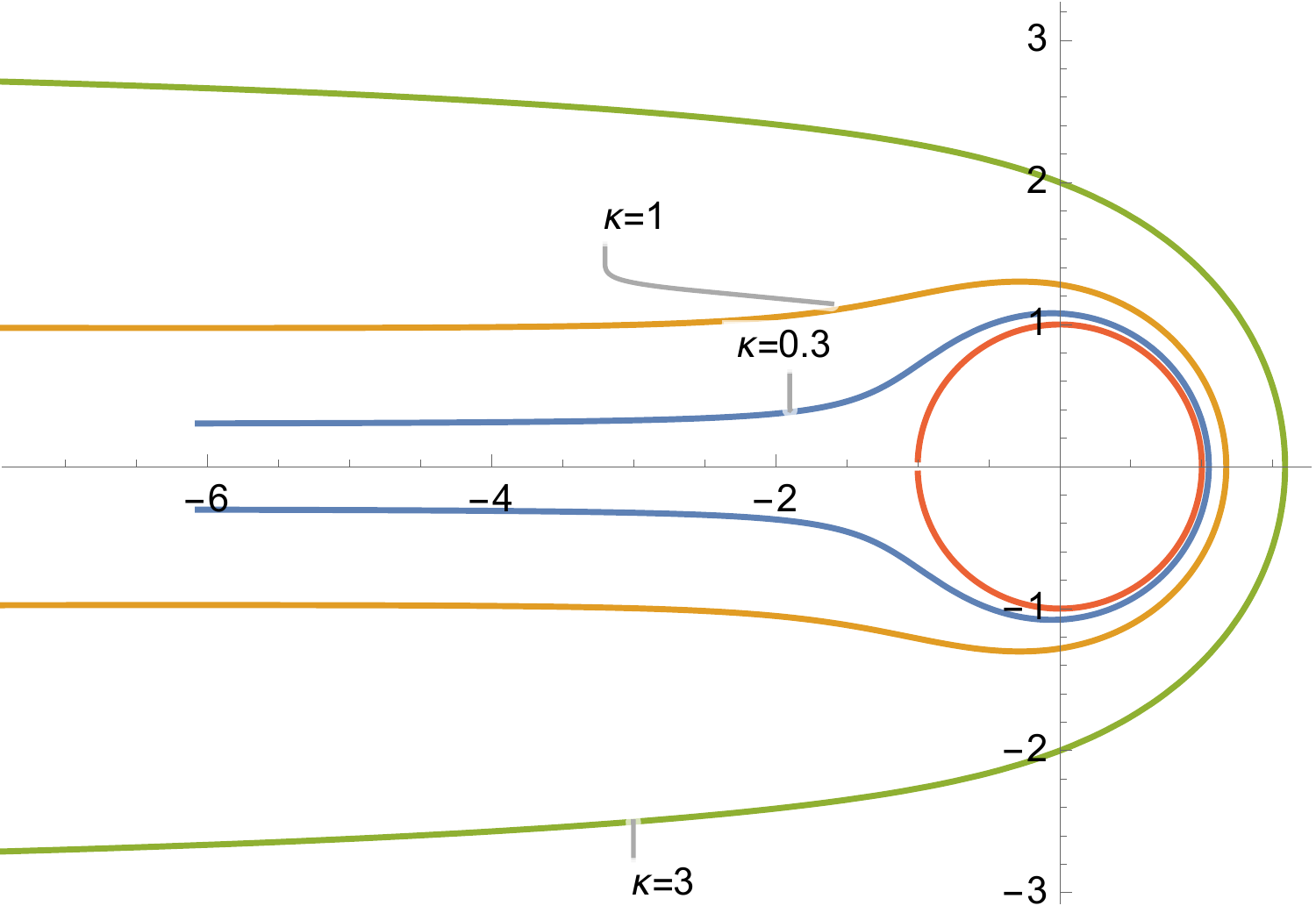} \quad \includegraphics*[width=0.35\textwidth]{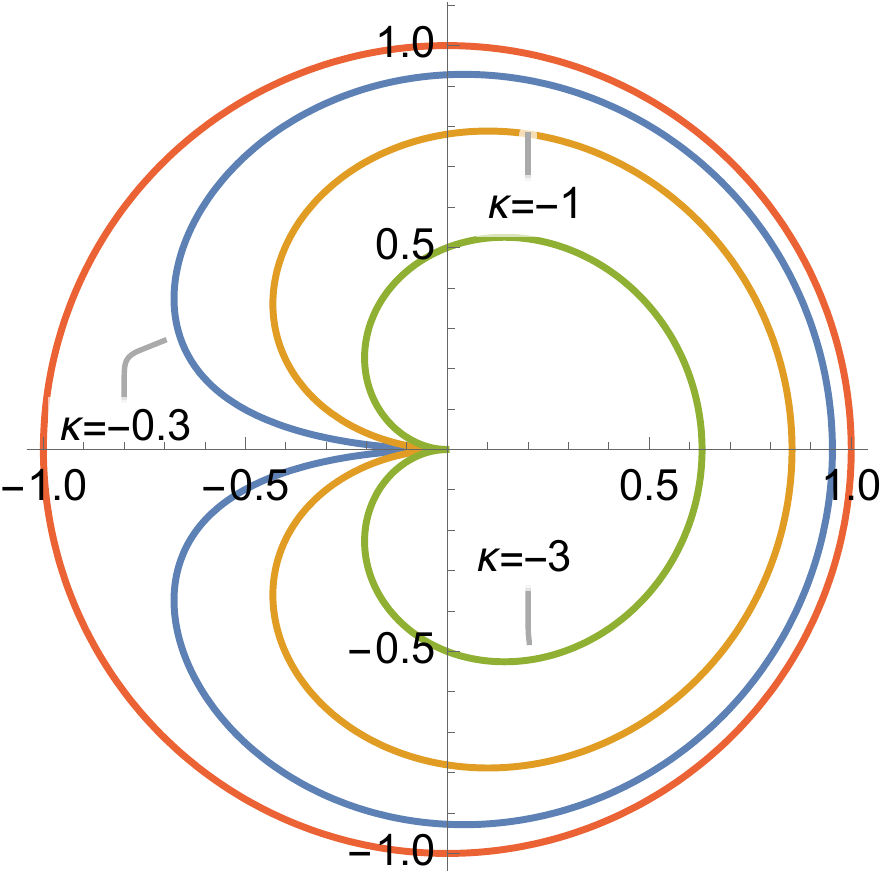} 
\end{center}
We see that the curves separate the $x$-plane into two regions, and, as was mentioned above,  we  choose the region that contains the point $x=+\infty$, and therefore the  semi-line $x\, \ge\, \x_+$ as the string theory physical region. For both signs of $\ka$ it is the region exterior to the one bounded by the curve \eqref{eq:rhophiv0}. 
Thus, for $\ka>0$ the string region does not contain the unit disc while for $\ka<0$ the string region  includes the unit circle and its boundary  is inside the unit disc. 
If we choose  the semi-line   $u\, \le\, \uu_+$ to be a cut of a $u$-plane and consider the string region  then the lower  edge of the cut is mapped to the lower part of the curve while the upper edge of the cut ($v=+0$) is mapped to the upper one. This is a long string theory cut which  in the limit $\ka\to0$  becomes a   cut from $-\infty$ to $+2$.

\item
Let us now consider the case $\nu=  \ka $
\bal
\label{eq:horlinevkapi}
 \nu=\ka :\qquad \left(\rho-\frac{1}{\rho} \right)\sin\p -{\ka\ov\pi}\p = \ka \,, \quad -\pi\le\p\le\pi
\eal
This equation also has two solutions.  The first one is
 \bal
 \phi =-\pi \,,\quad \rho\ge 0\,,\quad \Re (u) = -\rho-\frac{1}{\rho}   - {\kappa\ov\pi}\,\log \rho\, \le\, \Re(\uu_-^{-})
 \eal
 The semi-line $\Re (u)  \le\, \Re(\uu_-^{-})$ is the cut on the $u$-plane from $-\infty$ to $\uu_-^{-}$. 
The intervals $ \x_- \le x\le 0$ and  $x\le  \x_- $
on the lower edge of the cut $(-\infty,0)$ on the $x$-plane
are mapped to upper and lower edges  of the cut $(-\infty,\uu_-^{-})$, respectively. Since we have  chosen the principal branch of  $\log x$  on the $x$-plane, we cannot change the cut $(-\infty,\uu_-^{-})$ on the $u$-plane. If we would do so then we would have to change a branch of  $\log x$ correspondingly.  Clearly, the  interval $-\infty\le x \le 0$ is outside the string region for $\ka>0$ but inside it for $\ka<0$.  Thus, there is no cut $\Re (u)  \le\, \Re(\uu_-^{-})$  on the $u$-plane that is mapped to the string region for $\ka>0$ but it is there for $\ka<0$.
 The cut on the $u$-plane from $-\infty$ to $\uu_-^{-}$  in the limit $\ka\to0$ would become a long mirror theory cut from $-\infty$ to $-2$.  Combining it with the long string theory cut from $-\infty$ to $+2$, one gets the short string theory cut from $-2$ to $2$.

The second solution of \eqref{eq:horlinevkapi} is given by
\bal
\rho(\phi,\ka)= {\kappa\, (\phi +\pi) \ov 2\pi\sin \phi }+\sqrt{ 1+{\kappa^2\, (\phi+\pi)^2  \ov 4\pi^2\sin^2 \phi }}\,,\quad \rho(-\pi,\ka) = -\x_- \,.
\eal
It is represented by two disconnected curves located in the lower and upper half-planes,  see the figures below for 
$\ka=\pm 0.3, \pm 1, \pm 1.5 $.
\begin{center}
\includegraphics*[width=0.45\textwidth]{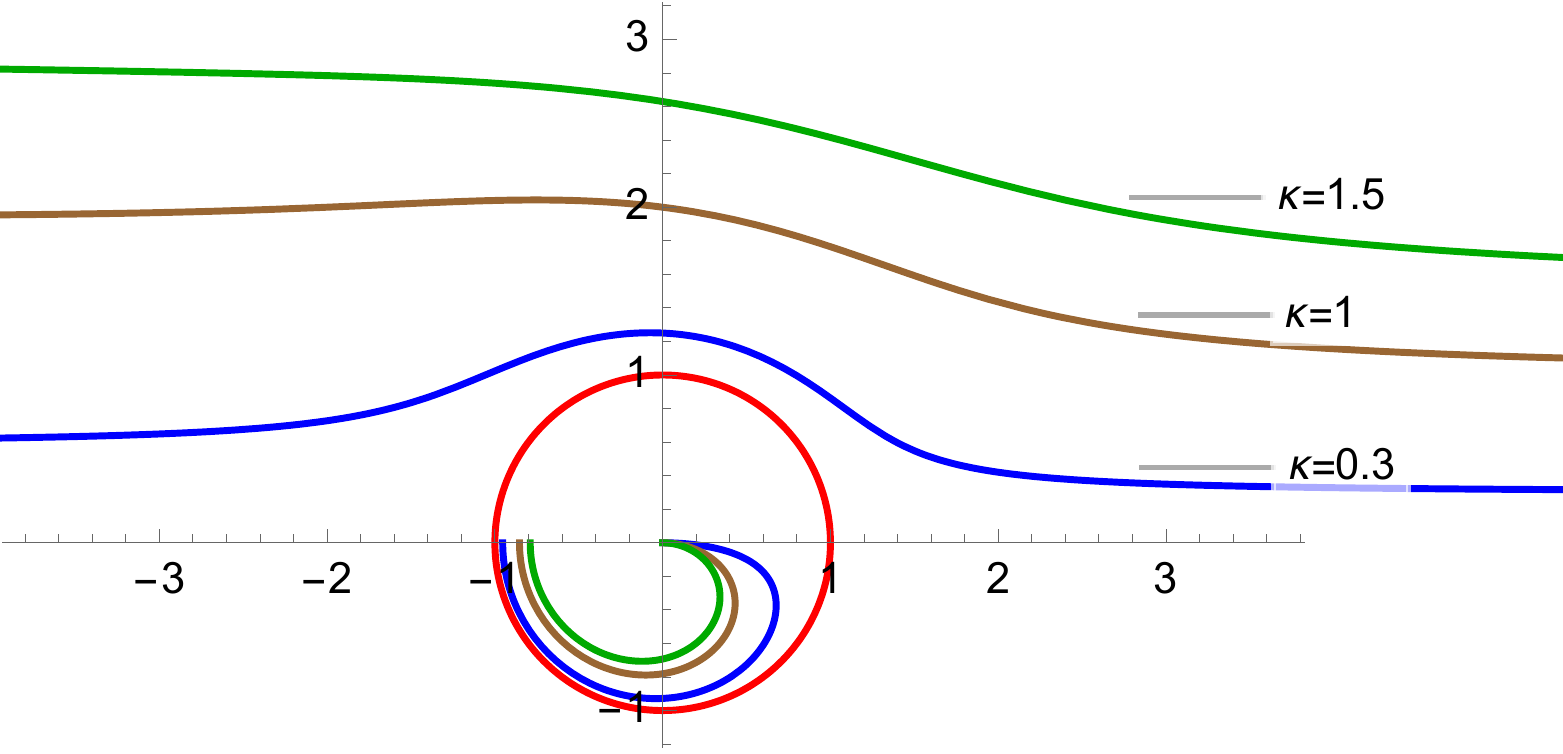} \quad \includegraphics*[width=0.4\textwidth]{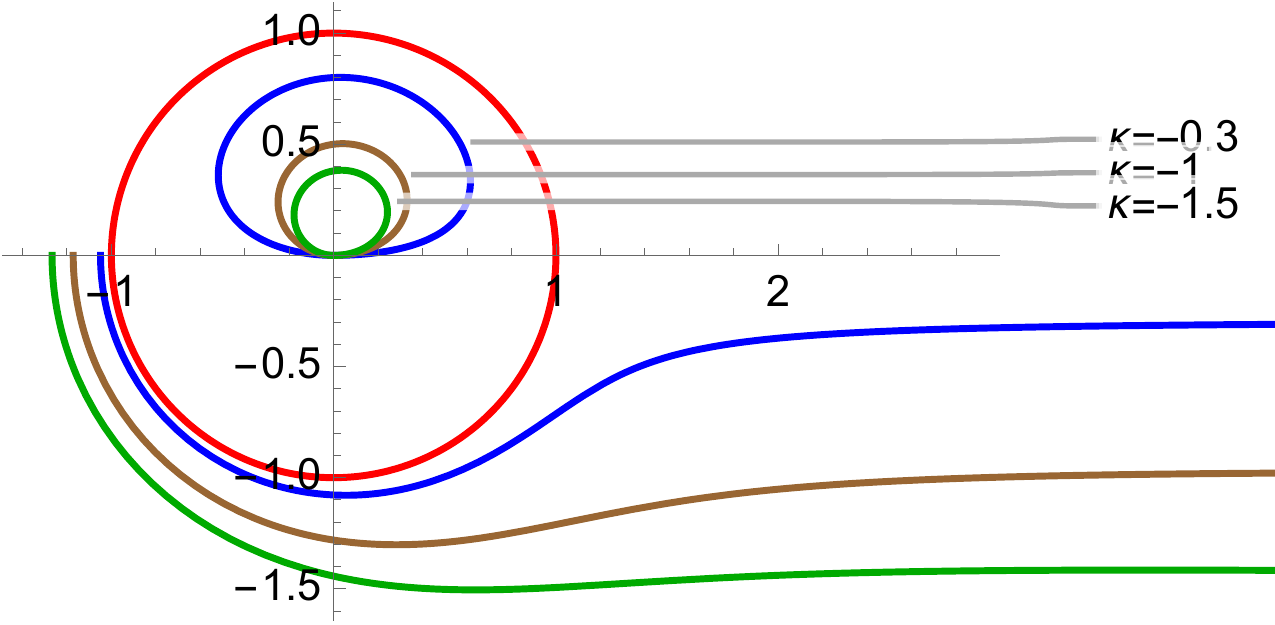} 
\end{center}
The curve  in the lower half-plane on each of the figures  ends at $\x_-$. 
It is   the image of the semi-line $\Re(u) \ge \Re(\uu_-^{-})$  of a $u$-plane. The union of the curve  with  the semi-line $x\le0$ is the image of the two edges  of the cut  from $-\infty$ to $\uu_-^{-}$, and the semi-line $\Re(u) \ge \Re(\uu_-^{-})$ of one and the same $u$-plane. On the other hand the curve in the upper half-plane is mapped to the whole line $\Im (u) = \ka $, and therefore it belongs to a $u$-plane which has no branch point at $u=\uu_-^{-}$.  For $\ka>0$ the curve in the upper half-plane is located in the string physical region, and therefore for $\ka>0$  the string $u$-plane has no branch point at $u=\uu_-^{-}$.  On the other hand for $\ka<0$ the curve is outside the string physical region, and therefore for $\ka<0$  the string $u$-plane has the branch point at $u=\uu_-^{-}$,  see the figures below for 
$\ka=\pm 1$ where the green curve is the image of the cut $(-\infty,\uu_+)$,  the blue curve   in the lower half-plane is the image  of the semi-line $\Re(u) \ge \Re(\uu_-^{-})$,    and the blue curve   in the upper half-plane is the image  of the line $\Im (u) = \ka $.
\begin{center}
\includegraphics*[width=0.5\textwidth]{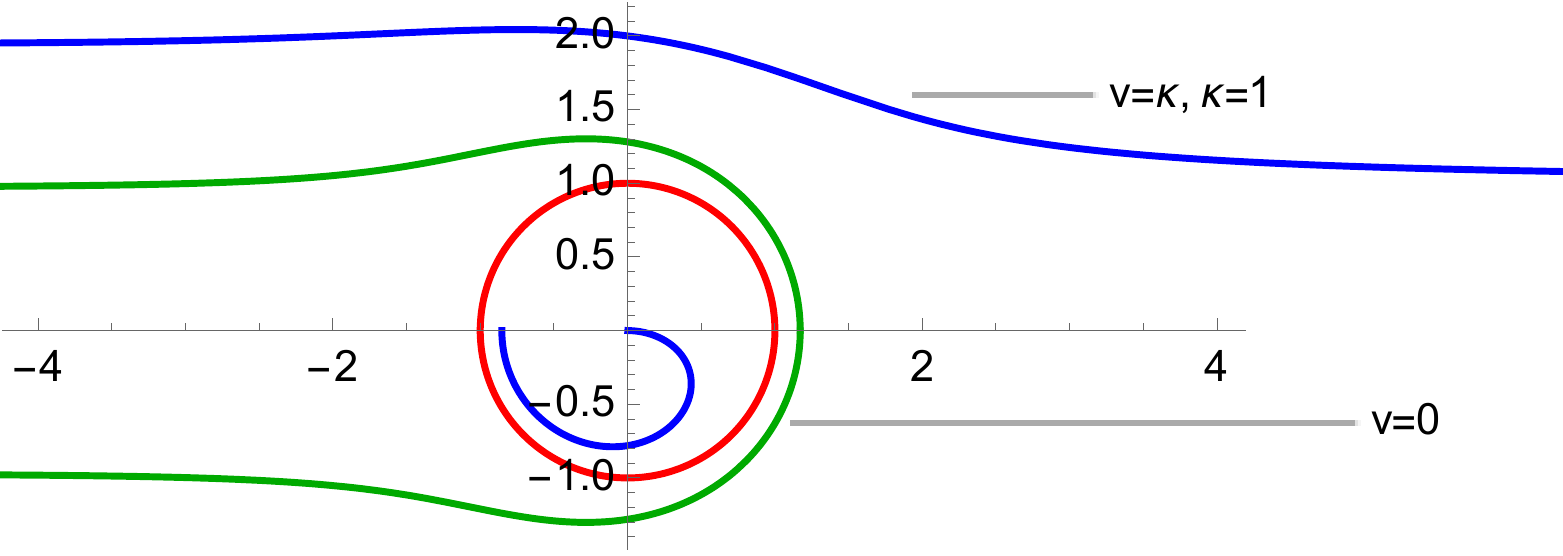} \quad \includegraphics*[width=0.4\textwidth]{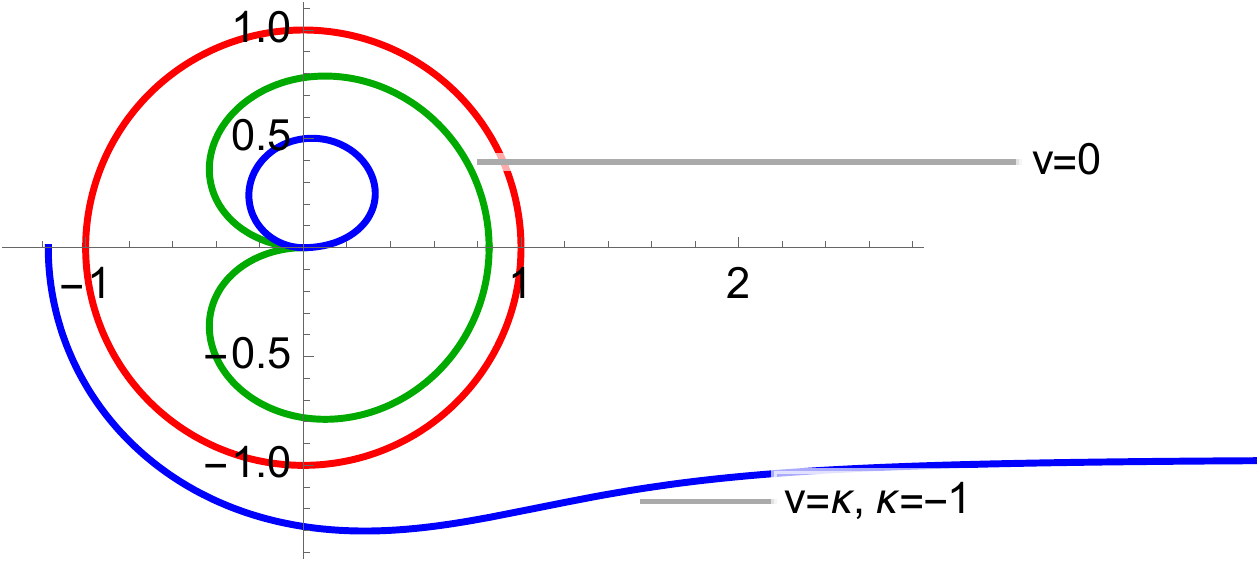} 
\end{center}

\item
The consideration  is immediately applied to $\nu=  -\ka $ because it is related to the case above by the reflection $\p\to-\p$. 
\eee
The images of the three string cuts from $-\infty$ to $\uu_+$, and from $-\infty$ to $\uu_-^{\pm}$ on the $x$-plane are shown in  the figures below for 
$\ka=\pm 1$.
\begin{center}
\includegraphics*[width=0.57\textwidth]{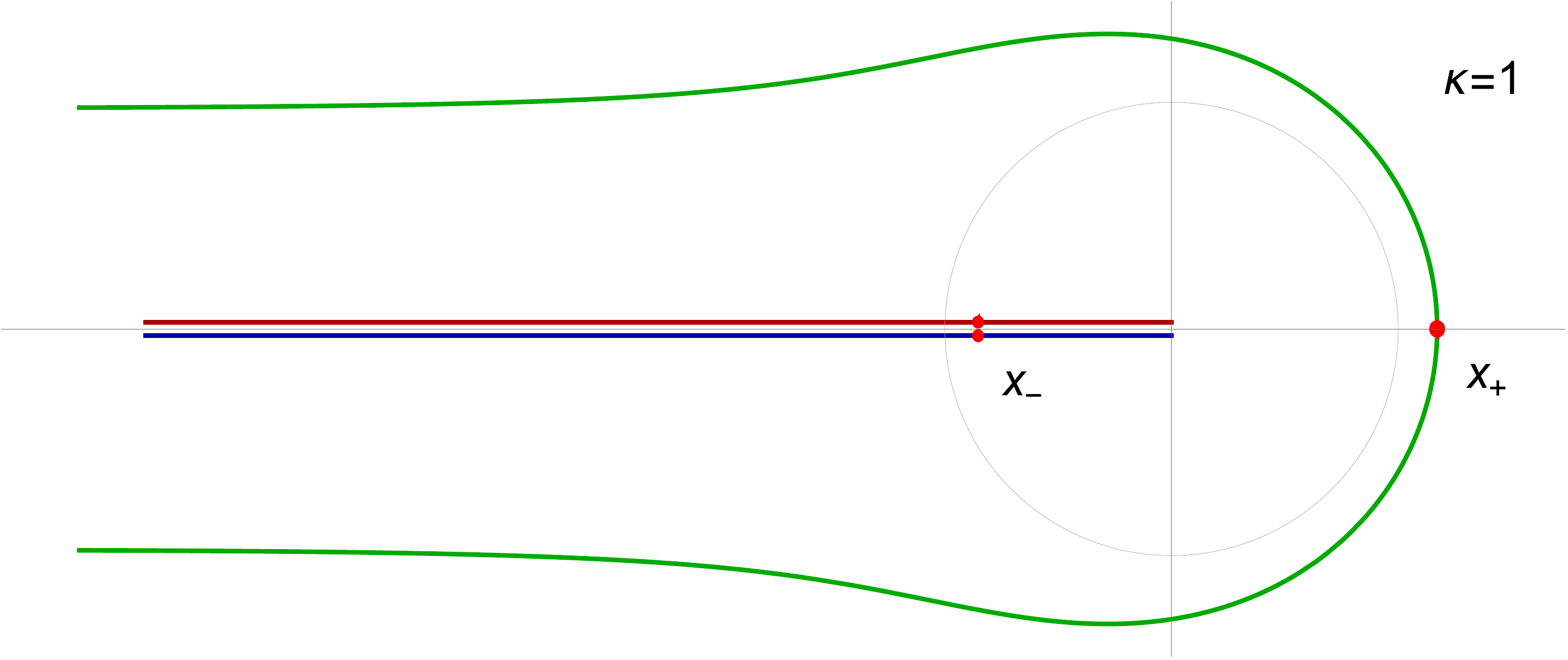} \quad \includegraphics*[width=0.39\textwidth]{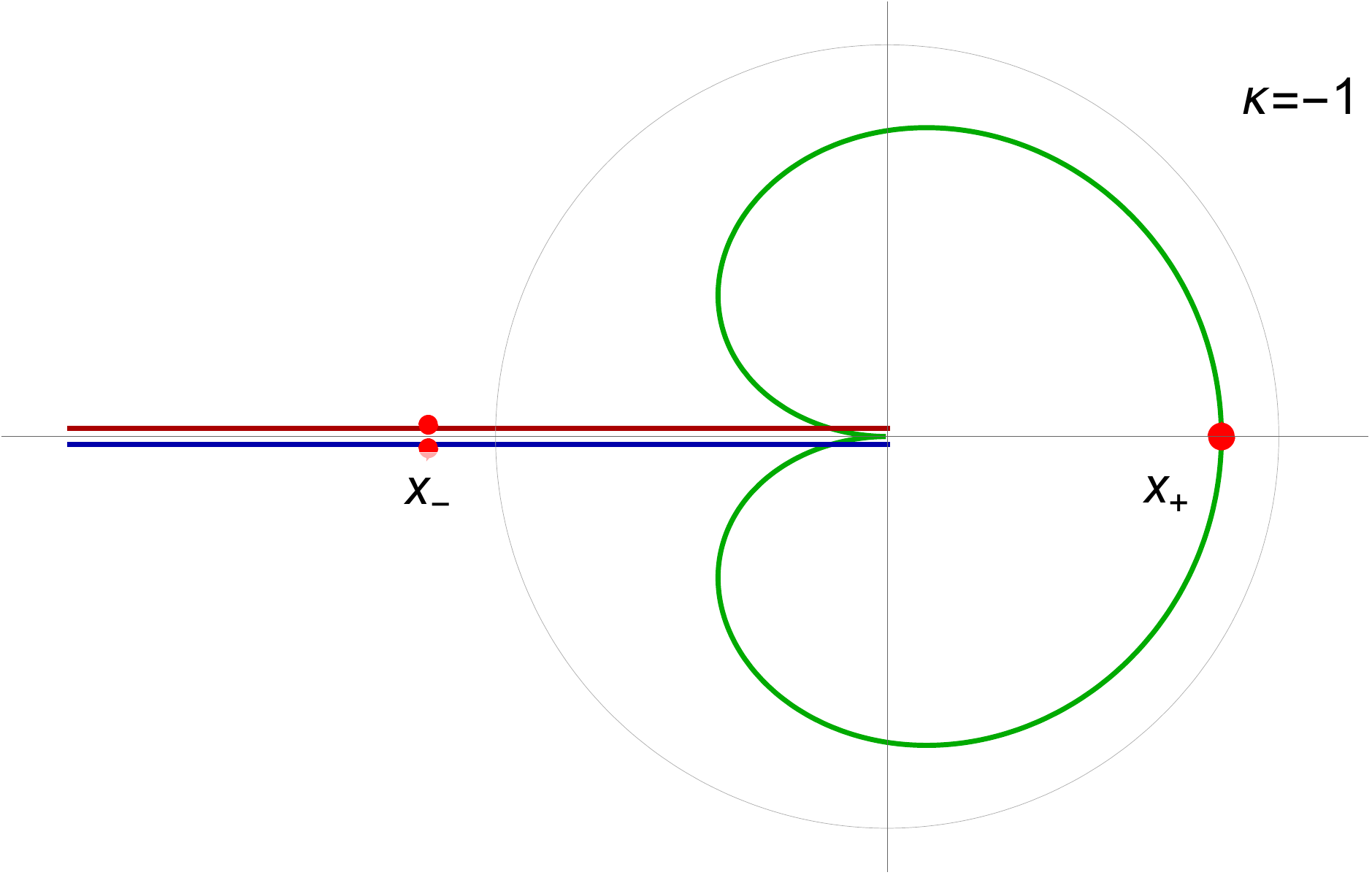} 
\end{center}
Since the semi-axes $x\le0$ is outside the string physical region for $\ka>0$, we conclude that
it is mapped to a $u$-plane which has only one cut from $-\infty$ to $\uu_+$.  For $\ka<0$  the string physical region includes all images of the cuts, and therefore, it is mapped to a $u$-plane which has all the three cuts. For both signs of $\ka$  we  define the principal branch of $x(u,\ka)$ to be the one on a $u$-plane with only one cut from $-\infty$ to $\uu_+$. 

Similarly, the images of three mirror cuts from  $\uu_+$ to $+\infty$, and from $-\infty$ to $\uu_-^{\pm}$ on the $x$-plane are shown in the figures below for 
$\ka=\pm 1$.
\begin{center}
\includegraphics*[width=0.45\textwidth]{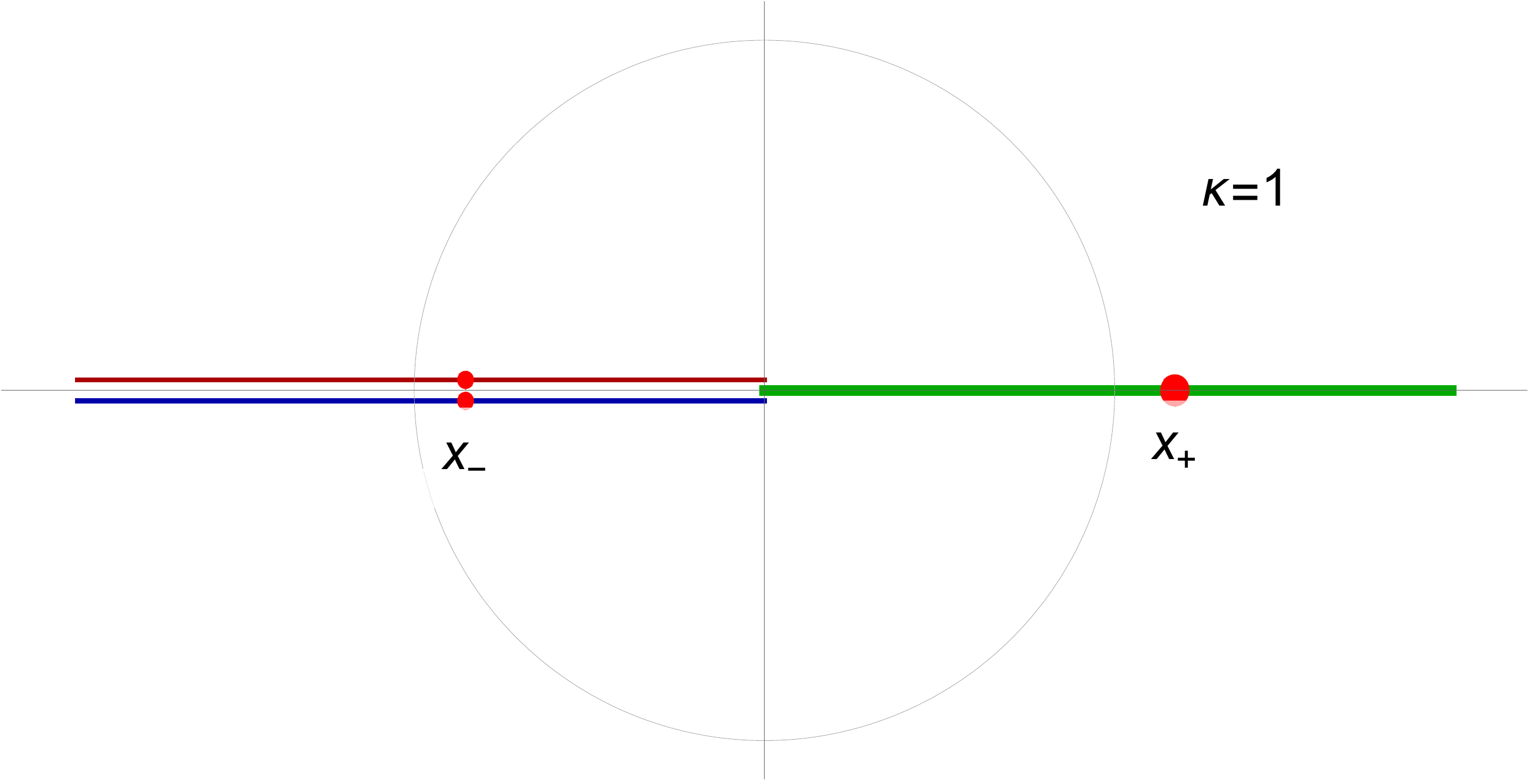} \quad \includegraphics*[width=0.48\textwidth]{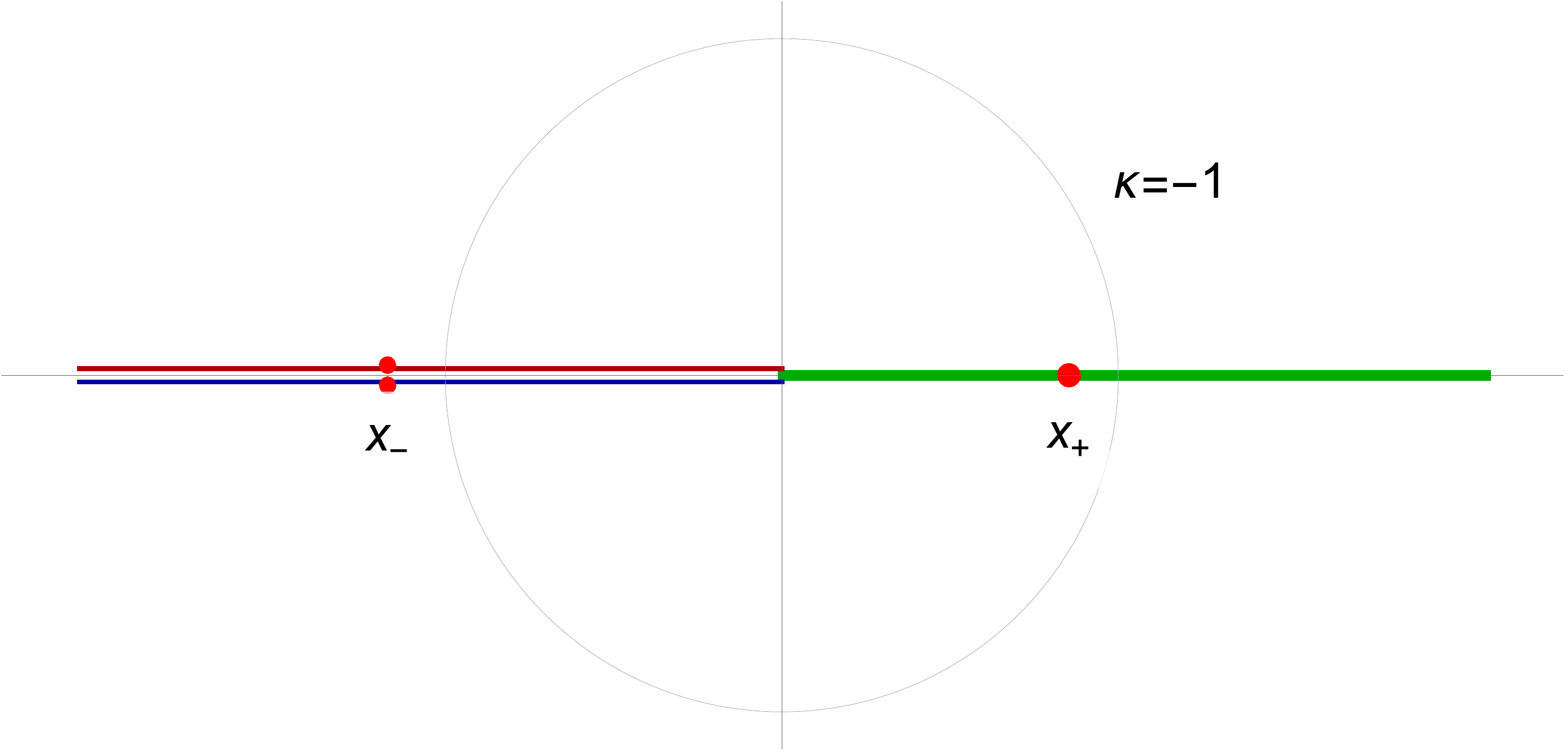} 
\end{center}
The only difference between the cases with $\ka>0$ and $\ka<0$ is the location of the images $\x_\pm$ of the  branch points. We define the mirror physical region as the one with $\Im (x)<0$. It is mapped to a $u$-plane with two  mirror cuts from  $\uu_+$ to $+\infty$, and from $-\infty$ to $\uu_-^{-}$.

Even though there is no analytic formula for $x(u,\ka)$, it is easy to describe each branch of the function parametrically by using the polar angle $\p$ in the $x$-plane and the imaginary part $\nu=\Im (u)$ in a $u$-plane. We find that for any $\ka$ the four branches of $x(u,\ka)$ analysed above can be described as
\bal\la{xuka}
\hspace{-1cm} x(u,\ka) = \rho(\phi,\nu)\,e^{i\p}\,,\quad u =  \mu(\phi,\nu) + i\,\nu\,,\
\eal
where $ \rho(\phi,\nu)$ and $ \mu(\phi,\nu) $ are given by \eqref{eq:rhophiv} and \eqref{eq:Reuphiv}, and the ranges of $\p$ and $\nu$ are as follows\footnote{.The branches $x^{(n)}(u,\ka)$ are also described by \eqref{xuka} with   the ranges of $\p$  shifted by $2\pi n$.}

\medskip

\noindent {\bf Ia.} The principle branch of $x(u,\ka)$  with one cut  from $-\infty$ to $\uu_+$ on the $u$-plane
\bal
\ka>0&:\quad -\pi\le\p\le0\ \ \text{for}\ \ \nu\le0\,;\quad  0\le\p\le\pi\ \ \text{for}\ \ \nu\ge0\,,
\\
\ka<0&:\quad -\pi\le\p\le0\ \ \text{for}\ \ \nu\ge0\,;\quad  0\le\p\le\pi\ \ \text{for}\ \ \nu\le0\,.
\eal
One also has to add a map from the semi-line  $[\uu_+\,,\,+\infty)$  to the semi-line $x\ge\x_+$ for $\ka>0$, and to the interval $(0,\x_+]$ for $\ka<0$.  This branch is defined on the string and   anti-string $u$-plane for $\ka>0$ and $\ka<0$, respectively.
Plots of images of several horizontal lines between $-1.5\ka$ and $+1.5\ka$  on the $u$-plane are shown below for $\ka=\pm1$.
\begin{center}
\includegraphics*[width=0.3\textwidth]{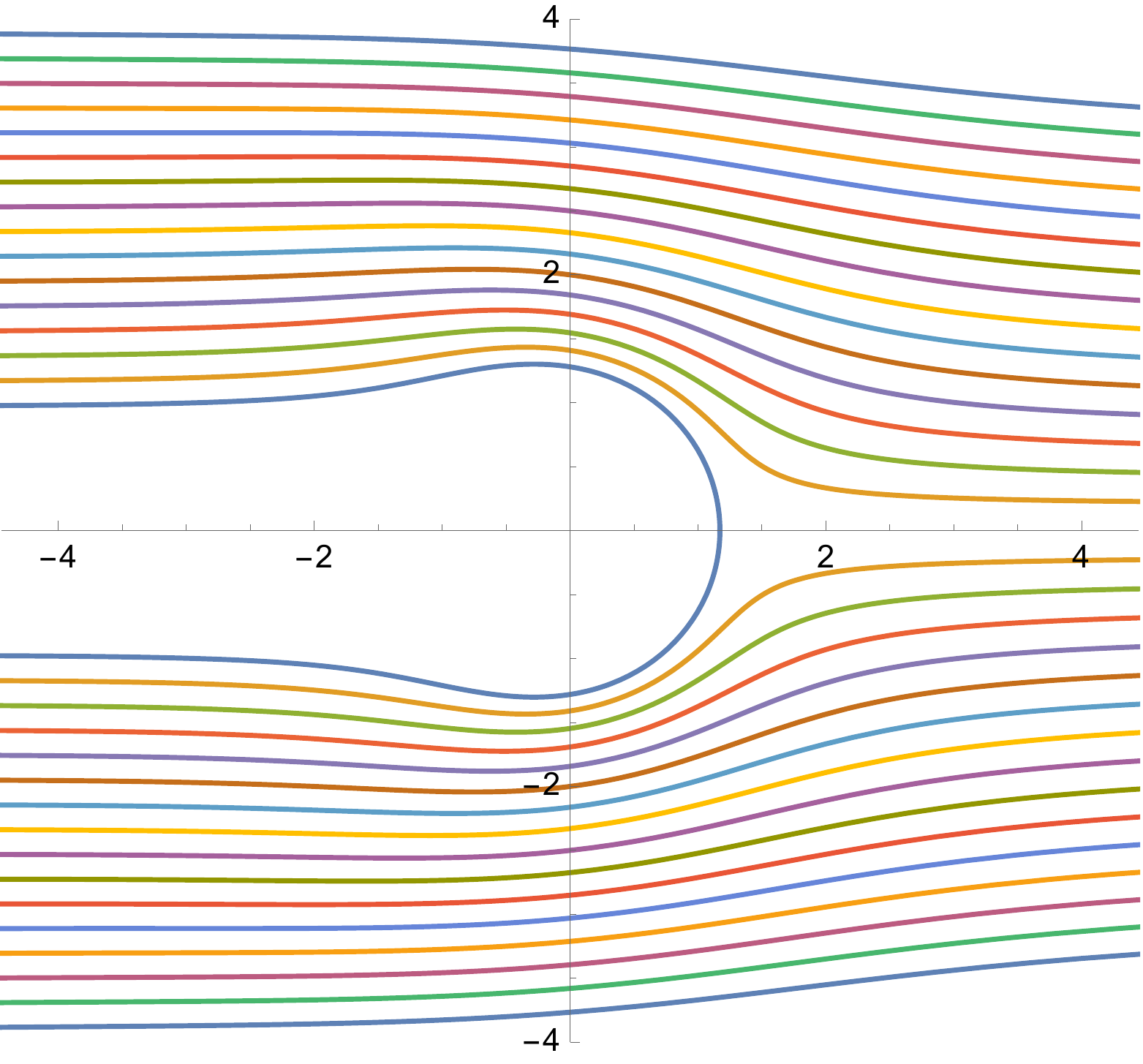} \qquad \includegraphics*[width=0.3\textwidth]{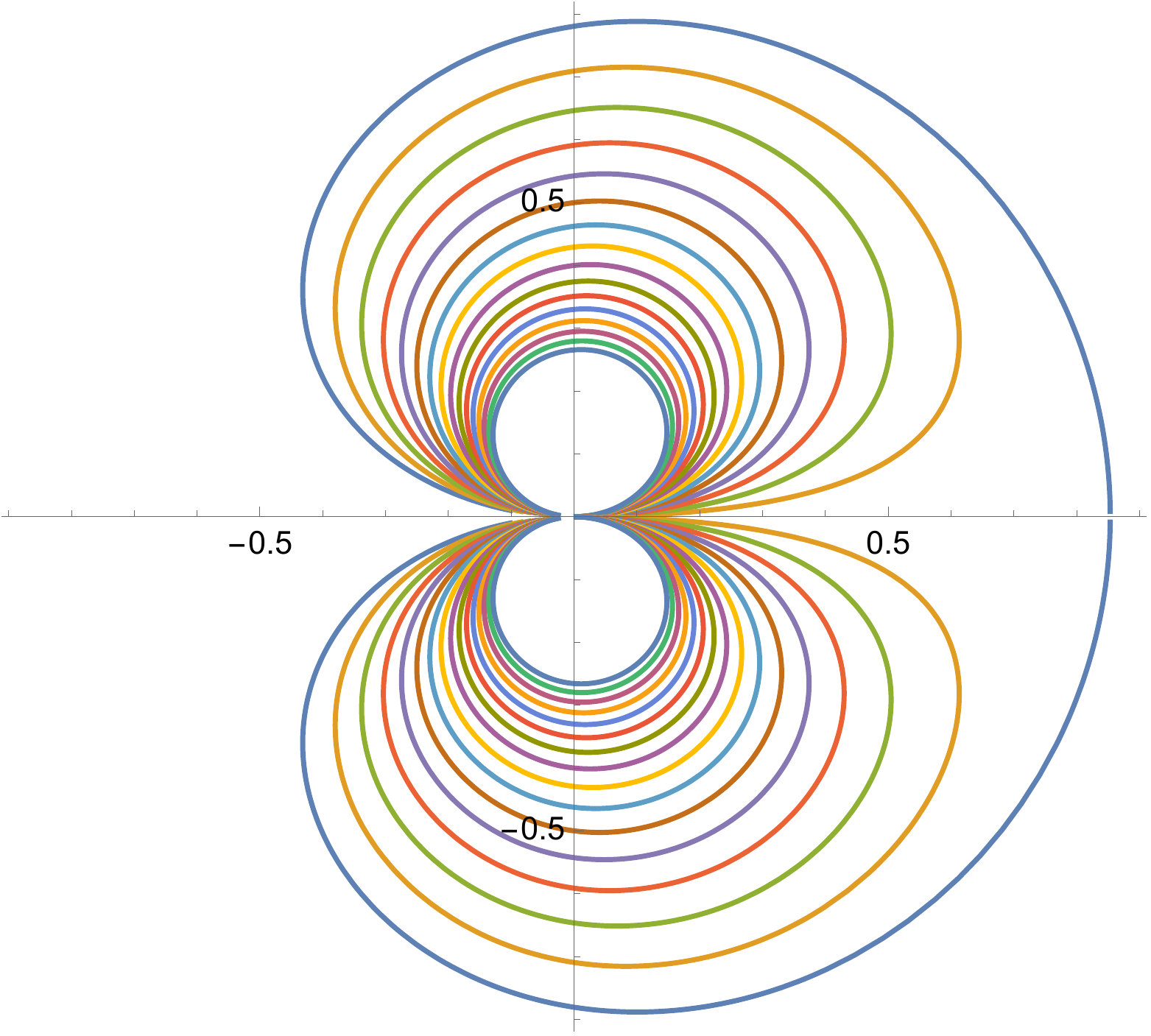} 
\end{center}

\medskip

\noindent {\bf Ib.}  The  branch of $x(u,\ka)$ with three cuts   on the $u$-plane
\bal
\ka>0&:\quad -\pi\le\p\le0\ \ \text{for}\ \ \nu\ge0\,;\quad  0\le\p\le\pi\ \ \text{for}\ \ \nu\le0\,,
\\
\ka<0&:\quad -\pi\le\p\le0\ \ \text{for}\ \ \nu\le0\,;\quad  0\le\p\le\pi\ \ \text{for}\ \ \nu\ge0\,.
\eal
One also has to add a map from the semi-line  $[\uu_+\,,\,+\infty)$  to  the interval $(0,\x_+]$ for $\ka>0$, and to  the semi-line $x\ge\x_+$ for $\ka<0$. This branch is defined on the anti-string and   string $u$-plane for $\ka>0$ and $\ka<0$, respectively.
Plots of images of several horizontal lines between $-1.5\ka$ and $+1.5\ka$  on the $u$-plane are shown below  for $\ka=\pm 1$.
\begin{center}
\includegraphics*[width=0.35\textwidth]{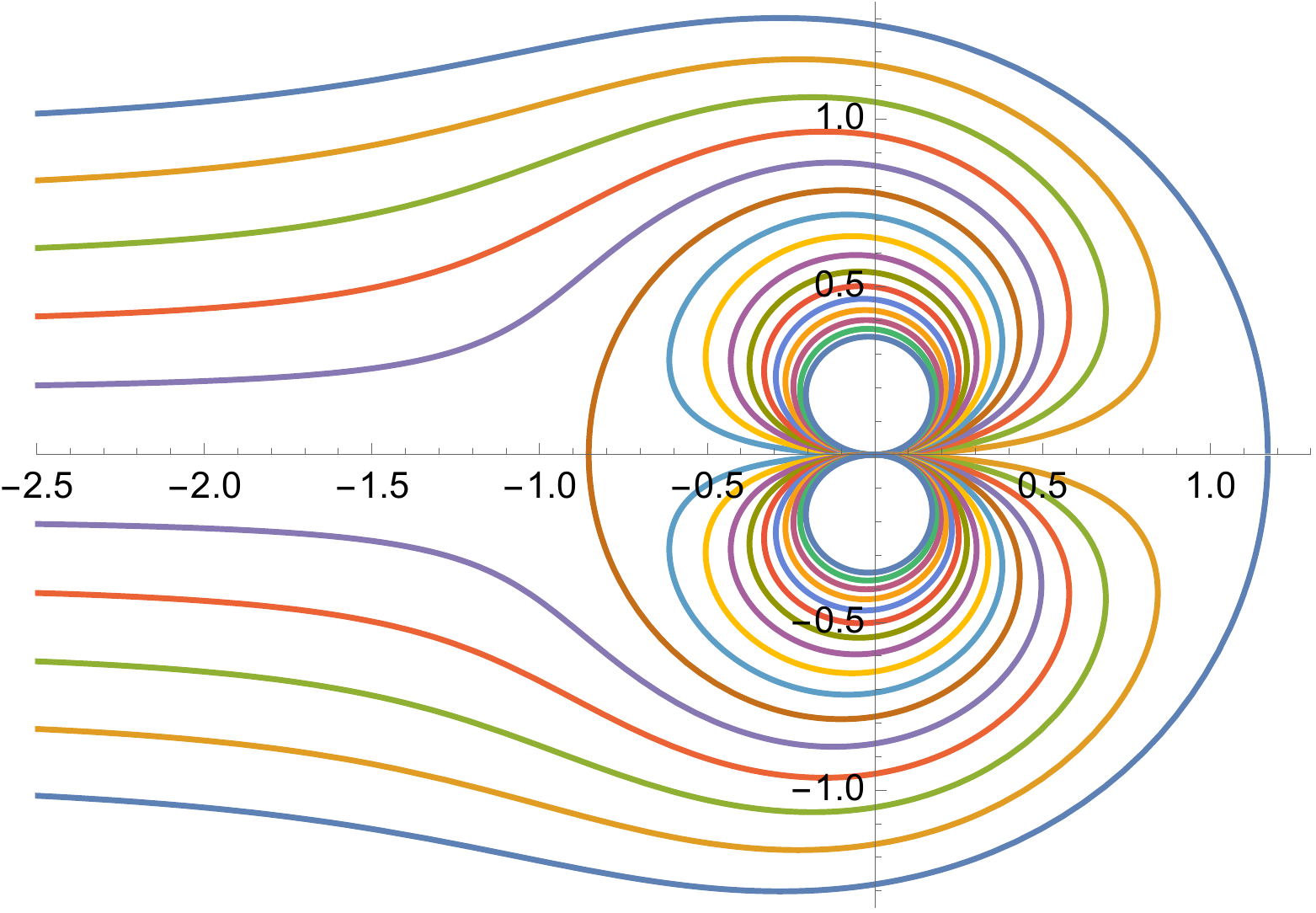} \qquad \includegraphics*[width=0.3\textwidth]{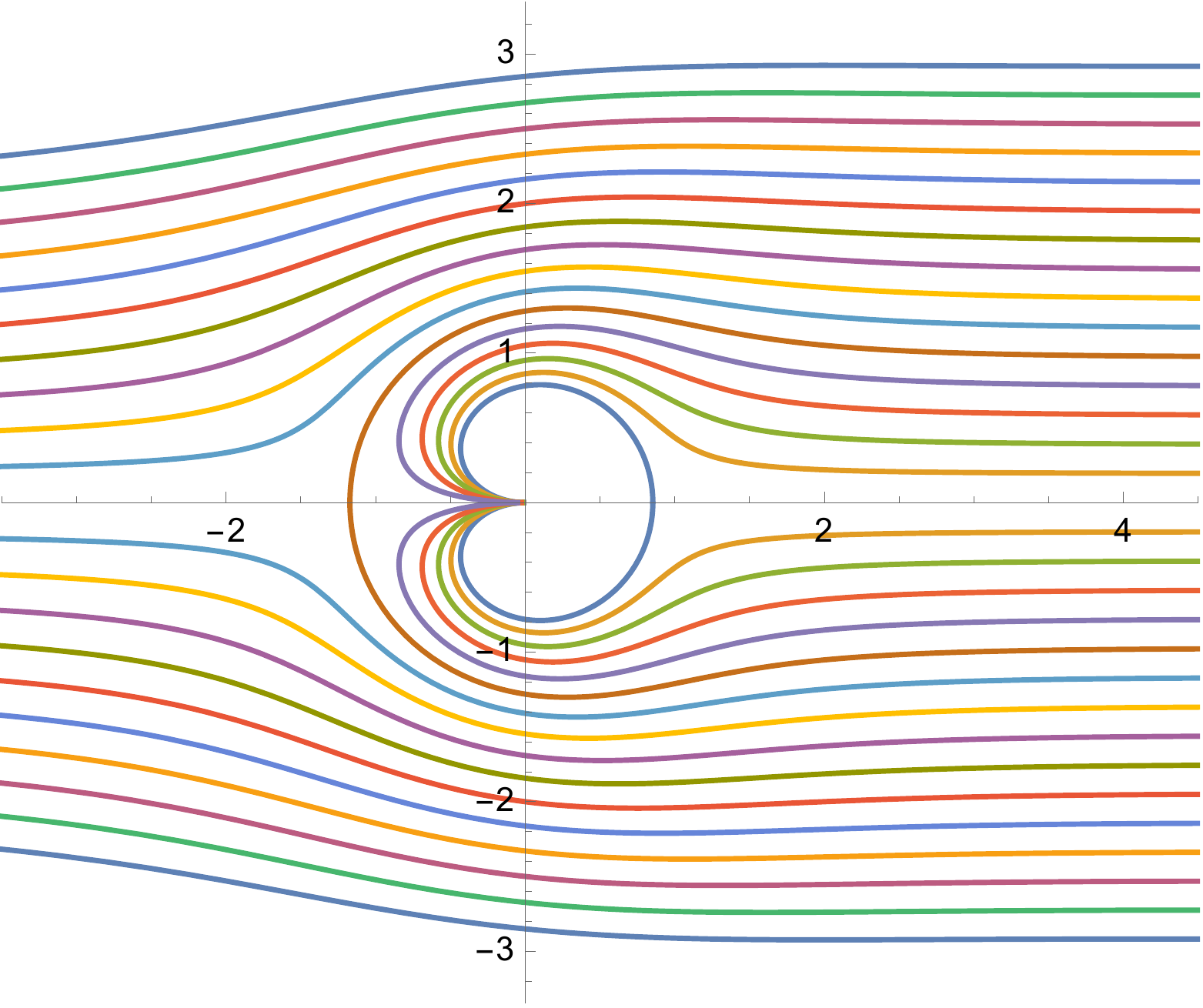} 
\end{center}
Obviously, for both branches Ia and Ib, $x(u,\ka)$ satisfies the string complex conjugation condition \eqref{stringcc}.

The two $u$-planes glued together are mapped by $x(u,\ka)$ to the  $x$-plane with the cut $(-\infty,0)$. Moving through  the cut $(-\infty, \uu_+)$, one gets from one $u$-plane to the other one which is still mapped to the same $x$-plane. It is easy to check by using  \eqref{eq:rhophiv} and \eqref{eq:Reuphiv}  that  $x(u,\ka)$ and $x(u,-\ka)$ on  any of the two branches are related according to the inversion symmetry
\bal
x(u,\ka) = {1\ov x(u,-\ka)}\,,
\eal
where $u$ belongs either to the $u$-plane with one cut or to the $u$-plane with three cuts.

If one moves through  the cut $(-\infty, \uu_-^{+})$ one gets to a $u$-plane with three cuts 
\bal
(-\infty, \uu_-^{-}-2i\ka)=(-\infty, \uu_-^{+})\,,\quad (-\infty, \uu_-^{+}-2i\ka)\quad  \text{and}\quad  (-\infty, \uu_+-2i\ka)\,.
\eal
This $u$-plane is mapped to another $x$-plane with $\log x = \ln x + 2i\pi$. The function $x_{\text{Ib}}^{(1)}(u,\ka) $ on this $u$-plane is given by
\bal
x_{\text{Ib}}^{(1)}(u,\ka) = x_{\text{Ib}}(u+2i\ka,\ka)\,.
\eal
Similarly, crossing the cut $(-\infty, \uu_-^{-})$ brings one to a $u$-plane with  cuts 
\bal
(-\infty, \uu_-^{+}+2i\ka)=(-\infty, \uu_-^{-})\,,\quad (-\infty, \uu_-^{-}+2i\ka)\quad  \text{and}\quad  (-\infty, \uu_+ +2i\ka)\,,
\eal
which is mapped to the $x$-plane with $\log x = \ln x - 2i\pi$ with the function $x_{\text{Ib}}^{(-1)}(u,\ka) $  given by
\bal
x_{\text{Ib}}^{(-1)}(u,\ka) = x_{\text{Ib}}(u-2i\ka,\ka)\,.
\eal
Note that in both cases if we cross the most lower cut then on the new $u$ plane it becomes the most upper one, and vice versa. In other words the cuts are reflected about the cut which has been crossed. This leads to a noticeable dependence of the result of the analytic continuation along a path around the branch point at $\infty$ where all the three cuts meet. Consider for definiteness $\ka>0$, and a path which begins at a point with $\Im(u) < -3\ka$ and goes up crossing the cut $(-\infty, \uu_-^{+})$. Once the point crosses the cut it  gets to the $u$-plane without any cut above  it, and since the lowest cut has $\Im(u)=-3\ka$, the point can be moved freely to its original coordinates on the  $u$-plane which is mapped to the $x$-plane with $\log x =\ln x +2i\pi$. If, however,  the path begins at a point with $-3\ka< \Im(u) < -\ka$, then the point would  have to cross the cut $(-\infty, \uu_-^{+})$ on its original $u$-plane, and also the cut $(-\infty, \uu_-^{+}-2i\ka)$ on the second $u$-plane, and it ends up on the $u$-plane with three cuts which is mapped to the $x$-plane with $\log x =\ln x +4i\pi$. Next, if the path begins at a point with $-\ka< \Im(u) < 0$, then the point crosses the cut $(-\infty, \uu_+)$ on its original $u$-plane with three cuts, and  gets to the $u$-plane with a single cut  which is mapped to the original $x$-plane with $\log x =\ln x$. Finally, if the path begins at a point with $0< \Im(u) < +\ka$, then the point  crosses the cut $(-\infty, \uu_-^{-})$,  and  gets to the $u$-plane with three cuts . The next cut it crosses on the new $u$-plane is $(-\infty, \uu_+ +2i\ka)$, and it gets the point to the $u$-plane with one cut which is mapped to the $x$-plane with $\log x =\ln x -2i\pi$.

\medskip

\noindent {\bf IIa.}  The mirror  branch of $x(u,\ka)$ with two cuts   on the $u$-plane
\bal
-\pi\le\p\le0\,,\quad  -\infty  < \nu< +\infty\,.
\eal
One also has to add a map from the cut  $(\uu_+\,,\,+\infty)$  to   the semi-line $x> 0$, and from the cut  $(-\infty\,,\,\uu_-^{-})$  to   the semi-line $x\le 0$ . Plots of images of several horizontal lines between $-1.5\ka$ and $+1.5\ka$  on the $u$-plane are shown below  for $\ka=\pm1$.
\begin{center}
\includegraphics*[width=0.35\textwidth]{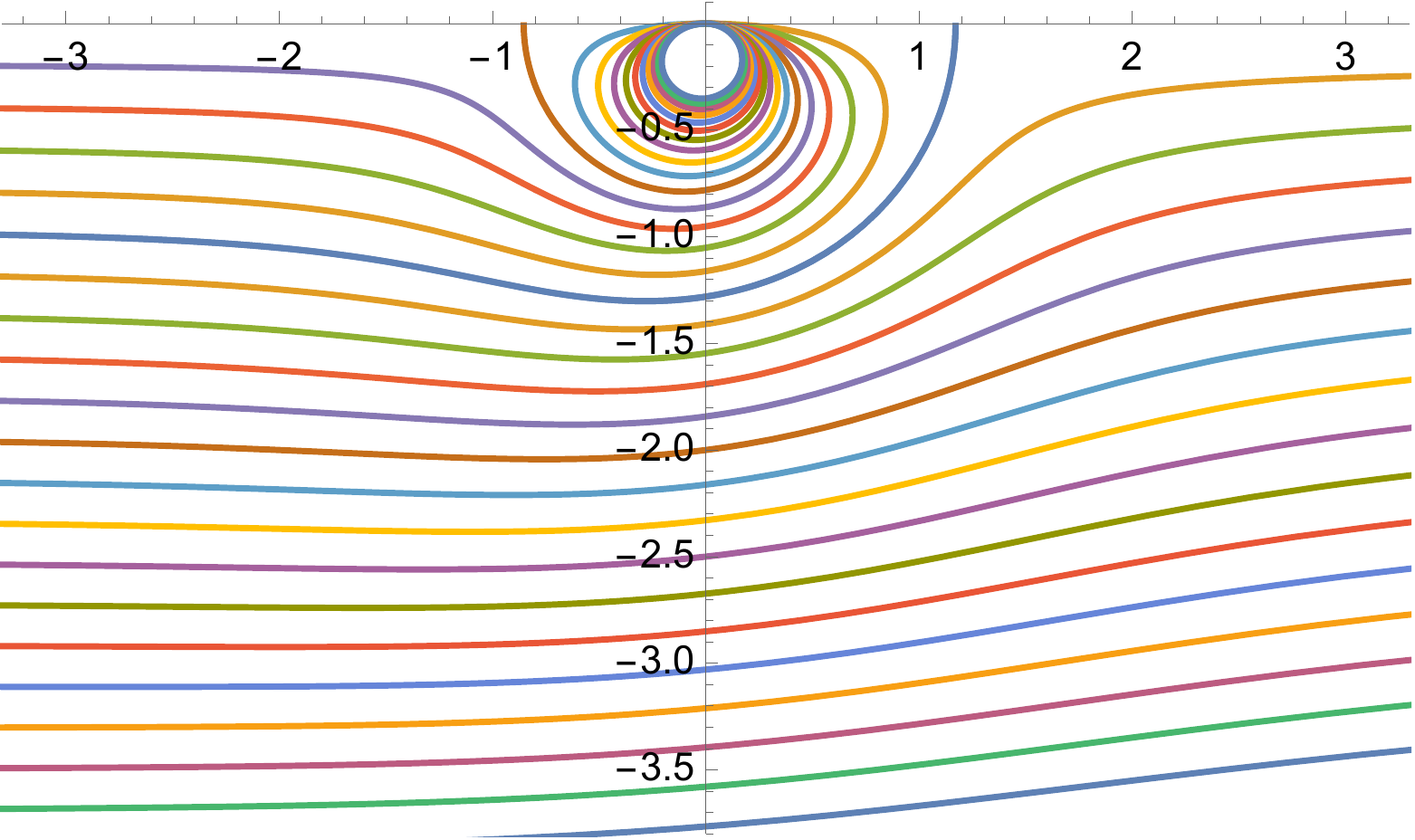} \qquad \includegraphics*[width=0.46\textwidth]{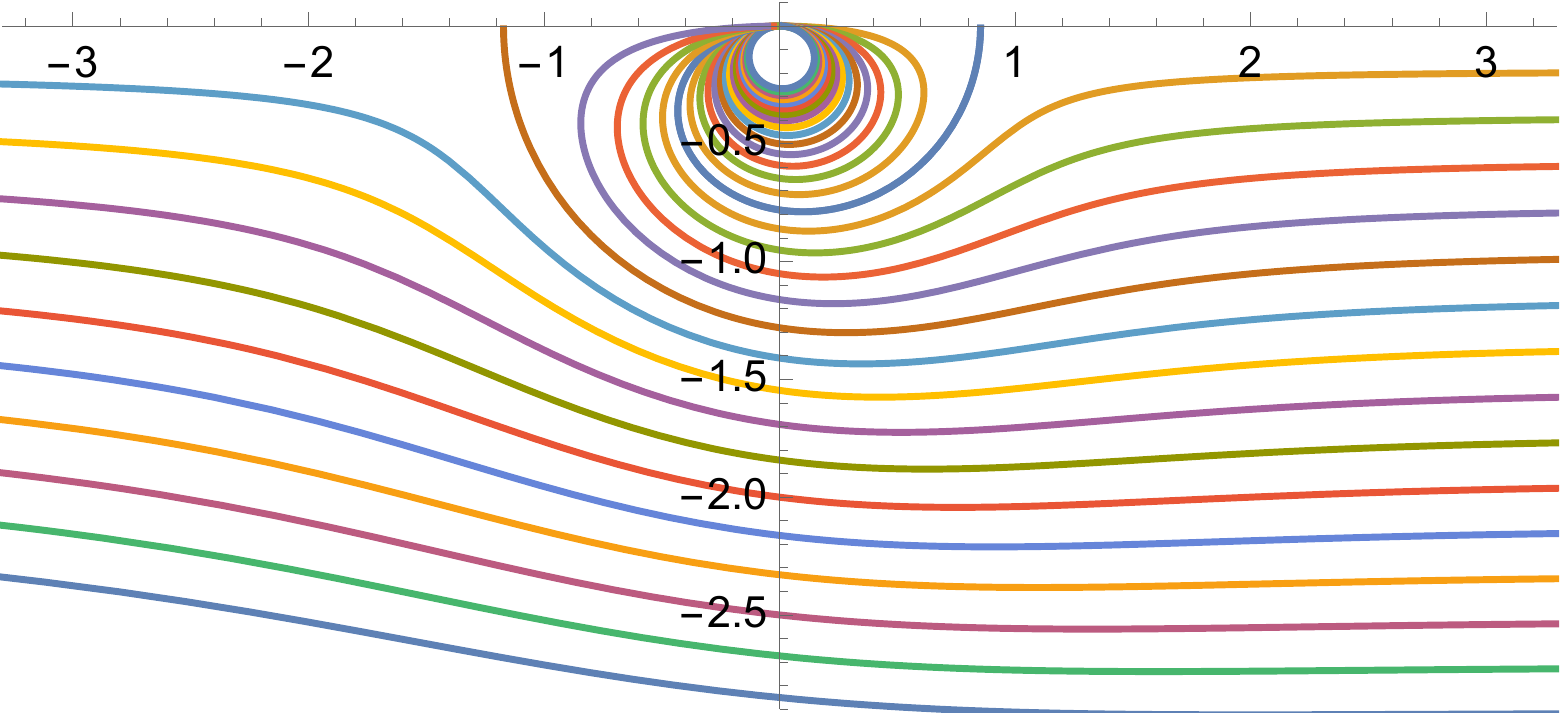} 
\end{center}

\medskip

\noindent {\bf IIb.}  The anti-mirror  branch of $x(u,\ka)$ with two cuts on the $u$-plane
\bal
0\le\p\le\pi\,,\quad  -\infty  < \nu< +\infty\,.
\eal
One also has to add a map from the cut  $(\uu_+\,,\,+\infty)$  to   the semi-line $x> 0$, and from the cut  $(-\infty\,,\,\uu_-^{+})$  to   the semi-line $x< 0$ . Plots of images of several horizontal lines between $-1.5\ka$ and $+1.5\ka$  on the $u$-plane are shown below  for $\ka=\pm1$.
\begin{center}
\includegraphics*[width=0.35\textwidth]{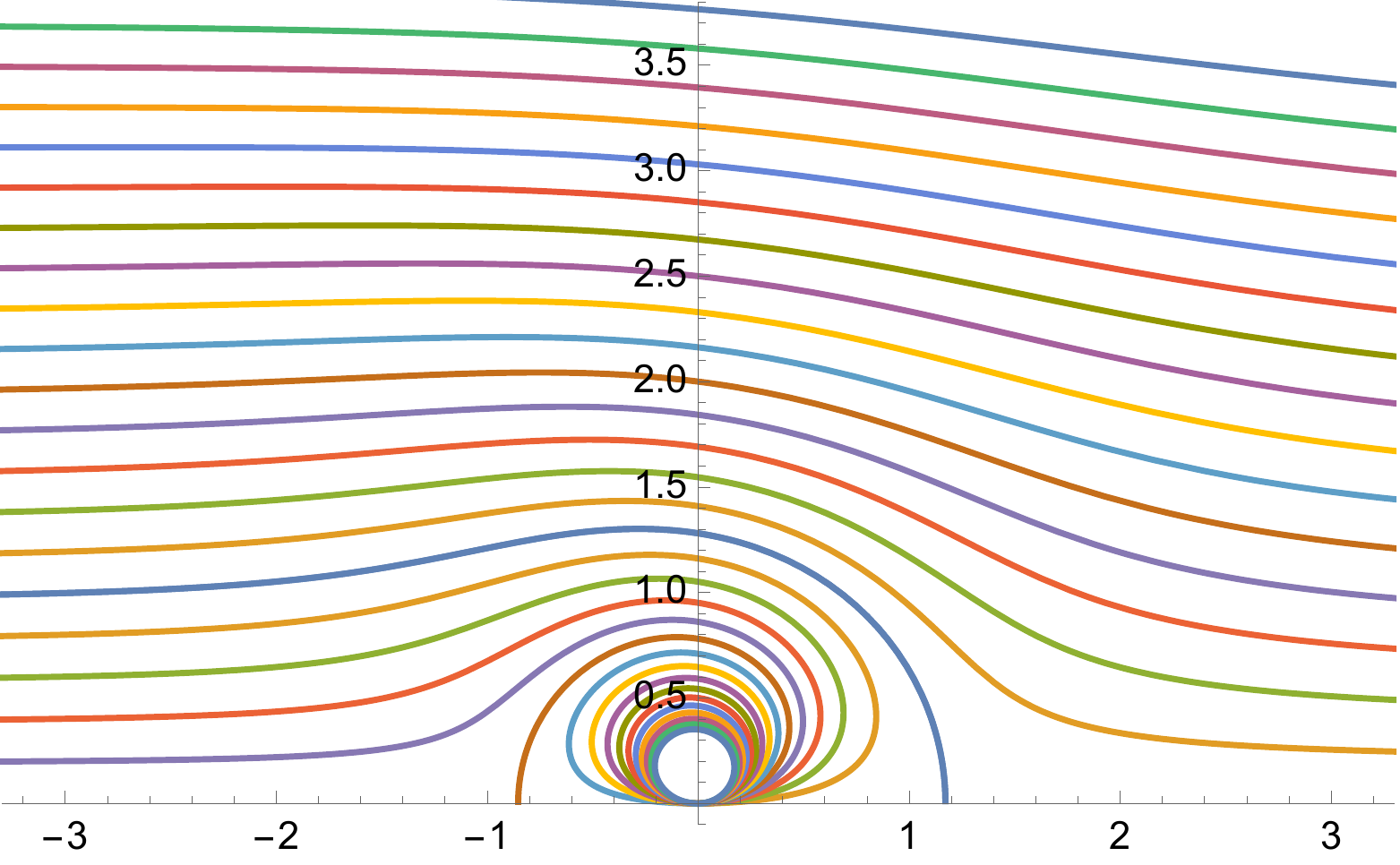} \qquad \includegraphics*[width=0.46\textwidth]{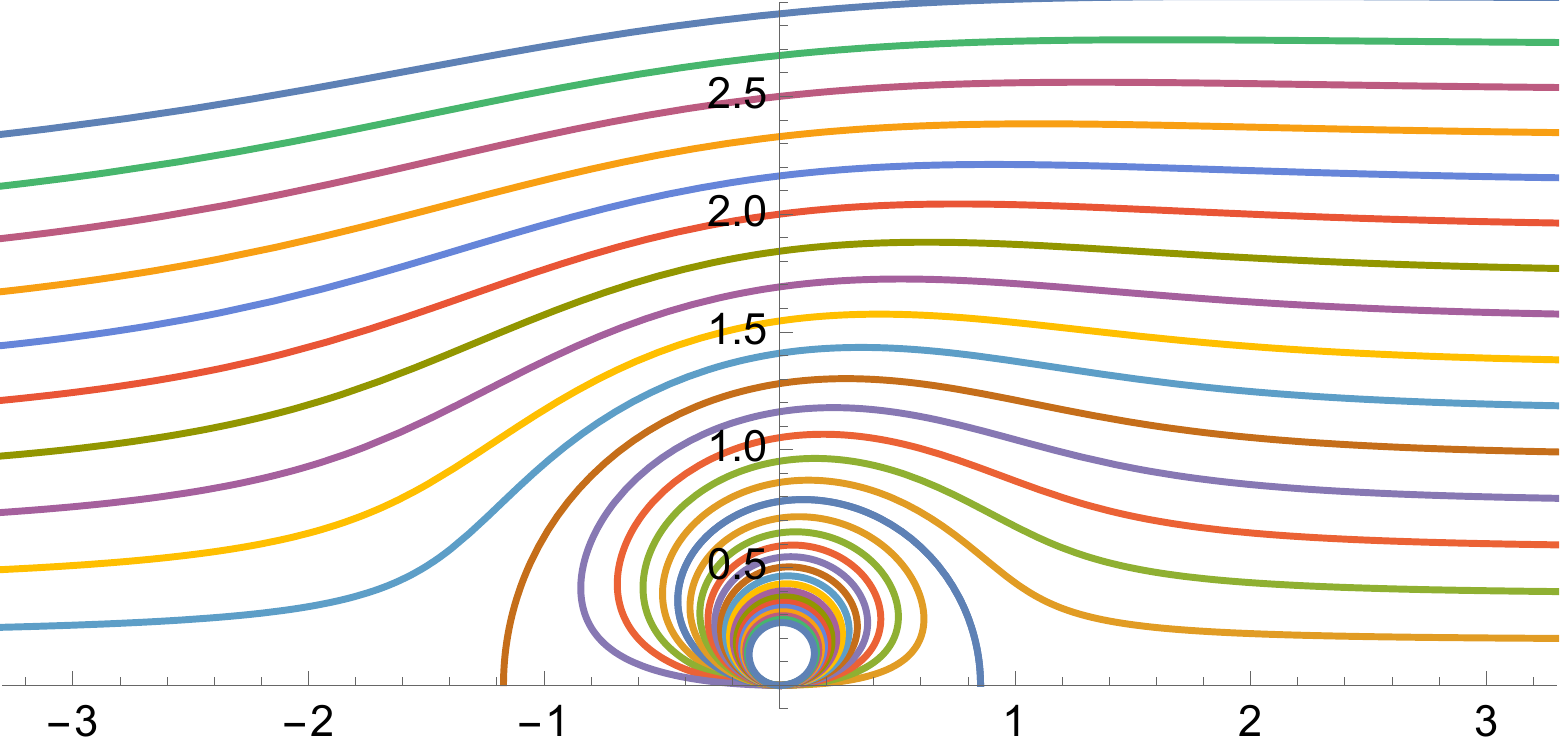} 
\end{center}
It is easy to check that for both branches IIa and IIb, $x(u,\ka)$ satisfies the mirror complex conjugation condition \eqref{mirrorcc}.  The analyses of what happens when one 
moves  through  the two cuts of any of the $u$-planes repeats the one for the (anti-)string $u$-planes.

\medskip
 
 To conclude this section let us note that, in addition to the limit $\ka\to0$,
  another interesting limit is the one where $\ka\to\infty$.  In this limit one has three options 
  \begin{enumerate}
      \item One keeps $x$ fixed but rescales $u$ as $u\to -\ka u/\pi$.
      Then, eq.\eqref{uplane} trivialises 
\bal
\label{uplanekainfty1}
u=\log x\quad \Rightarrow\quad x=e^u\,.
\eal
This is a relativistic limit, and the rescaled variable $u$ is identified with the rapidity variable $\theta$.
\item
 One rescales $x$ as $x\to |\ka| x/\pi$, and also rescales and shifts $u$ as $u\to |\ka| u/\pi -\ka/\pi\log |\ka|/\pi$. Then, eq.\eqref{uplane} takes the form
\bal
\label{uplanekainfty}
u=x- \text{sgn}(\kappa)\,\log x\,.
\eal
It is a well-known equation whose solutions can be given in terms of the Lambert (or productlog) $W$ function
\bal
 \ka<0:\, \quad x_n(u)=W_n\left(e^u\right)\,;\qquad 
  \ka>0:\, \quad x_n(u)=-W_n\left(-e^{-u}\right)\,,\quad n\in \bZ\,,
\eal
where the domain of $u$ depends on $n$, and it is a horizontal strip of width $2\pi$. 
In terms of  our description of the $\ka$-deformed Zhukovsky map, for $\ka<0$ a $u$-plane with two cuts $(-\infty ,-1\pm i\pi)$, and for $\ka>0$ a $u$-plane with one cut $(-\infty ,+1)$, is mapped to  the $x$-plane with the cut $(-\infty,0)$. 
\item One could instead  rescale $x$ as $x\to x \pi/|\ka| $, and also rescale and shift $u$ as $u\to |\ka| u/\pi +\ka/\pi\log |\ka|/\pi$. This leads to the equation $u=1/x- \text{sgn}(\kappa)\,\log x$ whose solutions are again expressed in terms of the Lambert $W$ function.
  \end{enumerate}

\section{S-matrix elements before the limit}
\label{app:Smatrices}

In this appendix, we report the S~matrices acting on the two-particle representations 
\begin{equation}
\begin{aligned}
    &\repr{B}{L}(m_1,p_1) \otimes \repr{B}{L}(m_2,p_2)  \hspace{1mm},\hspace{5mm} \repr{F}{R}(m_1,p_1) \otimes \repr{F}{R}(m_2,p_2) \,,\\
    &\repr{B}{L}(m_1,p_1) \otimes \repr{F}{R}(m_2,p_2) \hspace{1mm} , \hspace{5mm} \repr{F}{R}(m_1,p_1) \otimes \repr{B}{L}(m_2,p_2)\,,
    \end{aligned}
\end{equation}
 of the superalgebra $\mathfrak{su}(1|1)^2_{c.e.}$, normalising the scattering between highest-weight states to one. 
The two subscript indices (L or R) in the S~matrices label the choice of coefficients used to parameterise the associated supercharges, which can be functions of $x^{\pm}_{\L}$ or $x^{\pm}_{\R}$ respectively; the superscript indices label instead the highest weight states of the representations associated with the scattered particles and can be B or F (bosonic or fermionic).

\subsection{Left-left scattering}
The S-matrix acting on double-particle states in the representation $\repr{B}{L}(m_1,p_1) \otimes \repr{B}{L}(m_2,p_2)$ is determined by
\begin{equation}
\label{eq:scattering-structureLL}
\begin{aligned}
    \mathbf{S}_{\L\L}^{\B\B}\,|\phi^{\B}_{\L1}\,\phi^{\B}_{\L2}\rangle &= A_{\L\L}^{\B\B}\,|\phi^{\B}_{\L2}\,\phi^{\B}_{\L1}\rangle,\\
    \mathbf{S}_{\L\L}^{\B\B}\,|\phi^{\B}_{\L1}\,\varphi^{\F}_{\L2}\rangle &= B_{\L\L}^{\B\B}\,|\varphi^{\F}_{\L2}\,\phi^{\B}_{\L1}\rangle+ C_{\L\L}^{\B\B}\,|\phi^{\B}_{\L2}\,\varphi^{\F}_{\L1}\rangle,\\
    \mathbf{S}_{\L\L}^{\B\B}\,|\varphi^{\F}_{\L1}\,\phi^{\B}_{\L2}\rangle &= D_{\L\L}^{\B\B}\,|\phi^{\B}_{\L2}\,\varphi^{\F}_{\L1}\rangle+E_{\L\L}^{\B\B}\,|\varphi^{\F}_{\L2}\,\phi^{\B}_{\L1}\rangle,\\
    \mathbf{S}_{\L\L}^{\B\B}\,|\varphi^{\F}_{\L1}\,\varphi^{\F}_{\L2}\rangle &= F_{\L\L}^{\B\B}\,|\varphi^{\F}_{\L2}\,\varphi^{\F}_{\L1}\rangle\,,
\end{aligned}
\end{equation}
with coefficients:
\begin{equation}
\begin{aligned}
    A_{\L\L}^{\B\B}&=1\,,&
    B_{\L\L}^{\B\B}&=e^{-\frac{i}{2} p_1} \frac{x^+_{\text{\tiny L}1} - x^+_{\text{\tiny L}2}}{x^-_{\text{\tiny L}1} - x^+_{\text{\tiny L}2}}\,,\\
    C_{\L\L}^{\B\B}&=e^{\frac{i}{2}(p_2 - p_1)} \frac{x^-_{\text{\tiny L}1} - x^+_{\text{\tiny L}1}}{x^-_{\text{\tiny L}1} - x^+_{\text{\tiny L}2}} \ \frac{\eta_{\text{\tiny L}2}}{\eta_{\text{\tiny L}1}}\,,&
    D_{\L\L}^{\B\B}&=e^{\frac{i}{2}p_2} \frac{x^-_{\text{\tiny L}1} - x^-_{\text{\tiny L}2}}{x^-_{\text{\tiny L}1} - x^+_{\text{\tiny L}2}}\,,\\
    E_{\L\L}^{\B\B}&=\frac{x^-_{\text{\tiny L}1} - x^+_{\text{\tiny L}1}}{x^-_{\text{\tiny L}1} - x^+_{\text{\tiny L}2}} \ \frac{\eta_{\text{\tiny L}2}}{\eta_{\text{\tiny L}1}}\,,&\qquad
    F_{\L\L}^{\B\B}&=-e^{\frac{i}{2}(p_2 - p_1)} \frac{x^+_{\text{\tiny L}1} - x^-_{\text{\tiny L}2}}{x^-_{\text{\tiny L}1} - x^+_{\text{\tiny L}2}}\,.
\end{aligned}
\end{equation}

\subsection{Right-right scattering}

Using the convention~\eqref{eq:scattering-structureLL} also for the remaing S~matrices we obtain 

\begin{equation}
\begin{aligned}
    A_{\R\R}^{\F\F}&=1\,,&
    B_{\R\R}^{\F\F}&=-e^{\frac{i}{2}p_1} \frac{x^-_{\text{\tiny R}1} - x^-_{\text{\tiny R}2}}{x^+_{\text{\tiny R}1} - x^-_{\text{\tiny R}2}}\,,\\
    C_{\R\R}^{\F\F}&=e^{\frac{i}{2}(p_1 - p_2)} \frac{x^+_{\text{\tiny R}1} - x^-_{\text{\tiny R}1}}{x^+_{\text{\tiny R}1} - x^-_{\text{\tiny R}2}} \ \frac{\eta_{\text{\tiny R}2}}{\eta_{\text{\tiny R}1}}\,,&
    D_{\R\R}^{\F\F}&=-e^{-\frac{i}{2}p_2} \frac{x^+_{\text{\tiny R}1} - x^+_{\text{\tiny R}2}}{x^+_{\text{\tiny R}1} - x^-_{\text{\tiny R}2}}\,,\\
    E_{\R\R}^{\F\F}&=\frac{x^+_{\text{\tiny R}1} - x^-_{\text{\tiny R}1}}{x^+_{\text{\tiny R}1} - x^-_{\text{\tiny R}2}} \ \frac{\eta_{\text{\tiny R}2}}{\eta_{\text{\tiny R}1}}\,,&\qquad
    F_{\R\R}^{\F\F}&=-e^{\frac{i}{2}(p_1-p_2)} \frac{x^-_{\text{\tiny R}1} - x^+_{\text{\tiny R}2}}{x^+_{\text{\tiny R}1} - x^-_{\text{\tiny R}2}}\,.
\end{aligned}
\end{equation}

\subsection{Left-right scattering}

\begin{equation}
\begin{aligned}
    A_{\L\R}^{\B\F}&=1\,,&
    B_{\L\R}^{\B\F}&=e^{-\frac{i}{2} p_1} \frac{x^+_{\text{\tiny L}1}x^-_{\text{\tiny R}2} - 1}{x^-_{\text{\tiny L}1} x^-_{\text{\tiny R}2} - 1}\,,\\
    C_{\L\R}^{\B\F}&=e^{-\frac{i}{2}(p_1 + p_2)} \frac{x^-_{\text{\tiny L}1} - x^+_{\text{\tiny L}1}}{x^-_{\text{\tiny L}1} x^-_{\text{\tiny R}2} - 1} \  \frac{\eta_{\text{\tiny R}2}}{\eta_{\text{\tiny L}1}}\,,&
    D_{\L\R}^{\B\F}&=-e^{-\frac{i}{2}p_2} \frac{x^-_{\text{\tiny L}1} x^+_{\text{\tiny R}2} - 1}{x^-_{\text{\tiny L}1} x^-_{\text{\tiny R}2} - 1}\,,\\
    E_{\L\R}^{\B\F}&=\frac{x^+_{\text{\tiny L}1}-x^-_{\text{\tiny L}1}}{x^-_{\text{\tiny L}1} x^-_{\text{\tiny R}2} - 1} \ \frac{\eta_{\text{\tiny R}2}}{\eta_{\text{\tiny L}1}}\,,&\qquad
    F_{\L\R}^{\B\F}&=e^{-\frac{i}{2}(p_1 + p_2)} \frac{x^+_{\text{\tiny L}1} x^+_{\text{\tiny R}2} - 1}{x^-_{\text{\tiny L}1} x^-_{\text{\tiny R}2} - 1}\,.
\end{aligned}
\end{equation}

\subsection{Right-left scattering}

\begin{equation}
\begin{aligned}
    A_{\R\L}^{\F\B}&=1\,,&
    B_{\R\L}^{\F\B}&=-e^{\frac{i}{2}p_1} \frac{x^-_{\text{\tiny R}1} x^+_{\text{\tiny L}2} - 1}{x^+_{\text{\tiny R}1} x^+_{\text{\tiny L}2} - 1}\,,\\
    C_{\R\L}^{\F\B}&=e^{\frac{i}{2}(p_1+p_2)} \frac{x^+_{\text{\tiny R}1}-x^-_{\text{\tiny R}1}}{x^+_{\text{\tiny R}1}x^+_{\text{\tiny L}2}-1} \ \frac{\eta_{\text{\tiny L}2}}{\eta_{\text{\tiny R}1}}\,,&
    D_{\R\L}^{\F\B}&=e^{\frac{i}{2}p_2} \frac{x^+_{\text{\tiny R}1} x^-_{\text{\tiny L}2} - 1}{x^+_{\text{\tiny R}1} x^+_{\text{\tiny L}2} - 1}\,,\\
    E_{\R\L}^{\F\B}&=\frac{x^-_{\text{\tiny R}1}-x^+_{\text{\tiny R}1}}{x^+_{\text{\tiny R}1}x^+_{\text{\tiny L}2}-1} \ \frac{\eta_{\text{\tiny L}2}}{\eta_{\text{\tiny R}1}}\,,&\qquad
    F_{\R\L}^{\F\B}&=e^{\frac{i}{2}(p_1+p_2)} \frac{x^-_{\text{\tiny R}1} x^-_{\text{\tiny L}2}-1}{x^+_{\text{\tiny R}1}x^+_{\text{\tiny L}2}-1}\,.
\end{aligned}
\end{equation}

\section{Relativistic S~matrix from symmetries}
\label{app:relativisticS}

Using the representations constructed above we may try to fix the two-particle S~matrix for every value of~$m\in\mathbb{Z}$ and~$\theta\in\mathbb{R}$. Moreover, we may impose the following conditions
\begin{enumerate}
    \item The S~matrix obeys the Yang-Baxter equation;
    \item The S~matrix obeys physical unitarity and braiding unitarity, up to specifying an appropriate pre-factor; 
    \item The S~matrix obeys crossing symmetry, up to specifying an appropriate pre-factor.
\end{enumerate}
We will comment later on whether this coincides with a suitable limit of the S~matrix of appendix~\ref{sec:fulltheory}.

The precise form of the S-matrix will depend on  whether we pick a bosonic or fermionic highest weight state and on each representation (the other cases differ by some minus signs). For simplicity, let us consider the case where all representations have a bosonic highest-weight state, so that an explicit basis for the two-particle Hilbert space is
\begin{equation}
    \Big(
    |\phi^{\B}_{1}\,\phi^{\B}_{2}\rangle,\,
    |\phi^{\B}_{1}\,\varphi^{\F}_{2}\rangle,\,
    |\varphi^{\F}_{1}\,\phi^{\B}_{2}\rangle,\,
    |\varphi^{\F}_{1}\,\varphi^{\F}_{2}\rangle
    \Big)\,,
\end{equation}
where $1$ and~$2$ refer to $(m_1,\theta_1)$ and $(m_2,\theta_2)$, respectively. In this way we will find
\begin{equation}
\label{eq:scattering-structure_rel_limit}
\begin{aligned}
    \mathbf{S}^{\B\B}_{12}\,|\phi^{\B}_{1}\,\phi^{\B}_{2}\rangle &= A_{12}^{\B\B}\,|\phi^{\B}_{2}\,\phi^{\B}_{1}\rangle,\\
    \mathbf{S}^{\B\B}_{12}\,|\phi^{\B}_{1}\,\varphi^{\F}_{2}\rangle &= B_{12}^{\B\B}\,|\varphi^{\F}_{2}\,\phi^{\B}_{1}\rangle+ C_{12}^{\B\B}\,|\phi^{\B}_{2}\,\varphi^{\F}_{1}\rangle,\\
    \mathbf{S}^{\B\B}_{12}\,|\varphi^{\F}_{1}\,\phi^{\B}_{2}\rangle &= D_{12}^{\B\B}\,|\phi^{\B}_{2}\,\varphi^{\F}_{1}\rangle+E_{12}^{\B\B}\,|\varphi^{\F}_{2}\,\phi^{\B}_{1}\rangle,\\
    \mathbf{S}^{\B\B}_{12}\,|\varphi^{\F}_{1}\,\varphi^{\F}_{2}\rangle &= F_{12}^{\B\B}\,|\varphi^{\F}_{2}\,\varphi^{\F}_{1}\rangle\,,
\end{aligned}
\end{equation}
where the superscript ``BB'' indicates the highest-weight state.
We will also distinguish the case of massive ($m\neq0$~mod$k$) and massless ($m=0$~mod$k$) representations. In what follows it will be useful to use the short-hands
\begin{equation}
    \mathscr{S}_1 =\text{sgn}\big[\sin\frac{\pi m_1}{k}\big],\qquad
    \mathscr{S}_2 =\text{sgn}\big[\sin\frac{\pi m_2}{k}\big],\qquad (m_i\neq0~\text{mod}k).
\end{equation}

\paragraph{Massive-massive scattering.}
By normalising the highest-weight scattering to one, we find
\begin{equation}
\begin{aligned}
    A_{12}^{\B\B}=&\,1\,,\quad &
    B_{12}^{\B\B}=&\, \frac{\mathscr{S}_1 e^{\frac{i m_2 \pi}{k}+\theta}-\mathscr{S}_2 e^{\frac{i m_1 \pi}{k}}}{\mathscr{S}_1 e^{\frac{i (m_1+m_2) \pi}{k}+\theta}-\mathscr{S}_2}   \,,\\
    C_{12}^{\B\B}=&\,\frac{i e^{\frac{i m_1 \pi}{k} + \frac{\theta}{2}}\sqrt{\mu(m_1) \mu(m_2)}}{\mathscr{S}_1 e^{\frac{i (m_1+m_2)\pi}{k} + \theta}  - \mathscr{S}_2 }   \,,\quad &
    D_{12}^{\B\B}=&\, \frac{\mathscr{S}_1 e^{\frac{i m_1 \pi}{k}+\theta}-\mathscr{S}_2 e^{\frac{i m_2 \pi}{k}}}{\mathscr{S}_1 e^{\frac{i (m_1+m_2) \pi}{k}+\theta}-\mathscr{S}_2}   \,,\\
    E_{12}^{\B\B}=&\,\frac{i e^{\frac{i m_2 \pi}{k} + \frac{\theta}{2}}\sqrt{\mu(m_1) \mu(m_2)}}{\mathscr{S}_1 e^{\frac{i (m_1+m_2)\pi}{k} + \theta}  - \mathscr{S}_2 }   \,,\quad &
    F_{12}^{\B\B}=&\, \frac{-\mathscr{S}_1 e^{\theta}+\mathscr{S}_2 e^{\frac{i (m_1+m_2) \pi}{k}}}{\mathscr{S}_1 e^{\frac{i (m_1+m_2) \pi}{k}+\theta}-\mathscr{S}_2}    \,.
\end{aligned}
\end{equation}
Notice that this expression is not analytic in $m_1,m_2$ and depends on~$\mathscr{S}_1,\mathscr{S}_2$. It simplifies further when assuming a definite value for $\mathscr{S}_1,\mathscr{S}_2$. For instance, we have for $\mathscr{S}_1=\mathscr{S}_2=+1$
\begin{equation}
\label{S_matrix_coefficients_massive_massive_rel_limit}
\begin{aligned}
    A_{12}^{\B\B}=&\,1\,,\ &
    B_{12}^{\B\B}=&\, \frac{\sinh \Bigl(\frac{\theta}{2} - \frac{i \pi}{2 k} (m_1-m_2) \Bigr)}{\sinh \Bigl(\frac{\theta}{2} + \frac{i \pi}{2 k} (m_1+m_2) \Bigr)}   \,,\\
    C_{12}^{\B\B}=&\, \frac{i \sqrt{\mu(m_1) \mu(m_2)}}{2 \sinh \Bigl(\frac{\theta}{2} + \frac{i \pi}{2k} (m_1+m_2) \Bigr)} e^{\frac{i \pi}{2 k} (m_1-m_2)}   \,,\ &
    D_{12}^{\B\B}=&\, \frac{\sinh \Bigl(\frac{\theta}{2} + \frac{i \pi}{2 k} (m_1-m_2) \Bigr)}{\sinh \Bigl(\frac{\theta}{2} + \frac{i \pi}{2 k} (m_1+m_2) \Bigr)}   \,,\\
    E_{12}^{\B\B}=&\, \frac{i \sqrt{\mu(m_1) \mu(m_2)}}{ 2 \sinh \Bigl(\frac{\theta}{2} + \frac{i \pi}{2k} (m_1+m_2) \Bigr)} e^{-\frac{i \pi}{2 k} (m_1-m_2)}   \,,\quad &
    F_{12}^{\B\B}=&\, - \frac{\sinh \Bigl(\frac{\theta}{2} - \frac{i \pi}{2 k} (m_1+m_2) \Bigr)}{\sinh \Bigl(\frac{\theta}{2} + \frac{i \pi}{2 k} (m_1+m_2) \Bigr)}  \,,
\end{aligned}
\end{equation}
while for $\mathscr{S}_1=-\mathscr{S}_2=+1$ we find
\begin{equation}
\begin{aligned}
    A_{12}^{\B\B}=&\,1\,,\quad&
    B_{12}^{\B\B}=&\, \frac{\cosh \Bigl(\frac{\theta}{2} - \frac{i \pi}{2 k} (m_1-m_2) \Bigr)}{\cosh \Bigl(\frac{\theta}{2} + \frac{i \pi}{2 k} (m_1+m_2) \Bigr)}  \,,\\
    C_{12}^{\B\B}=&\, \frac{i \sqrt{\mu(m_1) \mu(m_2)}}{2 \cosh \Bigl(\frac{\theta}{2} + \frac{i \pi}{2k} (m_1+m_2) \Bigr)} e^{\frac{i \pi}{2 k} (m_1-m_2)}   \,,\quad&
    D_{12}^{\B\B}=&\, \frac{\cosh \Bigl(\frac{\theta}{2} + \frac{i \pi}{2 k} (m_1-m_2) \Bigr)}{\cosh \Bigl(\frac{\theta}{2} + \frac{i \pi}{2 k} (m_1+m_2) \Bigr)}   \,,\\
    E_{12}^{\B\B}=&\, \frac{i \sqrt{\mu(m_1) \mu(m_2)}}{ 2 \cosh \Bigl(\frac{\theta}{2} + \frac{i \pi}{2k} (m_1+m_2) \Bigr)} e^{-\frac{i \pi}{2 k} (m_1-m_2)}   \,,\quad&
    F_{12}^{\B\B}=&\, - \frac{\cosh \Bigl(\frac{\theta}{2} - \frac{i \pi}{2 k} (m_1+m_2) \Bigr)}{\cosh \Bigl(\frac{\theta}{2} + \frac{i \pi}{2 k} (m_1+m_2) \Bigr)}   \,.
\end{aligned}
\end{equation}

\paragraph{Different statistics.}
Were we to consider a different statistics for the highest-weight state we would expect to find similar minus signs on some matrix elements. In fact, because of the monodromy property~\eqref{eq:monodromy} of the two-particle representation, we have for the two-particle S~matrix
\begin{equation}
\label{eq:monodromySapp}
\begin{aligned}
    \mathbf{S}^{\B/\F, *}(m_1+k,\theta_1;m_2,\theta_2)&=\mathbf{S}^{\F/\B, *}(m_1,\theta_1;m_2,\theta_2)\,,\\
    \mathbf{S}^{*, \B/\F}(m_1,\theta_1;m_2+k,\theta_2)&=\mathbf{S}^{*, \F/\B}(m_1,\theta_1;m_2,\theta_2)\,,
\end{aligned}
\end{equation}
provided of course that the normalisation may be chosen appropriately. In other words, shifting $m_i$ by~$k$ is equivalent to flipping the statistics of the $i$-th particle.
Because this consideration relies only on the form of the coproduct that gives~\eqref{eq:monodromy}, it also applies to massless excitations.

\paragraph{Mixed-mass scattering.}
We may also consider the scattering of excitations of mixed-mass. In this case, the massless particle may be moving to the left or to the right. For a process to be physical, it is necessary to require that the particles are ordered so that for their velocities we have $v_1>v_2$. (More general processes can be considered to discuss unitarity, of course.)
With a slight abuse of notation let us set
\begin{equation}
    \mathscr{S}_i=\begin{cases}
        +1\,,&\qquad m_i=0~\text{mod}(2k),\\
        -1\,,&\qquad m_i=k~\text{mod}(2k).
    \end{cases}
\end{equation}
Using the same notation as in~\eqref{eq:scattering-structure_rel_limit} we find that
\begin{equation}
\begin{aligned}
    A_{12}^{\B_+\B}=&\,1\,,\quad &
    B_{12}^{\B_+\B}=&\, \mathscr{S}_1   \,,\quad&
    C_{12}^{\B_+\B}=&\,0   \,,\\
    D_{12}^{\B_+\B}=&\, e^{-\frac{i\pi m_2}{k}}   \,,\quad&
    E_{12}^{\B_+\B}=&\,0\,,\quad&
    F_{12}^{\B_+\B}=&\, -\mathscr{S}_1 e^{-\frac{i\pi m_2}{k}}  \,.
\end{aligned}
\end{equation}
and
\begin{equation}
\begin{aligned}
    A_{12}^{\B\B_-}=&\,1\,,\quad&
    B_{12}^{\B\B_-}=&\, e^{-\frac{i\pi m_1}{k}}   \,,\quad&
    C_{12}^{\B\B_-}=&\, 0  \,,\\
    D_{12}^{\B\B_-}=&\, \mathscr{S}_2  \,,\quad&
    E_{12}^{\B\B_-}=&\, 0 \,,\quad&
    F_{12}^{\B\B_-}=&\, -\mathscr{S}_2 e^{-\frac{i\pi m_1}{k}}   \,.
\end{aligned}
\end{equation}
where the plus and minus subscripts indicate the chirality of the massless particle. We see that the scattering is particularly simple, without any rotation in isotopic space.  The S-matrix elements of the inverse processes can be found by imposing braiding unitarity.

\paragraph{Massless scattering, opposite chirality}
In this case there is only one physical process due to the condition on the velocities, $v_1>v_2$. We find
\begin{equation}
\begin{aligned}
    A_{12}^{\B_+\B_-}=&\,1\,,\quad&
    B_{12}^{\B_+\B_-}=&\, \mathscr{S}_1   \,,\quad&
    C_{12}^{\B_+\B_-}=&\,0   \,,\\
    D_{12}^{\B_+\B_-}=&\,  \mathscr{S}_2   \,,\quad&
    E_{12}^{\B_+\B_-}=&\,0   \,,\quad&
    F_{12}^{\B_+\B_-}=&\, -\mathscr{S}_1 \mathscr{S}_2   \,.
\end{aligned}
\end{equation}

\paragraph{Massless scattering, same chirality.}
Let us now come to the case of two massless particles that have the same chirality. This is not a perturbative scattering process, as $v_1=v_2$, but it is very interesting to consider it nonetheless. 
By imposing the commutation with the supercharges we find several solutions. However, demanding unitarity, crossing symmetry, as well as that the Yang-Baxter equation is satisfied, we find that for all values of~$\mathscr{S}_1,\mathscr{S}_2$ there is a one-parameter family of solutions. The solution takes the form
\begin{equation}
\label{eq:masslessS_after_limit_appendix}
\begin{aligned}
    A_{12}^{\B_+\B_+}=&\,1\,,\\
    B_{12}^{\B_+\B_+}=&\, \mathscr{S}_1 \frac{ i+\mathscr{S}_2 \cot \frac{\alpha}{2} -e^\theta  (i+\mathscr{S}_1 \cot \frac{\alpha}{2})}{i + \mathscr{S}_2 \cot \frac{\alpha}{2} + e^\theta (i-\mathscr{S}_1 \cot \frac{\alpha}{2}) }   \,,\\
    C_{12}^{\B_+\B_+}=&\, \frac{2 i \mathscr{S}_1 \mathscr{S}_2 e^\frac{\theta}{2} }{i + \mathscr{S}_2 \cot \frac{\alpha}{2} + e^\theta (i-\mathscr{S}_1 \cot \frac{\alpha}{2}) }   \,,\\
    D_{12}^{\B_+\B_+}=&\, \frac{\cot \frac{\alpha}{2} - i \mathscr{S}_2 + \mathscr{S}_2 e^\theta (i - \mathscr{S}_1 \cot \frac{\alpha}{2}) }{i + \mathscr{S}_2 \cot \frac{\alpha}{2} + e^\theta (i-\mathscr{S}_1 \cot \frac{\alpha}{2}) }  \,,\\
    E_{12}^{\B_+\B_+}=&\, \frac{2 i e^{\frac{\theta}{2}}}{i + \mathscr{S}_2 \cot \frac{\alpha}{2}+e^\theta (i-\mathscr{S}_1 \cot \frac{\alpha}{2})}\,,\\
    F_{12}^{\B_+\B_+}=&\, \frac{\mathscr{S}_2 (i - \mathscr{S}_2 \cot \frac{\alpha}{2}) + \frac{i}{2} e^\theta (\mathscr{S}_1+\mathscr{S}_2)+ \frac{1}{2} \cot \frac{\alpha}{2} e^\theta (\mathscr{S}_1 \mathscr{S}_2+1)}{\mathscr{S}_1(i+\mathscr{S}_2 \cot \frac{\alpha}{2}) + \frac{i}{2} e^\theta (\mathscr{S}_1+\mathscr{S}_2)- \frac{1}{2} \cot \frac{\alpha}{2} e^\theta (\mathscr{S}_1 \mathscr{S}_2+1)}   \,.
\end{aligned}
\end{equation}
and it depends on a real parameter
\begin{equation}
    \alpha \in [0, 2 \pi]\,.
\end{equation}
Once again, flipping the signs~$\mathscr{S}_i$ is tantamount to swapping the statistics of the $i$-th~particle.

\subsection{Dressing factors and crossing equations}
\label{sec:relativistic:crossing_appendix}
In the previous subsections we have normalised all S-matrix elements~$A^{**}_{12}$ as
\begin{equation}
    A^{\B\B}(m_1,m_2;\theta_{12})=A^{\B_+\B}(m_1,m_2;\theta_{12})=\dots=A^{\B_+\B_+}(m_1,m_2;\theta_{12})=1\,.
\end{equation}
It is easy to imagine that this choice, while convenient, is not compatible with crossing. Indeed, let us introduce dressing factors for each block of the S~matrix so that
\begin{equation}
\label{eq:normalization_A_elements_appendix}
    A^{\B\B}(m_1,m_2;\theta_{12})=\sigma(m_1, m_2; \theta_{12})^{-1}\,,\quad\dots\,,\quad A^{\B_+\B_+}(m_1,m_2;\theta_{12})=\sigma(m_1^+, m_2^+; \theta_{12})^{-1}\,.
\end{equation}
The crossing equations will yield new constraints for the functions~$\sigma(m_1, m_2; \theta_{12})^{-1}$.

There are several ways to derive the crossing equations. One way which is particularly transparent physically is to construct an excitation $Z(m,\theta;m',\theta')$ which emerges from the tensor product of two of our representations, and is a \textit{singlet}  of the Zamolodchikov-Faddeev algebra of the theory (see \textit{e.g.}~\cite{Sfondrini:2014via} for a review). This means that the singlet has to be annihilated by all supercharges of the theory. Finally, we will require that it has bosonic statistics. Based on these requirements, consistency of the Zamolodchikov-Faddeev algebra indicates that the operator creating such a singlet must commute with all other ZF operators. This provides a way to derive the crossing equation.

Clearly the first step in this process is to determine  whether such a singlet exists at all.
The singlet representation of the algebra~\eqref{eq:lcalgebrasmall} is annihilated by all central charges. Hence it must be
\begin{equation}
    \genl{E}\,\left|Z(m,\theta;m',\theta')\right\rangle=
    \genl{M}\,\left|Z(m,\theta;m',\theta')\right\rangle=
    \genl{C}\,\left|Z(m,\theta;m',\theta')\right\rangle=0\,.
\end{equation}
The vanishing of the first two supercharges imposes that
\begin{equation}
    \theta'= \theta\pm i\pi\,,
\end{equation}
as we expect in order to obtain the crossing equations. The second imposes that
\begin{equation}
\label{eq:crossingm}
    m'=-m~\text{mod}k\,.
\end{equation}
This fact immediately implies that \textit{generically, particles of mass $m$ cannot be their own anti-particles}. Let us consider, for definiteness, a representation of mass $m$ with $0<m<k$ with bosonic highest-weight state. Eq.~\eqref{eq:crossingm} indicates that its antiparticles live in the representation with either $m'=-m$ or $m'=k-m$.%
\footnote{%
We are not discussing the cases $m'=2k-m$,  $m'=3k-m$, \textit{etc.}, because we have seen that all of our construction is trivially $2k$-periodic. 
} 
But which one is it? To answer this question, let us observe that the singlet must be constructed out of a linear combination of highest- and lowest-states, otherwise it cannot be annihilate by all supercharges. Schematically,
\begin{equation}
    \left|Z(m,\vartheta;m',\vartheta')\right\rangle=
    \left|\phi^*(m,\vartheta)\,\varphi^*(m',\vartheta')\right\rangle +
    c(m,m')\,
    \left|\varphi^*(m,\vartheta)\,\phi^*(m',\vartheta')\right\rangle\,,
\end{equation}
and an explicit computation yields the coefficient $c(m,m')$. Because we want $|Z\rangle$ to  behave as a boson when considering the scattering with a third particle, we need to consider the form of the coproduct on a three-particle state. From~\eqref{eq:coproductlimit} have that, schematically
\begin{equation}
\genl{q}_{(123)}=
    \genl{q}_{(1)}\otimes \genl{1}\otimes \genl{1}+
    e^{i\frac{\pi m_1}{k}}\,\Sigma\otimes\genl{q}_{(2)}\otimes \genl{1}+
    e^{i\frac{\pi (m_1+m_2)}{k}}\,\Sigma\otimes\Sigma\otimes\genl{q}_{(3)}\,,
\end{equation}
where the subscript indicates on which mass and rapidity the representation depends. If the first and second particle make up a singlet $|Z_{(12)}\rangle$ and the third particle is some generic $|X_{(3)}\rangle$, we have
\begin{equation}
    \genl{q}_{(123)}\big|Z_{(12)}\otimes X_{(3)}\big\rangle =  (-1)^{F_{(12)}}e^{i\frac{\pi(m+m')}{k}}\big|Z_{(12)}\otimes (\genl{q}_{(3)}X_{(3)})\big\rangle\,.
\end{equation}
For the singlet to have bosonic statistics we need that
\begin{equation}
    (-1)^{F_{(12)}}e^{i\frac{\pi(m+m')}{k}}=+1\,,
\end{equation}
where $(-1)^{F_{(12)}}$ is the naive fermion sign of the singlet's components.
This gives two possibilities in the case $0\leq m<k$ with bosonic highest weight (which we choose for definiteness):
\begin{enumerate}
    \item $m'=k-m$, so that $m+m'=k$. In this case, the highest-weight state of the two representations must have the same statistics (\textit{i.e.}, the ``prime'' representation must also have a bosonic a highest-weight state in this example) and
\begin{equation}
\label{eq:singlet_1_appendix}
    \left|Z\right\rangle=
    \left|\phi^{\B}(m,\vartheta)\,\varphi^{\F}(k-m,\vartheta')\right\rangle +
    c(m,k-m)\,
    \left|\varphi^{\F}(m,\vartheta)\,\phi^{\B}(k-m,\vartheta')\right\rangle,
\end{equation}
so that $F_{(12)}=+1$.
\item $m'=-m$, so that $m+m'=0$ instead. In this case we should take the opposite statistics, for the ``prime'' representations, which gives in this case
\begin{equation}
    \left|Z\right\rangle=
    \left|\phi^{\B}(m,\vartheta)\,\varphi^{\B}(-m,\vartheta')\right\rangle +
    c(m,-m)\,
    \left|\varphi^{\F}(m,\vartheta)\,\phi^{\F}(-m,\vartheta')\right\rangle,
\end{equation}
so that $F_{(12)}=0$.
This case is actually related to the previous due to the monodromy condition~\eqref{eq:monodromy}, see also~\eqref{eq:monodromySapp}. In fact, it yields the same crossing equations as it should.
\end{enumerate}
This discussion is perfectly compatible with the construction of the singlets before the limit, \textit{cf.}~\cite{Sfondrini:2014via}.

For massless particles, the two equivalent constructions of the singlet reduce to
\begin{equation}
\label{eq:massless_singlet_1_half_theory}
    \left|Z_0\right\rangle= \left|\phi^{\B}(0,\vartheta)\,\varphi^{\B}(0,\vartheta')\right\rangle +
    c(0,0)\,
    \left|\varphi^{\F}(0,\vartheta)\,\phi^{\F}(0,\vartheta')\right\rangle \,,
\end{equation}
for $m=0$ and to
\begin{equation}
\label{eq:massless_singlet_2_half_theory}
    \left|Z_k\right\rangle= \left|\phi^{\B}(k,\vartheta)\,\varphi^{\B}(-k,\vartheta')\right\rangle +
    c(k,-k)\,
    \left|\varphi^{\F}(k,\vartheta)\,\phi^{\F}(-k,\vartheta')\right\rangle
\end{equation}
for $m=k$.

Recall that the fundamental particles of the theory live in the tensor product of two representations of the relativistic limit of $su(1,1)^2_{c.e.}$, as shown in~\eqref{eq:fundamentalrepr};  for this reason, in the derivation of the crossing equations of the full model, we should consider the tensor product of two singlets of the form specified above.
Imposing that these singlets trivially commute with all the fundamental particles and their bound states we obtain the crossing equations in the limit.

\paragraph{Massive-massive and mixed-mass crossing equations.}
For the massive-massive and mixed-mass scattering taking the relativistic limit of the crossing equations of the full theory is equivalent to taking the limit of the theory first and then defining the crossing equations from scratch as explained above.
This is also the case for the scattering between massless particles with opposite chirality. 
This is expected because in all these cases the  the matrix part of the S~matrix is completely constrained after the limit. 
These crossing equations have been discussed in section~\ref{sec:relativistic}.

\paragraph{Same chirality massless crossing equations.}

The situation is different if we consider the scattering of massless particles with the same chirality: in this case, two among the four supercharges composing the algebra in~\eqref{eq:lcalgebrasmall} vanish (what supercharges depend on whether we consider the scattering between chiral-chiral particles or antichiral-antichiral particles) and the S~matrix remains partially unconstrained after the limit. This is clear by the fact that we obtain a one-parameter family of solutions for the S-matrix elements after the limit, as shown in~\eqref{eq:masslessS_after_limit_appendix}. 

In the following, we will consider the case where both massless particles are chiral (i.e. $M_1>0$ and $M_2>0$). The scattering of antichiral particles can be studied similarly. 
Let us consider the crossing equations for half representations first.
Normalising the S-matrix elements as in~\eqref{eq:normalization_A_elements_appendix} we find the following crossing equations
\begin{subequations}
\label{massless_crossing_equations}
\begin{align}
\label{massless_crossing_equation_1}
&\sigma(0^+, 0^+; \theta)^{-1} \sigma(0^+, k^+; \theta-i \pi)^{-1} = e^{-i \frac{\alpha}{2}}   \frac{\sinh \Bigl( \frac{\theta}{2} - i \frac{\alpha}{2} \Bigl)}{\sinh \frac{\theta}{2}} \, ,\\
\label{massless_crossing_equation_2}
&\sigma(0^+, k^+; \theta)^{-1} \sigma(0^+, 0^+; \theta-i \pi)^{-1} = e^{-i \frac{\alpha}{2}} \frac{\sinh \Bigl( \frac{\theta}{2} + i \frac{\pi}{2} \Bigl)}{\sinh \Bigl( \frac{\theta}{2} + \frac{i}{2} (\alpha+ \pi) \Bigl)}\, ,\\
\label{massless_crossing_equation_3}
&\sigma(k^+, 0^+; \theta)^{-1} \sigma(k^+, k^+; \theta-i \pi)^{-1} = e^{i \frac{\alpha}{2}} \frac{\sinh \Bigl( \frac{\theta}{2}+i \frac{\pi}{2} \Bigl)}{\sinh \Bigl( \frac{\theta}{2} - \frac{i}{2} (\alpha+\pi)  \Bigl)} \, ,\\
\label{massless_crossing_equation_4}
&\sigma(k^+, k^+; \theta)^{-1} \sigma(k^+, 0^+; \theta-i \pi)^{-1} = e^{i \frac{\alpha}{2}} \frac{\sinh \Bigl( \frac{\theta}{2}-\frac{i}{2} (2 \pi-\alpha) \Bigl)}{\sinh \frac{\theta}{2}} \,,
\end{align}
\end{subequations}
which are satisfied by
\begin{subequations}
\label{solutions_massless_crossing}
\begin{align}
&\sigma(0^+, 0^+; \theta)^{-1}  = - \frac{\sinh \Bigl( \frac{\theta}{2} - \frac{i \alpha}{2} \Bigl)}{\sinh \Bigl( \frac{\theta}{2} + \frac{i \alpha}{2} \Bigl)} \frac{R(\theta- i \alpha) R(\theta+ i  \alpha)}{R^2(\theta)} \,, \\
& \sigma(0^+, k^+; \theta)^{-1} = -e^{-i\frac{\alpha}{2}} \times  \frac{R(\theta+ i  \pi) R(\theta- i  \pi) }{R(\theta +i (\pi-\alpha)) R(\theta -i (\pi-\alpha))   } \,,\\
&\sigma(k^+, k^+; \theta)^{-1}  =   \frac{R(\theta- i \alpha) R(\theta+ i  \alpha)}{R^2(\theta)} \,,  \\
& \sigma(k^+, 0^+; \theta)^{-1} = -e^{+i\frac{\alpha}{2}} \times   \frac{R(\theta+ i  \pi) R(\theta- i  \pi) }{R(\theta +i (\pi-\alpha)) R(\theta -i (\pi-\alpha))   }  \,.
\end{align}
\end{subequations}
It is easy to check that all dressing factors written above satisfy both unitarity and braiding unitarity. Moreover, all the chiral-chiral massless S~matrices normalized with these factors have no poles in the physical strip $(0, i \pi)$. This is true for any value of $\alpha \in [0, 2 \pi]$.

If we are interested in finding the dressing factor of the full model, defined by the normalisation in~\eqref{chiral_chiral_normalization_massless_particles}, then we should consider the massless singlets for the full theory. These singlets can be constructed out of the half-theory singlets defined in~\eqref{eq:massless_singlet_1_half_theory} and~\eqref{eq:massless_singlet_2_half_theory}; they can be written as
\begin{equation}
\begin{split}
&| \Omega^{0 k} \rangle \simeq \left| {\color{red}Z_0} \, {\color{blue}Z_k}\right\rangle \in \Bigl({\color{red} \repr{B}{rel}(0, \theta)} \otimes {\color{blue} \repr{B}{rel}(k, \theta)} \Bigl) \otimes \Bigl({\color{red}\repr{B}{rel}(k, \theta + i \pi)} \otimes {\color{blue}\repr{B}{rel}(0, \theta+i \pi)} \Bigl) \\
&| \Omega^{k 0} \rangle  \simeq \left| {\color{blue}Z_k} \, {\color{red}Z_0}\right\rangle \in \Bigl({\color{blue} \repr{B}{rel}(k, \theta)} \otimes {\color{red} \repr{B}{rel}(0, \theta)} \Bigl) \otimes \Bigl({\color{blue}\repr{B}{rel}(0, \theta + i \pi)} \otimes {\color{red}\repr{B}{rel}(k, \theta+i \pi)} \Bigl) \,.
\end{split}
\end{equation}
The different colours show how the singlets split between the different representations of the full algebra. Requiring that these singlets scatter trivially with all massless particles in the representations $\repr{B}{rel}(0, \theta) \otimes \repr{B}{rel}(k, \theta) \simeq \repr{B}{rel}(0, \theta) \otimes \repr{F}{rel}(0, \theta)$ and $\repr{B}{rel}(k, \theta) \otimes \repr{B}{rel}(0, \theta)\simeq \repr{F}{rel}(0, \theta) \otimes \repr{B}{rel}(0, \theta)$ we obtain the following crossing equations
\begin{equation}
\label{eq:first_alpha_dependent_crossing_massless}
\begin{split}
    \Bigl( \sigma^{\circ \circ}(\theta+i \pi) \Bigl)^{-2} \Bigl( \sigma^{\circ \circ}(\theta) \Bigl)^{-2}=\frac{\cosh \Bigl( \frac{\theta}{2} - i \frac{\alpha}{2} \Bigl) \cosh \Bigl( \frac{\theta}{2} + i \frac{\alpha}{2} \Bigl)}{\cosh^2{\frac{\theta}{2}}} \,,
\end{split}
\end{equation}
\begin{equation}
\label{eq:second_alpha_dependent_crossing_massless}
\begin{split}
    \Bigl( \sigma^{\circ \circ}(\theta) \Bigl)^{-2} \Bigl( \sigma^{\circ \circ}(\theta+i \pi) \Bigl)^{-2}=\frac{\sinh^2{\frac{\theta}{2}}}{\sinh \Bigl( \frac{\theta}{2} - i \frac{\alpha}{2} \Bigl) \sinh \Bigl( \frac{\theta}{2} + i \frac{\alpha}{2} \Bigl)} \,.
\end{split}
\end{equation}
Equations~\eqref{eq:first_alpha_dependent_crossing_massless} and~\eqref{eq:second_alpha_dependent_crossing_massless} are obtained by multiplying~\eqref{massless_crossing_equation_1} and~\eqref{massless_crossing_equation_4}, and \eqref{massless_crossing_equation_2} and~\eqref{massless_crossing_equation_3} respectively, 
and noting that the normalisation~\eqref{chiral_chiral_normalization_massless_particles} requires to set
\begin{equation}
\Bigl(\sigma^{\circ \circ}(\theta) \Bigl)^{-2}=\sigma(0^+, 0^+; \theta)^{-1}  \sigma(k^+, k^+; \theta)^{-1} = - \sigma(0^+, k^+; \theta)^{-1} \sigma(k^+, 0^+; \theta)^{-1} \,.
\end{equation}
The minus sign in the second equality above is necessary to take into account fermionic exchanges in the full model. Note indeed that particles in the representations $\rho^{\B}(0, \theta_1) \otimes \rho^{\F}(0, \theta_1)$ and $\rho^{\F}(0, \theta_2) \otimes \rho^{\B}(0, \theta_2)$ are fermions and their scattering produces a minus sign which is not taken into account in the scattering between particles in half representations.

If we assume that the dressing factor is the same for the scattering between all the massless representations, we obtain an overconstrained system of equations. To solve this system we need to require the RHS of~\eqref{eq:first_alpha_dependent_crossing_massless} and~\eqref{eq:second_alpha_dependent_crossing_massless} to be the same.
This is possible only if $\alpha=0$, $\alpha=2 \pi$ or $\alpha=\pi$.
The point $\alpha=\pi$ corresponds to the limit of the full theory and the solution is given in~\eqref{eq:solutions_massless_crossing_limit_from_original_theory}.
On the other hand, if $\alpha=0$ or $\alpha= 2 \pi$ all the massless-massless dressing factors can be set equal to $1$ and the whole S~matrix is in fact trivial.
It is worth remarking that the conclusion about the allowed values of~$\alpha$ would not have changed even if we had allowed for a non-trivial rotation in the $su(2)_\circ$ space, as that is factorised with respect to the internal $su(1|1)$~structure.

\bibliographystyle{JHEP}
\bibliography{refs}

\makeatletter \@ifundefined{Sphere}{\newcommand{\Sphere}{\text{S}}}{}
  \@ifundefined{AdS}{\newcommand{\AdS}{\text{AdS}}}{}
  \@ifundefined{CFT}{\newcommand{\CFT}{\text{CFT}}}{}
  \@ifundefined{CP}{\newcommand{\CP}{\text{CP}}}{}
  \@ifundefined{Torus}{\newcommand{\Torus}{\text{T}}}{}
  \@ifundefined{superN}{\newcommand{\superN}{\mathcal{N}}}{}
  \@ifundefined{grpOSp}{\newcommand{\grpOSp}{\mathrm{OSp}}}{}
  \@ifundefined{grpPSU}{\newcommand{\grpPSU}{\mathrm{PSU}}}{}
  \@ifundefined{grpSU}{\newcommand{\grpSU}{\mathrm{SU}}}{}
  \@ifundefined{grpU}{\newcommand{\grpU}{\mathrm{U}}}{}
  \@ifundefined{grpD}{\newcommand{\grpD}{\mathrm{D}}}{}
  \@ifundefined{grpSL}{\newcommand{\grpSL}{\mathrm{SL}}}{}
  \@ifundefined{grpSp}{\newcommand{\grpSp}{\mathrm{Sp}}}{}
  \@ifundefined{grpUSp}{\newcommand{\grpUSp}{\mathrm{USp}}}{}
  \@ifundefined{grpSO}{\newcommand{\grpSO}{\mathrm{SO}}}{}
  \@ifundefined{grpO}{\newcommand{\grpO}{\mathrm{O}}}{}
  \@ifundefined{algOSp}{\newcommand{\algOSp}{\mathfrak{osp}}}{}
  \@ifundefined{algPSU}{\newcommand{\algPSU}{\mathfrak{psu}}}{}
  \@ifundefined{algSU}{\newcommand{\algSU}{\mathfrak{su}}}{}
  \@ifundefined{algSp}{\newcommand{\algSp}{\mathfrak{sp}}}{}
  \@ifundefined{algSL}{\newcommand{\algSL}{\mathfrak{sl}}}{}
  \@ifundefined{algGL}{\newcommand{\algGL}{\mathfrak{gl}}}{}
  \@ifundefined{algU}{\newcommand{\algU}{\mathfrak{u}}}{}
  \@ifundefined{algSO}{\newcommand{\algSO}{\mathfrak{so}}}{}
  \@ifundefined{algO}{\newcommand{\algO}{\mathfrak{o}}}{}
  \@ifundefined{Integers}{\newcommand{\Integers}{\mathbb{Z}}}{}
  \@ifundefined{Reals}{\newcommand{\Reals}{\mathbb{R}}}{} \makeatother

\providecommand{\href}[2]{#2}\begingroup\raggedright\begin{thebibliography}{10}

\bibitem{Maldacena:1997re}
J.M.~Maldacena, \emph{The large {N} limit of superconformal field theories and
  supergravity}, {\emph{Adv. Theor. Math. Phys.} {\bfseries 2} (1998) 231}
  [\href{https://arxiv.org/abs/hep-th/9711200}{{\ttfamily hep-th/9711200}}].

\bibitem{Larsen:1999uk}
F.~Larsen and E.J.~Martinec, \emph{{$\grpU(1)$} charges and moduli in the
  {D1}-{D5} system}, {\emph{JHEP} {\bfseries 9906} (1999) 019}
  [\href{https://arxiv.org/abs/hep-th/9905064}{{\ttfamily hep-th/9905064}}].

\bibitem{OhlssonSax:2018hgc}
O.~Ohlsson~Sax and B.~Stefa{\'n}ski, \emph{{Closed strings and moduli in
  AdS$_{3}$/CFT$_{2}$}},
  \href{https://doi.org/10.1007/JHEP05(2018)101}{\emph{JHEP} {\bfseries 05}
  (2018) 101} [\href{https://arxiv.org/abs/1804.02023}{{\ttfamily
  1804.02023}}].

\bibitem{Maldacena:2000hw}
J.M.~Maldacena and H.~Ooguri, \emph{Strings in {$\AdS_{3}$} and {$\grpSL(2,R)$}
  {WZW} model. {I}}, \href{https://doi.org/10.1063/1.1377273}{\emph{J. Math.
  Phys.} {\bfseries 42} (2001) 2929}
  [\href{https://arxiv.org/abs/hep-th/0001053}{{\ttfamily hep-th/0001053}}].

\bibitem{Giribet:2018ada}
G.~Giribet, C.~Hull, M.~Kleban, M.~Porrati and E.~Rabinovici,
  \emph{{Superstrings on AdS$_{3}$ at $k = 1$}},
  \href{https://doi.org/10.1007/JHEP08(2018)204}{\emph{JHEP} {\bfseries 08}
  (2018) 204} [\href{https://arxiv.org/abs/1803.04420}{{\ttfamily
  1803.04420}}].

\bibitem{Eberhardt:2018ouy}
L.~Eberhardt, M.R.~Gaberdiel and R.~Gopakumar, \emph{{The Worldsheet Dual of
  the Symmetric Product CFT}},
  \href{https://doi.org/10.1007/JHEP04(2019)103}{\emph{JHEP} {\bfseries 04}
  (2019) 103} [\href{https://arxiv.org/abs/1812.01007}{{\ttfamily
  1812.01007}}].

\bibitem{Eberhardt:2021vsx}
L.~Eberhardt, \emph{{A perturbative CFT dual for pure NS\textendash{}NS
  AdS$_{3}$ strings}}, \href{https://doi.org/10.1088/1751-8121/ac47b2}{\emph{J.
  Phys. A} {\bfseries 55} (2022) 064001}
  [\href{https://arxiv.org/abs/2110.07535}{{\ttfamily 2110.07535}}].

\bibitem{Cagnazzo:2012se}
A.~Cagnazzo and K.~Zarembo, \emph{{B}-field in {$\AdS_{3}/\CFT_{2}$}
  correspondence and integrability},
  \href{https://doi.org/10.1007/JHEP11(2012)133,
  10.1007/JHEP04(2013)003}{\emph{JHEP} {\bfseries 1211} (2012) 133}
  [\href{https://arxiv.org/abs/1209.4049}{{\ttfamily 1209.4049}}].

\bibitem{Sfondrini:2014via}
A.~Sfondrini, \emph{Towards integrability for {$\AdS_{3}/\CFT_{2}$}},
  \href{https://doi.org/10.1088/1751-8113/48/2/023001}{\emph{J. Phys.}
  {\bfseries A48} (2015) 023001}
  [\href{https://arxiv.org/abs/1406.2971}{{\ttfamily 1406.2971}}].

\bibitem{Arutyunov:2009ga}
G.~Arutyunov and S.~Frolov, \emph{Foundations of the {$\AdS_{5} \times
  \Sphere^5$} superstring. part {I}},
  \href{https://doi.org/10.1088/1751-8113/42/25/254003}{\emph{J. Phys. A}
  {\bfseries A42} (2009) 254003}
  [\href{https://arxiv.org/abs/0901.4937}{{\ttfamily 0901.4937}}].

\bibitem{Beisert:2010jr}
N.~Beisert, C.~Ahn, L.F.~Alday, Z.~Bajnok, J.M.~Drummond, L.~Freyhult et~al.,
  \emph{{Review of AdS/CFT Integrability: An Overview}},
  \href{https://doi.org/10.1007/s11005-011-0529-2}{\emph{Lett. Math. Phys.}
  {\bfseries 99} (2012) 3} [\href{https://arxiv.org/abs/1012.3982}{{\ttfamily
  1012.3982}}].

\bibitem{Hoare:2013ida}
B.~Hoare and A.~Tseytlin, \emph{Massive {S}-matrix of {$\AdS_{3} \times
  \Sphere^3 \times \Torus^4$} superstring theory with mixed 3-form flux},
  \href{https://doi.org/10.1016/j.nuclphysb.2013.04.024}{\emph{Nucl. Phys.}
  {\bfseries B873} (2013) 395}
  [\href{https://arxiv.org/abs/1304.4099}{{\ttfamily 1304.4099}}].

\bibitem{Lloyd:2014bsa}
T.~Lloyd, O.~Ohlsson~Sax, A.~Sfondrini and B.~Stefa{\'n}ski, jr., \emph{The
  complete worldsheet {S} matrix of superstrings on {$\AdS_{3} \times \Sphere^3
  \times \Torus^4$} with mixed three-form flux},
  \href{https://doi.org/10.1016/j.nuclphysb.2014.12.019}{\emph{Nucl. Phys.}
  {\bfseries B891} (2015) 570}
  [\href{https://arxiv.org/abs/1410.0866}{{\ttfamily 1410.0866}}].

\bibitem{Hoare:2013pma}
B.~Hoare and A.A.~Tseytlin, \emph{On string theory on {$\AdS_{3} \times
  \Sphere^3 \times \Torus^4$} with mixed 3-form flux: tree-level {S}-matrix},
  \href{https://doi.org/10.1016/j.nuclphysb.2013.05.005}{\emph{Nucl. Phys.}
  {\bfseries B873} (2013) 682}
  [\href{https://arxiv.org/abs/1303.1037}{{\ttfamily 1303.1037}}].

\bibitem{Hoare:2013lja}
B.~Hoare, A.~Stepanchuk and A.~Tseytlin, \emph{Giant magnon solution and
  dispersion relation in string theory in {$\AdS_{3} \times \Sphere^3 \times
  \Torus^4$} with mixed flux},
  \href{https://doi.org/10.1016/j.nuclphysb.2013.12.011}{\emph{Nucl. Phys.}
  {\bfseries B879} (2014) 318}
  [\href{https://arxiv.org/abs/1311.1794}{{\ttfamily 1311.1794}}].

\bibitem{Arutyunov:2004vx}
G.~Arutyunov, S.~Frolov and M.~Staudacher, \emph{{B}ethe ansatz for quantum
  strings}, \href{https://doi.org/10.1088/1126-6708/2004/10/016}{\emph{JHEP}
  {\bfseries 0410} (2004) 016}
  [\href{https://arxiv.org/abs/hep-th/0406256}{{\ttfamily hep-th/0406256}}].

\bibitem{Janik:2006dc}
R.A.~Janik, \emph{The {$\AdS_{5} \times \Sphere^5$} superstring worldsheet
  {S}-matrix and crossing symmetry},
  \href{https://doi.org/10.1103/PhysRevD.73.086006}{\emph{Phys. Rev.}
  {\bfseries D73} (2006) 086006}
  [\href{https://arxiv.org/abs/hep-th/0603038}{{\ttfamily hep-th/0603038}}].

\bibitem{Beisert:2006ez}
N.~Beisert, B.~Eden and M.~Staudacher, \emph{Transcendentality and crossing},
  \href{https://doi.org/10.1088/1742-5468/2007/01/P01021}{\emph{J. Stat. Mech.}
  {\bfseries 0701} (2007) P01021}
  [\href{https://arxiv.org/abs/hep-th/0610251}{{\ttfamily hep-th/0610251}}].

\bibitem{Frolov:2021fmj}
S.~Frolov and A.~Sfondrini, \emph{{New dressing factors for AdS3/CFT2}},
  \href{https://doi.org/10.1007/JHEP04(2022)162}{\emph{JHEP} {\bfseries 04}
  (2022) 162} [\href{https://arxiv.org/abs/2112.08896}{{\ttfamily
  2112.08896}}].

\bibitem{Baggio:2018gct}
M.~Baggio and A.~Sfondrini, \emph{{Strings on NS-NS Backgrounds as Integrable
  Deformations}}, \href{https://doi.org/10.1103/PhysRevD.98.021902}{\emph{Phys.
  Rev.} {\bfseries D98} (2018) 021902}
  [\href{https://arxiv.org/abs/1804.01998}{{\ttfamily 1804.01998}}].

\bibitem{Fontanella:2019ury}
A.~Fontanella, O.~Ohlsson~Sax, B.~Stefa\'nski and A.~Torrielli, \emph{{The
  effectiveness of relativistic invariance in AdS$_{3}$}},
  \href{https://doi.org/10.1007/JHEP07(2019)105}{\emph{JHEP} {\bfseries 07}
  (2019) 105} [\href{https://arxiv.org/abs/1905.00757}{{\ttfamily
  1905.00757}}].

\bibitem{Fendley:1991ve}
P.~Fendley and K.A.~Intriligator, \emph{{Scattering and thermodynamics of
  fractionally charged supersymmetric solitons}},
  \href{https://doi.org/10.1016/0550-3213(92)90365-I}{\emph{Nucl. Phys. B}
  {\bfseries 372} (1992) 533}
  [\href{https://arxiv.org/abs/hep-th/9111014}{{\ttfamily hep-th/9111014}}].

\bibitem{Fendley:1992dm}
P.~Fendley and K.A.~Intriligator, \emph{{Scattering and thermodynamics in
  integrable N=2 theories}},
  \href{https://doi.org/10.1016/0550-3213(92)90523-E}{\emph{Nucl. Phys. B}
  {\bfseries 380} (1992) 265}
  [\href{https://arxiv.org/abs/hep-th/9202011}{{\ttfamily hep-th/9202011}}].

\bibitem{Eden:2021xhe}
B.~Eden, D.l.~Plat and A.~Sfondrini, \emph{{Integrable bootstrap for
  AdS$_{3}$/CFT$_{2}$ correlation functions}},
  \href{https://doi.org/10.1007/JHEP08(2021)049}{\emph{JHEP} {\bfseries 08}
  (2021) 049} [\href{https://arxiv.org/abs/2102.08365}{{\ttfamily
  2102.08365}}].

\bibitem{Berenstein:2002jq}
D.E.~Berenstein, J.M.~Maldacena and H.S.~Nastase, \emph{Strings in flat space
  and {pp} waves from {$\superN = 4$} super {Y}ang {M}ills}, {\emph{JHEP}
  {\bfseries 0204} (2002) 013}
  [\href{https://arxiv.org/abs/hep-th/0202021}{{\ttfamily hep-th/0202021}}].

\bibitem{Borsato:2013qpa}
R.~Borsato, O.~Ohlsson~Sax, A.~Sfondrini, B.~Stefa{\'n}ski, jr. and
  A.~Torrielli, \emph{The all-loop integrable spin-chain for strings on
  {$\AdS_{3} \times \Sphere^3 \times \Torus^4$}: the massive sector},
  \href{https://doi.org/10.1007/JHEP08(2013)043}{\emph{JHEP} {\bfseries 1308}
  (2013) 043} [\href{https://arxiv.org/abs/1303.5995}{{\ttfamily 1303.5995}}].

\bibitem{Beisert:2005tm}
N.~Beisert, \emph{The {$\algSU(2|2)$} dynamic {$S$}-matrix}, {\emph{Adv. Theor.
  Math. Phys.} {\bfseries 12} (2008) 945}
  [\href{https://arxiv.org/abs/hep-th/0511082}{{\ttfamily hep-th/0511082}}].

\bibitem{Arutyunov:2006ak}
G.~Arutyunov, S.~Frolov, J.~Plefka and M.~Zamaklar, \emph{The off-shell
  symmetry algebra of the light-cone {$\AdS_{5} \times \Sphere^5$}
  superstring}, \href{https://doi.org/10.1088/1751-8113/40/13/018}{\emph{J.
  Phys.} {\bfseries A40} (2007) 3583}
  [\href{https://arxiv.org/abs/hep-th/0609157}{{\ttfamily hep-th/0609157}}].

\bibitem{Borsato:2012ud}
R.~Borsato, O.~Ohlsson~Sax and A.~Sfondrini, \emph{A dynamic {$\algSU(1|1)^2$}
  {S}-matrix for {$\AdS_{3}/\CFT_{2}$}},
  \href{https://doi.org/10.1007/JHEP04(2013)113}{\emph{JHEP} {\bfseries 1304}
  (2013) 113} [\href{https://arxiv.org/abs/1211.5119}{{\ttfamily 1211.5119}}].

\bibitem{Sfondrini:2020ovj}
A.~Sfondrini, \emph{{Long strings and symmetric product orbifold from the
  AdS$_3$Bethe equations}},
  \href{https://doi.org/10.1209/0295-5075/133/10004}{\emph{EPL} {\bfseries 133}
  (2021) 10004} [\href{https://arxiv.org/abs/2010.02782}{{\ttfamily
  2010.02782}}].

\bibitem{Borsato:2013hoa}
R.~Borsato, O.~Ohlsson~Sax, A.~Sfondrini, B.~Stefa{\'n}ski, jr. and
  A.~Torrielli, \emph{Dressing phases of {$\AdS_{3}/\CFT_{2}$}},
  \href{https://doi.org/10.1103/PhysRevD.88.066004}{\emph{Phys. Rev.}
  {\bfseries D88} (2013) 066004}
  [\href{https://arxiv.org/abs/1306.2512}{{\ttfamily 1306.2512}}].

\bibitem{Seibold:2022mgg}
F.K.~Seibold and A.~Sfondrini, \emph{{Transfer matrices for AdS3/CFT2}},
  \href{https://doi.org/10.1007/JHEP05(2022)089}{\emph{JHEP} {\bfseries 05}
  (2022) 089} [\href{https://arxiv.org/abs/2202.11058}{{\ttfamily
  2202.11058}}].

\bibitem{Frolov:2021zyc}
S.~Frolov and A.~Sfondrini, \emph{{Massless S matrices for AdS3/CFT2}},
  \href{https://doi.org/10.1007/JHEP04(2022)067}{\emph{JHEP} {\bfseries 04}
  (2022) 067} [\href{https://arxiv.org/abs/2112.08895}{{\ttfamily
  2112.08895}}].

\bibitem{Braden:1989bu}
H.W.~Braden, E.~Corrigan, P.E.~Dorey and R.~Sasaki, \emph{{Affine Toda Field
  Theory and Exact S Matrices}},
  \href{https://doi.org/10.1016/0550-3213(90)90648-W}{\emph{Nucl. Phys. B}
  {\bfseries 338} (1990) 689}.

\bibitem{Arinshtein:1979pb}
A.E.~Arinshtein, V.A.~Fateev and A.B.~Zamolodchikov, \emph{{Quantum s Matrix of
  the (1+1)-Dimensional Todd Chain}},
  \href{https://doi.org/10.1016/0370-2693(79)90561-6}{\emph{Phys. Lett. B}
  {\bfseries 87} (1979) 389}.

\bibitem{Frolov:2021bwp}
S.~Frolov and A.~Sfondrini, \emph{{Mirror thermodynamic Bethe ansatz for
  AdS3/CFT2}}, \href{https://doi.org/10.1007/JHEP03(2022)138}{\emph{JHEP}
  {\bfseries 03} (2022) 138}
  [\href{https://arxiv.org/abs/2112.08898}{{\ttfamily 2112.08898}}].

\bibitem{Brollo:2023pkl}
A.~Brollo, D.~le~Plat, A.~Sfondrini and R.~Suzuki, \emph{{The Tensionless Limit
  of Pure-Ramond-Ramond AdS3/CFT2}},
  \href{https://arxiv.org/abs/2303.02120}{{\ttfamily 2303.02120}}.

\bibitem{Frolov:2023wji}
S.~Frolov, A.~Pribytok and A.~Sfondrini, \emph{{Ground state energy of twisted
  $AdS_{3}\times S^{3}\times T^{4}$ superstring and the TBA}},
  \href{https://arxiv.org/abs/2305.17128}{{\ttfamily 2305.17128}}.

\bibitem{Dei:2018mfl}
A.~Dei and A.~Sfondrini, \emph{{Integrable spin chain for stringy
  Wess-Zumino-Witten models}},
  \href{https://doi.org/10.1007/JHEP07(2018)109}{\emph{JHEP} {\bfseries 07}
  (2018) 109} [\href{https://arxiv.org/abs/1806.00422}{{\ttfamily
  1806.00422}}].

\bibitem{Ekhammar:2021pys}
S.~Ekhammar and D.~Volin, \emph{{Monodromy bootstrap for SU(2|2) quantum
  spectral curves: from Hubbard model to AdS$_{3}$/CFT$_{2}$}},
  \href{https://doi.org/10.1007/JHEP03(2022)192}{\emph{JHEP} {\bfseries 03}
  (2022) 192} [\href{https://arxiv.org/abs/2109.06164}{{\ttfamily
  2109.06164}}].

\bibitem{Cavaglia:2021eqr}
A.~Cavagli\`a, N.~Gromov, B.~Stefa\'nski, Jr., Jr. and A.~Torrielli,
  \emph{{Quantum Spectral Curve for AdS$_{3}$/CFT$_{2}$: a proposal}},
  \href{https://doi.org/10.1007/JHEP12(2021)048}{\emph{JHEP} {\bfseries 12}
  (2021) 048} [\href{https://arxiv.org/abs/2109.05500}{{\ttfamily
  2109.05500}}].

\bibitem{Cavaglia:2022xld}
A.~Cavagli\`a, S.~Ekhammar, N.~Gromov and P.~Ryan, \emph{{Exploring the Quantum
  Spectral Curve for AdS${}_3$/CFT${}_2$}},
  \href{https://arxiv.org/abs/2211.07810}{{\ttfamily 2211.07810}}.

\end{thebibliography}\endgroup

\end{document}